\documentclass[12pt]{article}
\usepackage{color}
\usepackage{amsmath} 
\usepackage{amssymb} 
\usepackage[english]{babel}
\usepackage{graphics}
\usepackage[pdftex]{graphicx}
\usepackage{xspace}
\usepackage[gen]{eurosym}
\usepackage[font=footnotesize,format=plain,labelfont=bf,up,textfont=it,up]{caption}
\usepackage[left]{lineno}
\newcommand{\nue}{\ensuremath{\nu_{e}}\xspace}
\newcommand{\nubare}{\ensuremath{\overline{\nu}_{e}}\xspace}

\newcommand{\numu}{\ensuremath{\nu_{\mu}}\xspace}

\newcommand{\nubarmu}{\ensuremath{\overline{\nu}_{\mu}}\xspace}

\newcommand{\nutau}{\ensuremath{\nu_{\tau}}\xspace}
\newcommand{\numunue}{\ensuremath{\nu_\mu \rightarrow \nu_e}\xspace}

\newcommand{\nuenue}{\ensuremath{\nue \rightarrow \nue}\xspace}
\newcommand{\pnuenue}{\ensuremath{P(\nue \rightarrow \nue)}\xspace}

\newcommand{\pnumunumu}{\ensuremath{P(\nu_\mu \rightarrow \nu_\mu)}\xspace}

\newcommand{\numunutau}{\ensuremath{\numu \rightarrow \nutau}\xspace}

\newcommand{\pnumunue}{\ensuremath{P(\nu_\mu \rightarrow \nu_e)}\xspace}

\newcommand{\nubarmunubare}{\ensuremath{\overline{\nu}_\mu \rightarrow \overline{\nu}_e}\xspace}

\newcommand{\boss}[2]{\ensuremath{\rlap{\kern-2.5pt\ensuremath{\overset{\scriptscriptstyle(-)}{\phantom{#1}}}}{\ensuremath{{#1}_{#2}}}}}

\newcommand{\BARII}        {1}
\newcommand{\BARIU}        {2}
\newcommand{\BOLOGNAI}     {3}
\newcommand{\BOLOGNAU}     {4}
\newcommand{\FRASCATI}    {5}
\newcommand{\GENOVAI}    {6}
\newcommand{\LECCEI}        {7}
\newcommand{\LECCEU}        {8}
\newcommand{\LECCEUU}        {9}
\newcommand{\PADOVAI}     {10}
\newcommand{\PADOVAU}     {11}
\newcommand{\ROMAU}       {12}
\newcommand{\SALERNOU}    {13}
\newcommand{\CONTACT}    {a}
\newcommand{\NessieInstitutes}{
\BARII        . INFN-Bari, 70126 Bari, Italy \\
\BARIU        . Dipartimento di Fisica dell'Universit\`a  di Bari, 70126 Bari, Italy \\
\BOLOGNAI     . INFN-Bologna, 40127 Bologna, Italy \\
\BOLOGNAU     . Dipartimento di Fisica dell'Universit\`a  di Bologna, 40127 Bologna, Italy \\
\FRASCATI    . Laboratori Nazionali di Frascati dell'INFN, 00044 Frascati (Roma), Italy \\
\GENOVAI        . INFN-Genova, 16146 Genova, Italy \\
\LECCEI        . INFN-Lecce, 70126 Lecce, Italy \\
\LECCEU        . Dipartimento di Fisica dell'Universit\`a  del Salento, 73100 Lecce, Italy \\
\LECCEUU     . Dipartimento di Ingegneria dell'Innovazione dellÕUniversit\`a del Salento, 73100 Lecce, Italy\\
\PADOVAI      . INFN-Padova, 35131 Padova, Italy \\
\PADOVAU      . Dipartimento di Fisica dell'Universit\`a  di Padova, 35131 Padova, Italy \\
\ROMAU        . Dipartimento di Fisica dell'Universit\`a  di Roma ``La Sapienza" and INFN, 00185 Roma, Italy \\
\SALERNOU     . Dipartimento di Fisica dell'Universit\`a  di Salerno and INFN, 84084 Fisciano, Salerno, Italy \\
\CONTACT \, {\em Contact Person}\\
}
\newcommand{\NessieAuthorList}{
P.~Bernardini$^{\LECCEU, \LECCEI}$,
A.~Bertolin$^{\PADOVAI}$,
C.~Bozza$^{\SALERNOU}$,
R.~Brugnera$^{\PADOVAU, \PADOVAI}$,
A.~Cecchetti$^{\FRASCATI}$,
S.~Cecchini$^{\BOLOGNAI}$,
G.~Collazuol$^{\PADOVAU, \PADOVAI}$,
F.~Dal~Corso$^{\PADOVAI}$,
I.~De~Mitri$^{\LECCEU, \LECCEI}$
M.~De~Serio$^{\BARII}$,
D.~Di~Ferdinando$^{\BOLOGNAI}$,
U.~Dore$^{\ROMAU}$,
S.~Dusini$^{\PADOVAI}$,
P.~Fabbricatore$^{\GENOVAI}$,
C.~Fanin$^{\PADOVAI}$,
R.~A.~Fini$^{\BARII}$,
A.~Garfagnini$^{\PADOVAU, \PADOVAI}$,
G.~Grella$^{\SALERNOU}$,
U.~Kose$^{\PADOVAI}$,
M.~Laveder$^{\PADOVAU, \PADOVAI}$,
P.~Loverre$^{\ROMAU}$,
A.~Longhin$^{\FRASCATI}$,
G.~Marsella$^{\LECCEUU, \LECCEI}$
G.~Mancarella$^{\LECCEU, \LECCEI}$
G.~Mandrioli$^{\BOLOGNAI}$,
N.~Mauri$^{\FRASCATI}$,
E.~Medinaceli$^{\PADOVAU, \PADOVAI}$,
M.~Mezzetto$^{\PADOVAI}$,\\
M.~T.~Muciaccia$^{\BARIU, \BARII}$,
D.~Orecchini$^{\FRASCATI}$,
A.~Paoloni$^{\FRASCATI}$,
A.~Pastore$^{\BARIU, \BARII}$,
L.~Patrizii$^{\BOLOGNAI}$,
M.~Pozzato$^{\BOLOGNAU, \BOLOGNAI}$,
R.~Rescigno$^{\SALERNOU}$,
G.~Rosa$^{\ROMAU}$,
S.~Simone$^{\BARIU, \BARII}$,
M.~Sioli$^{\BOLOGNAU, \BOLOGNAI}$,
G.~Sirri$^{\BOLOGNAI}$,
M.~Spurio$^{\BOLOGNAU, \BOLOGNAI}$,
L.~Stanco$^{\PADOVAI, \CONTACT}$,
S.~Stellacci$^{\SALERNOU}$,
A.~Surdo$^{\LECCEI}$,
M.~Tenti$^{\BOLOGNAU, \BOLOGNAI}$,
and
V.~Togo$^{\BOLOGNAI}$.\\
}
  
\begin{document}


\renewcommand{\thefootnote}{\alph{footnote}}
  \thispagestyle{empty}
\title{\bf Prospect for Charge Current Neutrino Interactions Measurements at the CERN-PS\footnote{Also referenced as
CERN-SPSC-2011-030 and SPSC-P-343.}}
\date{}
\maketitle

\vspace{0.5 cm}

\centerline{\em The NESSiE Collaboration}

\author{\noindent \\ \NessieAuthorList }

\begin{flushleft}
\footnotesize{\NessieInstitutes }
\end{flushleft}

\newpage
\tableofcontents
  
\newpage

\section{Introduction}

Tensions in several phenomenological models grew with experimental results on neutrino/antineutrino oscillations 
at Short-Baseline (SBL) and with the recent, carefully recomputed, antineutrino fluxes from nuclear reactors.
At a refurbished SBL CERN-PS facility an experiment aimed to address the open issues has been proposed~\cite{Icarus-PS}, based on the technology 
of imaging in ultra-pure cryogenic Liquid Argon (LAr).

Motivated by this scenario a detailed study of the physics case was performed. We tackled specific physics models and we optimized
the neutrino beam through a full simulation. Experimental aspects not fully covered by the LAr detection, i.e. the 
measurements of the lepton charge on event-by-event basis and their energy over a wide range, were also investigated.
Indeed the muon leptons from Charged Current (CC) (anti-)neutrino interactions play an important role in disentangling 
different phenomenological scenarios provided their charge state is determined. Also,
the study of muon appearance/disappearance can benefit of the large statistics of CC muon events from the primary 
neutrino beam.

Results of our study are reported in detail in this proposal. We aim to design, construct and install two Spectrometers at 
``NEAR'' and ``FAR'' sites of the SBL CERN-PS, compatible with the already proposed LAr detectors. 
Profiting of the large mass of the two Spectrometers their stand-alone performances have also been exploited.

Some important practical constraints were assumed in order to draft the proposal on a
conservative, manageable basis, and maintain it sustainable in terms of time-scale and cost.
Well known technologies were considered as well as re-using parts of existing detectors 
(should they become available; if not it would imply  an increase of the costs with
no additional delay). 

The momentum and charge state measurements of muons in a wide energy range, from few hundreds $MeV$ to several 
$GeV$, over a $> \ 50\ m^2$ surface, is an extremely challenging task if constrained by a 10 (and not 100) millions~\euro\ budget
for construction and installation.  Running costs have to be kept at low level, too.


The experiment is identified throughout the proposal with the acronym NESSiE (Neutrino Experiment with SpectrometerS in Europe).

In the next Section the relevant physics pleas for the Spectrometer proposal are summarized. In Section~\ref{sec:physics} an extensive
discussion of the physics case and the possible phenomenological models are described. In Section~\ref{sec:beam} details of a complete
new simulation of the beam fluxes and horn designs are reported to work out a realistic situation and to be possibly used as contribution to the 
eventual design study group. In Section~\ref{sec:spect1} the choice and design of the Spectrometers are discussed. 
After a brief reminder of the
details of the Monte Carlo simulation for neutrino events (Section~\ref{sec:MC}) the obtained physics performances are outlined in
Section~\ref{sec:spect2}. Sections~\ref{sec:mech-struc} and ~\ref{sec:mag-SC}  deal with the technical definition of the mechanical structure and
electrical setting-up for the magnets. Sections~\ref{sec:rpc} and ~\ref{sec:hpt} show  the use of possible detectors either with a 
coarse resolution ($\sim$1 $cm$) or with a boosted one ($\sim$1 $mm$). Section~\ref{sec:bck} debates about background levels to be taken into account
for the data taking, described in Section~\ref{sec:daq}. The last two sections report about CERN setting-up, schedule and costs. Finally  conclusions are recapped.


\section{Aims and Landmarks}
The main physics topic of a new SLB CERN-PS experiment is to confirm or reject the so called neutrino 
anomalies and in case of a positive signal fully exploit the 
new physics connected with {\em sterile} neutrinos.
The reported anomalies are the \nubarmunubare transitions observed  by 
LSND~\cite{Aguilar:2001ty}, the \nue disappearance claimed by the Gallium source experiments for the Gallium 
solar neutrino reactions~\cite{Giunti:2010zu} and the \nubare 
disappearance of reactor antineutrinos~\cite{Mueller:2011nm}.

If such anomalies are not connected with new physics the  LAr plus Spectrometer experiments will report
null signal in the \nue and \nubare spectra, both in the absolute flux and energy shape, ruling out
the anomalies at exceeding $\sigma$'s. In fact as in some sterile neutrino models, discussed in the following, cancellations 
occur between \nue(\nubare) appearance, 
via \nubarmunubare transitions and \nue (\nubare) disappearance, a null result in the \nue sector by LAr
experiments would not be sufficient  to fully exclude {\em sterile} neutrino models.
In this case it will be crucial to precisely measure \numu disappearance since it would be
impossible for any model to fit both null results  
in the \nue and the \numu sectors.

We note that the best limit available so far on \numu disappearance in the $\Delta m^2$ range under study, 
published by CDHS~\cite{CDHS}, is based on the analysis of 3300 events at the Far detector, 
corresponding to just one day of data taking of the experiment we propose.

We therefore clinch that in case of null signal the conjunct experiments will be able to rule out all 
the existing anomalies as connected to the presence of sterile neutrinos. 

If instead the anomalies  are really hints of the existence of sterile neutrinos 
the experiments will be able to:
\begin{itemize}
	\item
provide a direct signature of non-sterile neutrinos through the detection of Neutral Current (NC) disappearance. In fact while
 CC disappearance may occur via 
active neutrino oscillations\footnote{A $\mu$-neutrino oscillating to a $\tau$-neutrino 
will indeed disappear because at these energies $\tau$ leptons cannot be 
created by neutrino interactions as they are below threshold.}, NC disappearance 
can only occur when active neutrinos oscillate to sterile ones.
It goes without saying that NC disappearance should be at the same rate of \numu charged 
current disappearance rate as measured by a spectrometer\footnote{It is amazing  that the evidence
 of sterile neutrinos via the disappearance of NC 
neutrino interactions, might be made with the same neutrino beam used  40 year ago 
to detect for the first time the existence of NC neutrino interactions.}.
\clearpage
	\item
 Characterize the sterile neutrino model, namely to determine:
\begin{itemize}
	\item
   all the parameters of 3+1 models (3 parameters: $U_{4e}$, $U_{4\mu}$, $\Delta m^2_{14}$) 
   or 3+2 models (7 parameters: $U_{4e}$, $U_{4\mu}$, $\Delta m^2_{14}$,
$U_{5e}$, $U_{5\mu}$, $\Delta m^2_{15}$, $\Delta_s$);
	\item
the number of steriles from the number of $\Delta m^2$ needed to fit the data that
 will  correspond to \numunue, \nubarmunubare appearances and 
\numu disappearance. Furthermore any manifestation of CP violation will be an evidence 
of at least 2 sterile neutrinos participating to oscillations.
	\item
 CP violation in the {\em sterile} sector from the comparison of \numunue and \nubarmunubare 
transitions.
\end{itemize}
\end{itemize}

Therefore in case sterile neutrinos indeed exist the experiments will fully exploit the very rich 
and exciting physics information related to them. 

From this concise  discussion that will be more extensively addressed in the following,
it clearly appears  that a spectrometer capable of measuring the muon momenta
 would complement the LAr experiment by allowing to:
\begin{itemize}
	\item
measure \numu disappearance in the entire available momentum range. This is a key information in 
rejecting/observing the anomalies over the whole expected parameter space of sterile neutrino oscillations;
	\item
measure the neutrino flux at the Near detector, in the full muon momentum range, which 
is decisive to keep the systematic errors at the lowest possible values.
\end{itemize}
The measurement of the sign of the muon charge would furthermore enable to: 
\begin{itemize}
\item
separate \numu from \nubarmu in the antineutrino beam where the \numu
contamination is relevant. 
Such a measure is crucial for a firm observation  of  any 
difference between \numunue and \nubarmunubare transitions,
a possible signature of CP violation;
\item
and finally reduce the data taking period by  a concurrent collection of 
both \numu  and \nubarmu events.
\end{itemize}

\clearpage

\section{Physics}\label{sec:physics}

\subsection{Introduction\label{sec:phys-intro}}
The recent re-computation of antineutrino fluxes from nuclear reactors~\cite{Mueller:2011nm} leds to the so called ``reactor neutrino 
anomaly''~\cite{Giunti:2010wz,Mention:2011rk}, i.e.
a  $\sim$ 3\% deficit of the reactor
antineutrino rates measured at short baselines by past reactor experiments, reinforcing 
the case of sterile neutrinos.

The latest fits of cosmological parameters are in favor of a number of neutrinos above
3 (see e.g.~\cite{Mangano:2006ur}), provided that masses of extra neutrinos are below $\sim$1 $eV/c^2$~\cite{Melchiorri-beyond}.

Besides reactor antineutrinos~\cite{Mention:2011rk,Giunti:2009zz} and cosmology, sterile neutrinos
were invoked to explain results
from beam dump experiments, LSND~\cite{Aguilar:2001ty}, accelerator neutrinos (and antineutrinos) as measured by the MiniBooNE 
experiment~\cite{AguilarArevalo:2007it,AguilarArevalo:2009xn} and source calibration data of 
the Gallium solar neutrino experiments~\cite{Giunti:2010zu}.
Global fits to neutrino oscillation data~\cite{Kopp:2011qd}
are performed by adding just one sterile neutrino to the three active, ``3+1'' models, or adding two sterile neutrinos, ``3+2''.
The latter models (see Fig.~\ref{fig:chisq-dmq}) 
provide a reasonable good fit to the data. In particular ``3+2'' models can 
introduce CP violation in the sterile sector, explaining the discrepancy
between the \nubare appearance detected by the LSND and the lack of
\nue appearance as reported by MiniBooNE.
``3+1'' models cannot introduce CP violation and when compared to
``3+2''  they are disfavored at 97.2\% C.L. according to~\cite{Kopp:2011qd}.
A way-out to improve the goodness of fit of the ``3+1'' model is to introduce the
CPT violation~\cite{Giunti:2010zs}.

Global fits do not provide a clear unique solution emerging from the data. Furthermore it should be noted that tension still exists 
between appearance and disappearance data~\cite{Kopp:2011qd}; for this reason even ``3+2'' global fits are not fully satisfactory.

In this Section we set the physics case of an accelerator neutrino experiment aiming at being the conclusive one on this topic.
We recall the capabilities of a LAr detector and we focus on the physics opportunities that a spectrometer could add to a liquid 
argon target. We first discuss measures like
the LSND/MiniBooNE evidence of \nubarmunubare transitions and the \nue , \nubare  disappearances, as well as
measures like \numu, \nubarmu disappearances. Then the possible signatures of sterile neutrinos, including NC 
disappearance, will be addressed and eventually the predictions based on some models will be exploited.

It is worth noting that, should the anomalies be confirmed, a short-baseline accelerator neutrino experiment would have an 
outstanding set of discoveries in its hands, namely:
\begin{itemize}
	\item
	the discovery of a new type of particles, the sterile neutrinos, interacting only via the gravitational force;
\item
	the proof that at least two sterile neutrinos exist;
\item
	the proof that the oscillations between active and sterile neutrinos violate CP.
\end{itemize}
\subsection{\nubarmunubare oscillations} 
The detection of \nubarmunubare oscillations has been claimed by the
LSND~\cite{Aguilar:2001ty} and the MiniBooNE~\cite{AguilarArevalo:2009xn} experiments. The two results may be in
agreement with each other. The KARMEN experiment ~\cite{Armbruster:2002mp},
which did not report
any evidence of these transitions, limited the LSND signal parameter space.
Furthermore MiniBooNE did not confirm the result in the neutrino sector~\cite{AguilarArevalo:2007it}.

A spectrometer would add very little to the detection of electrons corresponding to the signature of \nubarmunubare transitions,
which can be very well measured by a Liquid Argon detector. 
Indeed the Icarus Collaboration already reported in their previous proposal~\cite{Icarus-PS} an excellent sensitivity to such transitions.

Nevertheless a spectrometer can play a fundamental role in the determination
of \nubarmunubare transitions by measuring the muon charges. That corresponds to a notable constraint:
in the negative focussed beam, as discussed in Sect.~\ref{sec:beam}, the rate of \numu interactions is a sizable fraction of \nubarmu interactions.
It turns out that  it is mandatory to disentangle \numu and \nubarmu rates, possibly on an event-by-event basis,
at the Near detector, in order to reduce the systematic errors associated with the
prediction of the \numu and \nubarmu fluxes
in the Far detector.

Considering that the Icarus detector would collect the LSND statistics in about  10 days, it is evident that its sensitivity will be dominated by systematic 
errors and the information added by the Spectrometers looks very important, if not strictly mandatory to keep them as small as possible.

\subsection{\numu and \nubarmu disappearance} 
No experiment has so far reported evidence of \numu or \nubarmu disappearance
in the allowed $\Delta m^2$ region for sterile neutrinos. The limits set by
CDHS~\cite{CDHS}, MiniBooNE~\cite{MiniBooNE-numu} and atmospheric neutrino experiments~\cite{atmo}
are among the most severe constraints on sterile neutrino oscillations. 

We stress here that the CDHS limit is based upon the analysis of 3300 neutrino events collected in the
Far detector of a two-site detection setup at the CERN-PS neutrino beam. Such number of events corresponds
to the statistics collected by our experiment in just one day's data taking.

An improved measurement of \numu(\nubarmu) disappearance could severely challenge the sterile global fits in case of null result or provide a 
spectacular confirmation in case of signal observation.

The \numu and \nubarmu transitions are the main physics topics that a Spectrometer experiment could address.
The measurement of \numu and \nubarmu spectra at the Near detector in the full momentum range is mandatory to constrain systematic errors 
(the \numu and \nubarmu flux ratios at the Near and Far detectors are expected to be different, as discussed in Sect.~\ref{sec:beam}).

In Sect.~\ref{sec:spect2} the computed sensitivity of a spectrometer in measuring
\numu and \nubarmu disappearance is also discussed. It is important to note that with just 
3 years' running with the negative focussing beam the experiment could improve
the existing limits on \numu disappearance and provide a measurement of \nubarmu disappearance, 
so far never performed in the sterile $\Delta m^2$ region.
\subsection{NC disappearance} 
Sterile neutrinos come into play to accommodate the anomalies 
in a global phenomenological framework by introducing a fourth light neutrino. 
Since the number of active neutrinos is limited to three 
by the measurement of the $Z_0$ width~\cite{Alexander:1991vi}, additional 
light neutrinos cannot have electroweak couplings (sterile neutrinos).

The detection of \numu(\nubarmu) disappearance would be extremely important since it would also open the door to a 
sensitive search for the disappearance of NC events, which is a direct signature of the existence of sterile neutrinos.
Indeed none of the anomalies reported so far can be interpreted as a direct manifestation of sterile neutrinos.

For instance \numu may oscillate to \nutau that cannot produce the associated $\tau$ lepton being under threshold
for the CC interaction at the energies of the beam.

Instead NC interactions can ``disappear'' only if active neutrinos oscillate to sterile neutrinos. 
NC events, either \nue or \numu, can be efficiently detected by the Liquid Argon 
detector. However the transition rate measured with NC events has to agree with the \numu CC  disappearance rate 
once the \numunutau \, and \numunue contributions have been subtracted (these rates are anyway small
 at the L/E values of the present beam configuration). Indeed the NC disappearance is better measured by the double ratio:

\begin{equation}
	\frac{\frac{NC}{CC}_{Near}}{\frac{NC}{CC}_{Far}}
\end{equation}

The double ratio is the most robust experimental quantity to detect NC disappearance, once
$CC_{Near}$ and $CC_{Far}$ are precisely measured thanks to the Spectrometers,
at the Near and Far locations,  via the
disentanglement of  \numu and \nubarmu contributions. 

\subsection{Modelization} 
Anomalies and computed sensitivities were (and are) usually addressed with
empirical two neutrino oscillation formulas. Recently,
the interest in sterile neutrinos has been greatly reinforced because, after the appearance of
the reactor antineutrino anomaly~\cite{Mueller:2011nm},
the LSND/MiniBooNE anomaly and the Gallium source anomaly
can be accommodated together with all other existing measurements
of neutrino oscillations in a single model which incorporates two
sterile neutrinos~\cite{Kopp:2011qd}.
While this model can provide a good overall $\chi^2$ by fitting existing data,
tension still exists between appearance and disappearance oscillation results.

As introduced in Sect.~\ref{sec:phys-intro}
the main physics reason why the ``3+2'' models provide a better overall fit with respect
to the ``3+1'' models is that they settle the experimental conflict between the \nubarmunubare signal
as detected by LSND and the lack of \numunue oscillations as reported by MiniBooNE thanks to
the introduction of CP violation that requires at least two sterile neutrinos.

In the following, the discovery potential of a LAr+NESSiE experiment will be discussed in the context of 
``3+2" models by selecting their best fit points in the parameter space. For completeness the 
``3+1" models will also be considered.

\subsubsection{``3+2'' neutrino oscillations model\label{sec:formula}} 

 In a short-baseline (SBL) accelerator experiment, where $\Delta m^2_{21} \approx \Delta m^2_{31} \approx 0$,
 the relevant \nue appearance probability is given by~\cite{Maltoni:2007zf}: 

\begin{multline} \label{eq:5nu-prob}
    P_{\nu_\mu\to\nu_e} =
    4 \, |U_{e4}|^2 |U_{\mu 4}|^2 \, \sin^2 \phi_{41} +
    4 \, |U_{e5}|^2 |U_{\mu 5}|^2 \, \sin^2 \phi_{51}
    \\
    + 8 \,|U_{e4}U_{\mu 4}U_{e5}U_{\mu 5}| \,
    \sin\phi_{41}\sin\phi_{51}\cos(\phi_{54} - \delta) \,,
\end{multline}

\noindent with the definitions

\begin{equation} \label{eq:5nu-def}
    \phi_{ij} \equiv \frac{\Delta m^2_{ij}L}{4E} \,,
    \qquad \delta \equiv
    \arg\left(U_{e4}^* U_{\mu 4} U_{e5} U_{\mu 5}^* \right) \,. 
\end{equation}

\noindent where symbols have the usual meaning. Eq.~\eqref{eq:5nu-prob} holds for neutrinos; for
antineutrinos one has to replace $\delta \to - \delta$. 

The survival probability in the same SBL approximation is given by

\begin{equation}
    P_{\nu_\alpha\to\nu_\alpha} = 1
    - 4\left(1 - \sum_{i=4,5} |U_{\alpha i}|^2 \right)
    \sum_{i=4,5} |U_{\alpha i}|^2 \, \sin^2\phi_{i1} 
    - 4\, |U_{\alpha 4}|^2|U_{\alpha 5}|^2 \, \sin^2\phi_{54}
\end{equation}

\noindent where $\phi_{ij}$ is given in Eq.~\eqref{eq:5nu-def}. 

 
\begin{figure}[htbp]
  \centering 
  \includegraphics[width=0.67\textwidth]{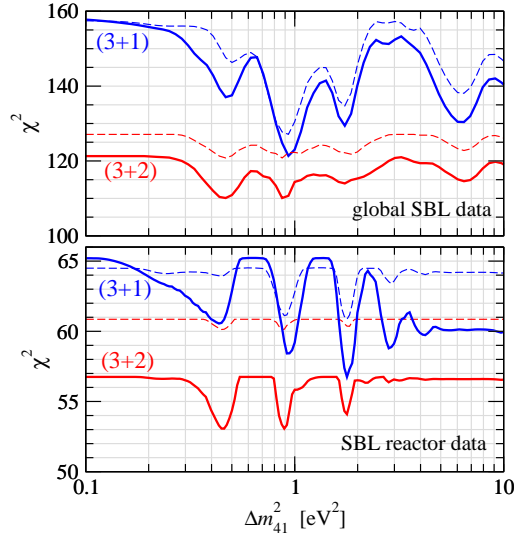}
    \caption{$\chi^2$ from global SBL data (upper panel) and from SBL
    reactor data alone (lower panel) for the 3+1 (blue) and 3+2 (red)
    scenarios. Dashed curves were computed using the old reactor
    antineutrino flux prediction~\cite{Schreckenbach:1985ep}, solid curves
    are for the new one~\cite{Mueller:2011nm}. All undisplayed parameters
    are minimized over.  The total number of data points is 137 (84) for the
    global (reactor) analysis.
    } \label{fig:chisq-dmq}
\end{figure}

The $\chi^2$ of the "3+2" model fit to the SBL oscillation data is shown in Fig.~\ref{fig:chisq-dmq}.
It appears that there is more than one solution for the oscillation parameters. In the following,
the four best fits indicated in~\cite{Kopp:2011qd} plus the best fit point of the preliminary updated analysis 
by Karagiorgi et al.~\cite{Karagiorgi} and the best fit by Giunti-Laveder~\cite{Giunti:2011gz} are considered
(see Tab.~\ref{tab:global-bfp}\footnote{It could be argued that the $\chi^2$ values reported in Tab.~\ref{tab:global-bfp}
are ``too good''. This happens because several data are basically not sensitive
to sterile oscillations (i.e. Chooz data~\cite{Chooz-final}), and are well fitted under any hypothesis.}). 

\begin{table}[htbp]
  \begin{center}
  \begin{tabular}{ccccccccc}
  & $\Delta m^2_{41}$ & $|U_{e4}|$ & $|U_{\mu 4}|$ & 
    $\Delta m^2_{51}$ & $|U_{e5}|$ & $|U_{\mu 5}|$ & 
    $\delta / \pi$ & $\chi^2$/dof\\
    \hline
    1) & 0.47 & 0.128 & 0.165 & 
    0.87 & 0.1380& 0.148 & 1.64 & $110.4/130$\\
    2)  &
    0.47 & 0.117 & 0.201 & 
    1.70 & 0.1150& 0.101 & 1.39 & $114.4/130$\\
    3)  &
    1.00 & 0.133 & 0.162 & 
    1.60 & 0.151 & 0.078 & 1.48 & $114.4/130$\\
    4)  &
    0.90 & 0.123 & 0.163 & 
    6.30 & 0.135 & 0.091 & 1.67 & $115.0/130$\\
	  \hline
    5) & 0.92 & 0.14 & 0.14 & 
   26.60 & 0.077 & 0.15 & 1.7 & 182.6/192\\
	  \hline
     6) & 0.90 & 0.158 & 0.152 & 
   1.61 & 0.130 & 0.078 & 1.51 & 91.6/100\\
  \end{tabular}
  \caption{Parameter values and $\chi^2$ at the global
  best fit points for the four best ``3+2'' fit points of \protect{\cite{Kopp:2011qd}},
  table entries 1-4, the best fit of Karagiorgi et al. \protect{\cite{Karagiorgi}}, 
  entry (5); ($\Delta m^2$'s in eV$^2$) and the best fit by 
  Giunti-Laveder~\cite{Giunti:2011gz}, entry (6).}
  \label{tab:global-bfp}
  \end{center}
\end{table}

\subsection{Probabilities at the Far detector\label{sec:prob}}
\begin{figure}[htbp]
  \includegraphics[width=0.9\textwidth]{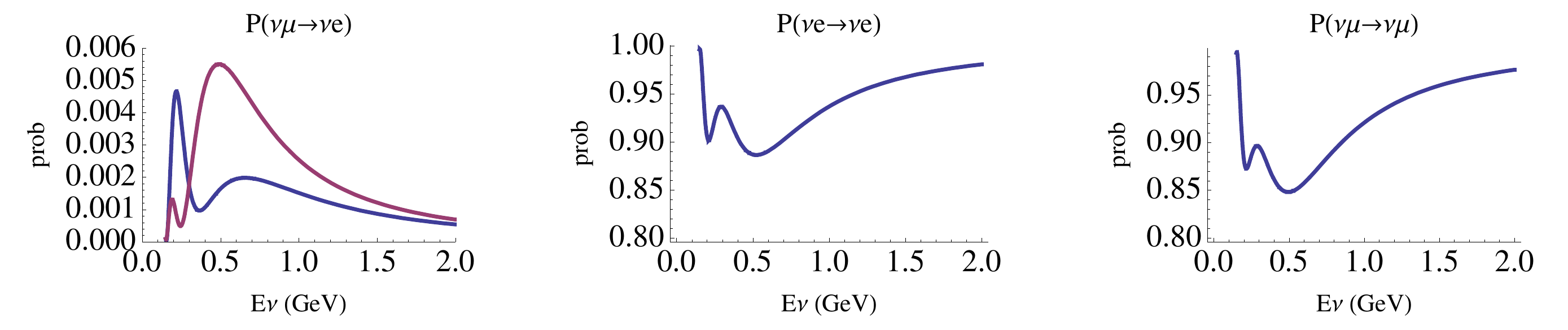}
  \includegraphics[width=0.9\textwidth]{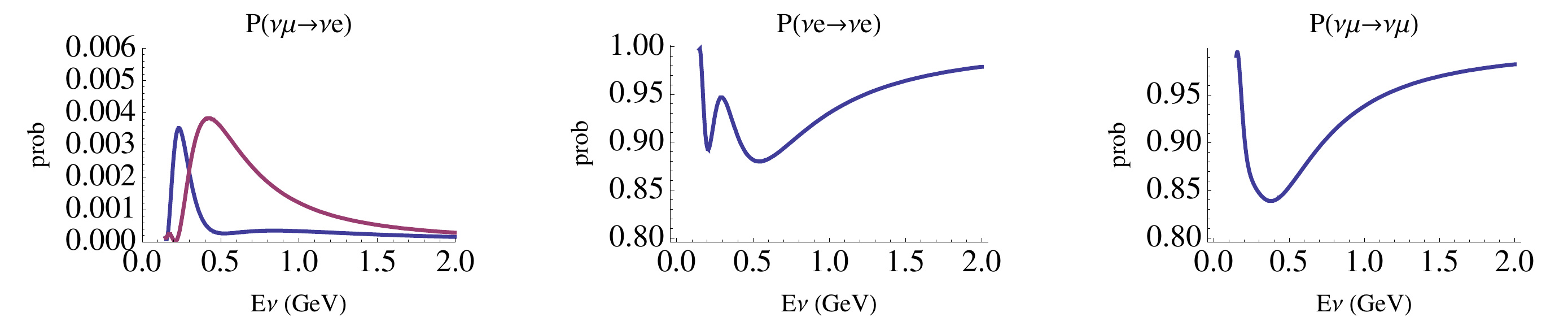}
  \includegraphics[width=0.9\textwidth]{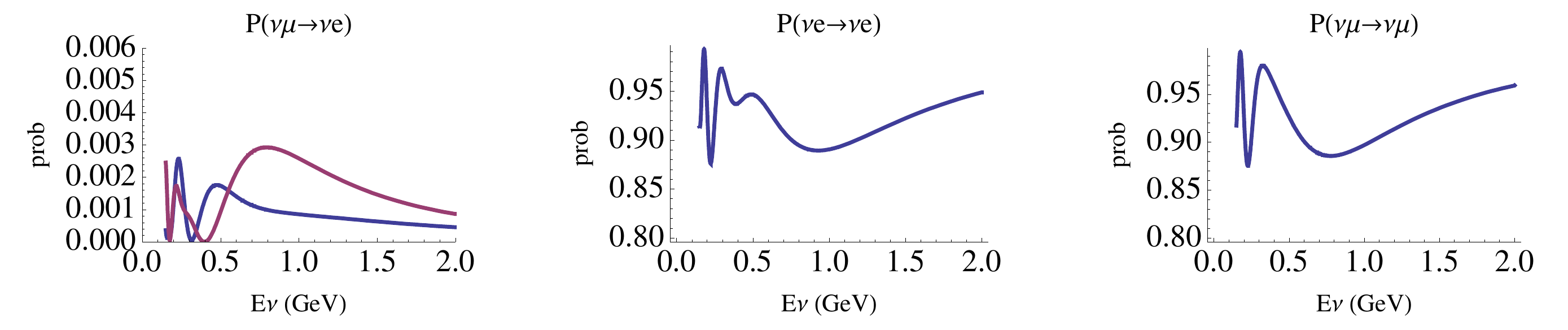}
  \includegraphics[width=0.9\textwidth]{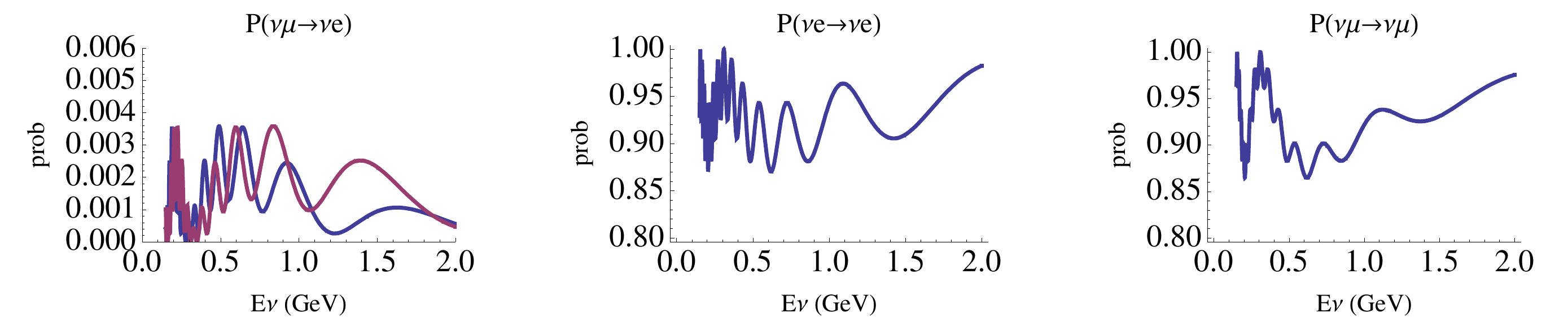}
  \includegraphics[width=0.9\textwidth]{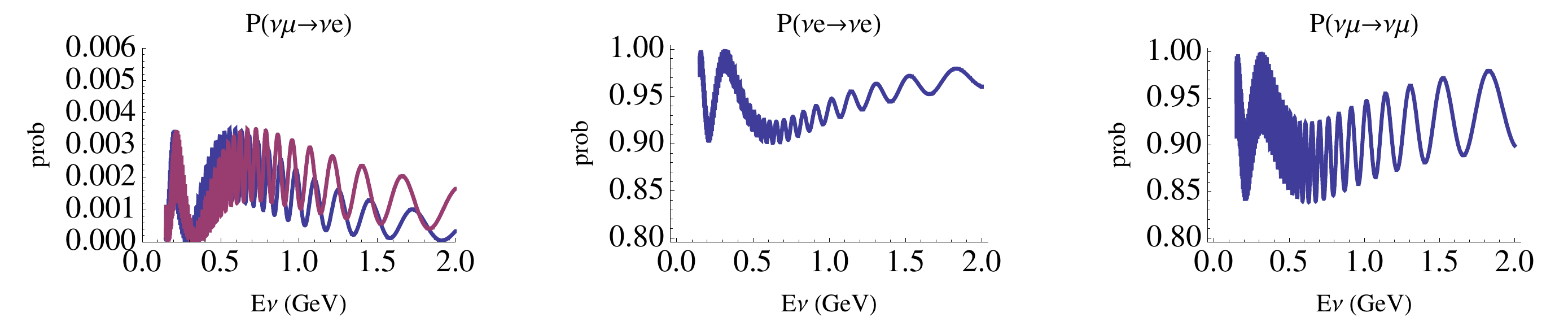}
  \includegraphics[width=0.9\textwidth]{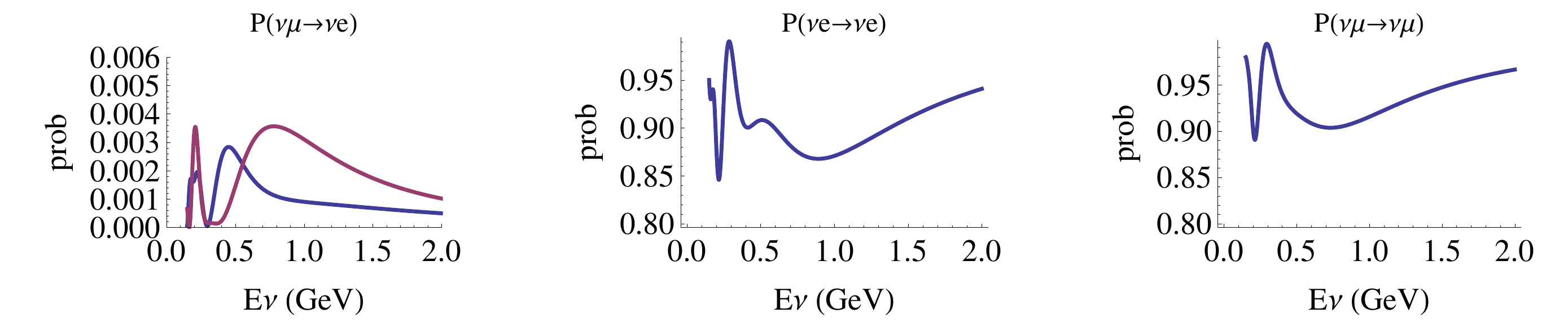}
    \caption{Oscillation probabilities computed at a baseline of 850 m for the
    four best fit points of ``3+2''~ \cite{Kopp:2011qd}, the best fit of~\cite{Karagiorgi} 
    and the best fit of~\cite{Giunti:2011gz}.  \pnumunue on the left (neutrinos in blue, 
    antineutrinos in magenta), \pnuenue at center and \pnumunumu on the right.}
  \label{fig:FirstProb}
\end{figure}

The oscillation probabilities \pnumunue, \pnuenue and \pnumunumu are computed at a distance of 850 m from the proton target
for the six best values reported in Tab.~\ref{tab:global-bfp} (see Fig.~\ref{fig:FirstProb}). Many interesting features occur within
the interplay of appearance/disappearance either for \numu or \nue sectors, the most exquisite distinctiveness being that
the scenario may be rather complicate. Therefore to disentangle that scenario we will need the best measurements of \numu and \nue 
as well as \nubarmu and \nubare in the widest available energy range.

In any case we like to underline some attractive outcomes from these probability computations, primarily for the ``3+2''  models:
\begin{itemize}
	\item
in the electron sector competition occurs between \nue disappearance and \nue appearance through the \numunue transitions. 
This is the {\em tricky} way by which ``3+2'' may cancel out \nue appearance in MiniBooNE. 
	\item
The different results in \nue and \nubare by MiniBooNE may be explained by the behavior  of the \nubarmunubare transitions which
take into account the CP violation introduced by the ``3+2'' models.
	\item
A low energy \nue excess emerges. The peak is entirely due to   the interference term of Eq.
\ref{eq:5nu-prob} and it would be a signature of the existence of not one, but two sterile neutrinos.\\
It should be noted that the MiniBooNE low energy excess detected in the neutrino run can be accommodated by the ``3+2" models.
However when disappearance data are also considered the peak of the interference term does not fit anymore the
measure~\cite{Kopp:2011qd}. 
	\item
A sizable \numu disappearance probability as large as 15\%  at the oscillation maximum is predicted 
  below 1$\div$2 GeV, depending on the different best fits.
\end{itemize}
\subsection{Neutrino Rates at the Far detector\label{sec:rates}}
Neutrino interaction rates are computed using  the neutrino flux discussed in Sect.~\ref{sec:beam}, the GENIE~\cite{ref_GEN} cross sections
and the above-mentioned probabilities. Event rates are normalized to
2 years' run of the positive focussing neutrino beam, with 30 kW proton beam power, and 3 years' run of negative focussing neutrino beam.
A neutrino efficiency of 100\% is considered, while energy resolution effects
and systematic errors are not included in the plots\footnote{Results with full simulation, including energy resolution effects
and systematic errors, are reported in Sect.~\ref{sec:spect2}.}.

\begin{figure}[htbp]
   \includegraphics[width=0.42\textwidth]{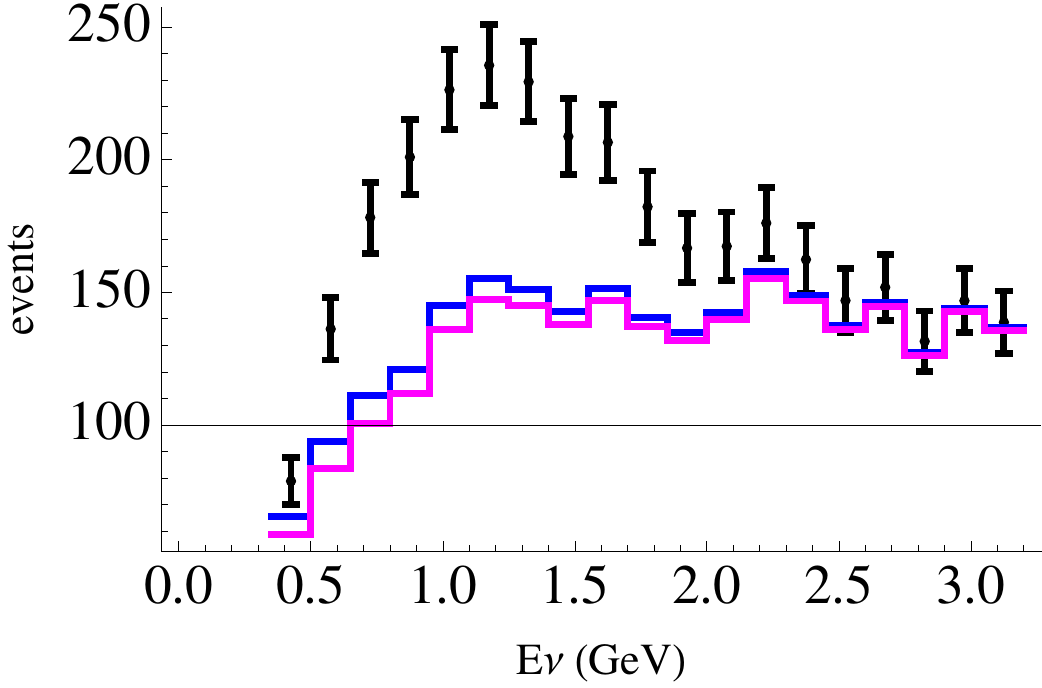}
   \includegraphics[width=0.42\textwidth]{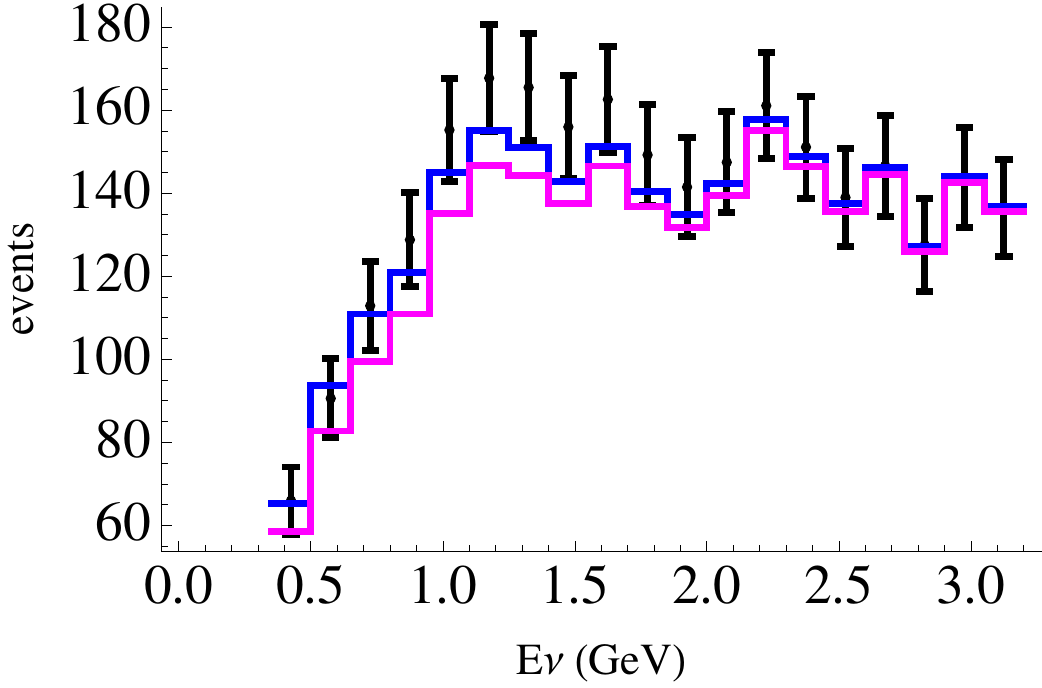}
   \includegraphics[width=0.42\textwidth]{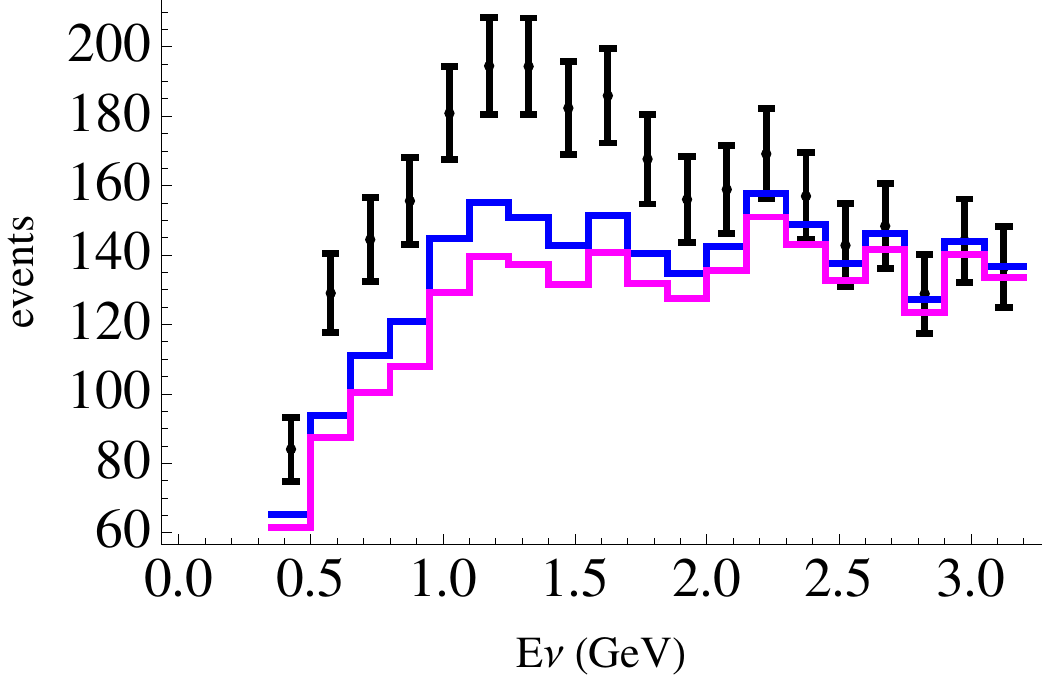}
   \includegraphics[width=0.42\textwidth]{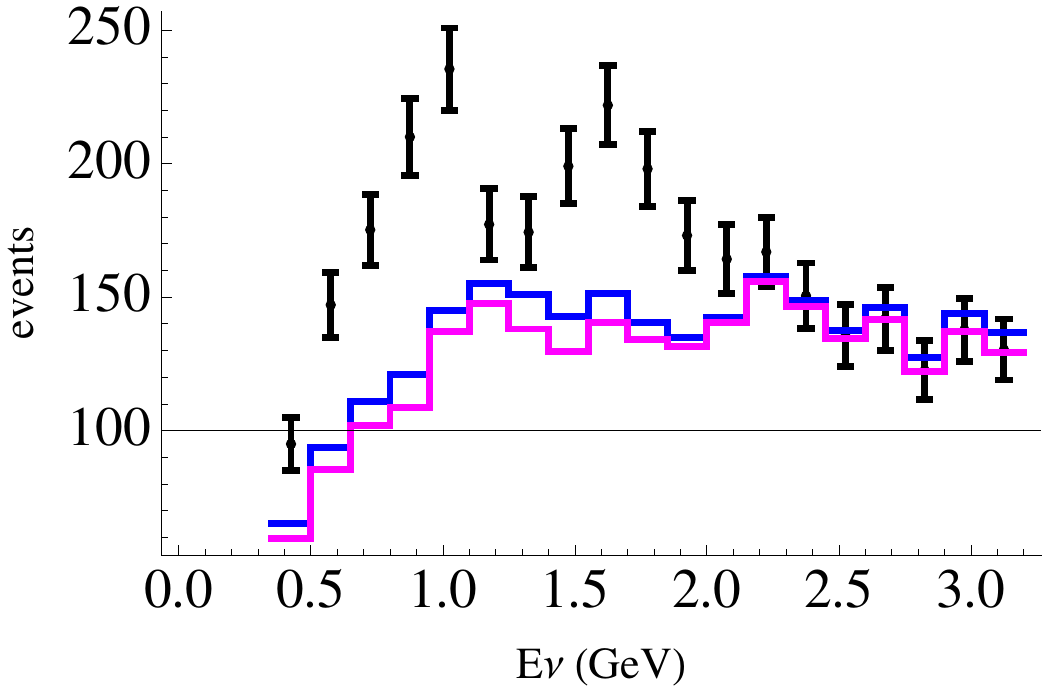}
   \includegraphics[width=0.42\textwidth]{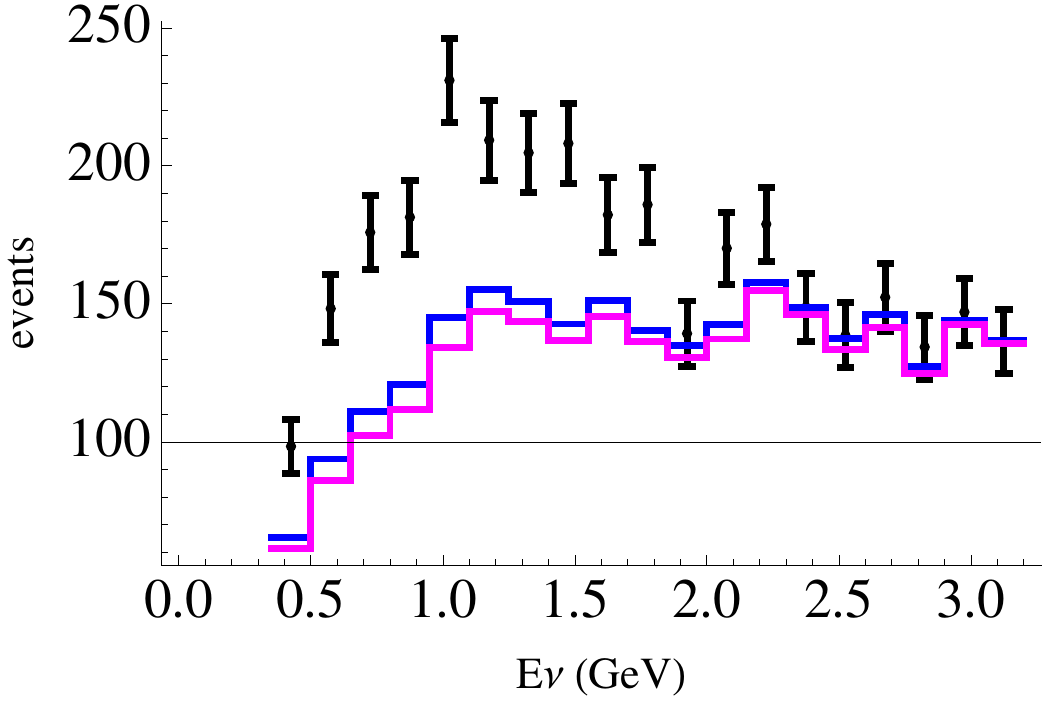}
   \includegraphics[width=0.42\textwidth]{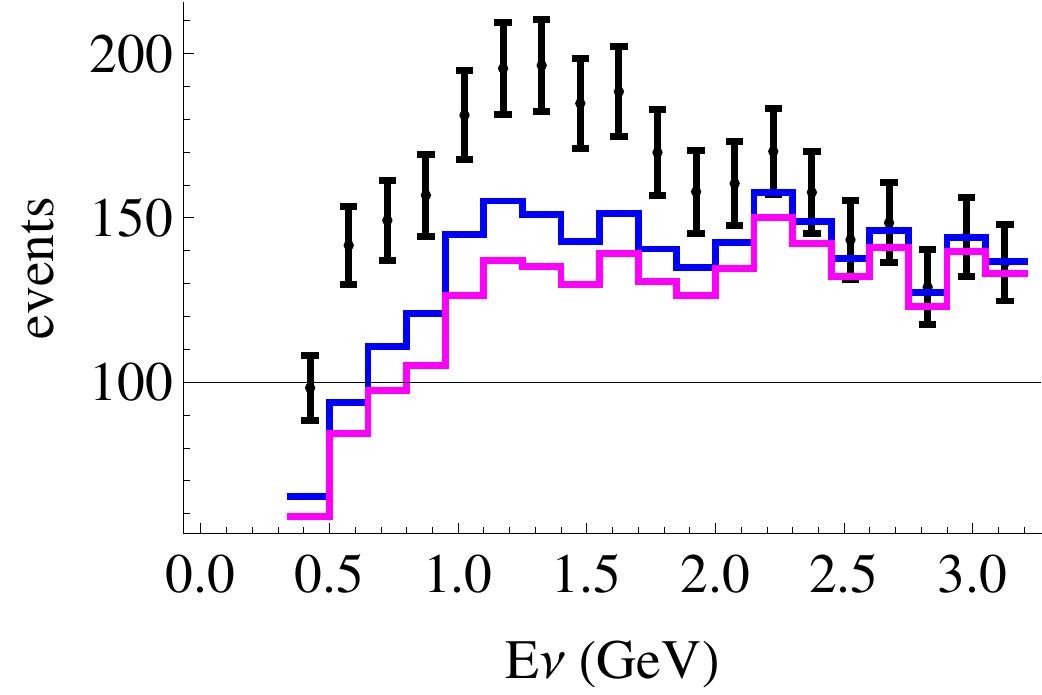}
    \caption{Electron neutrino spectra for the six test points described in the text. Blue 
    lines corresponds to no oscillations, magenta to disappearance only, 
    black points with statistical error bars corresponds to disappearance plus 
    appearance events. }
     \label{fig:Nue}
\end{figure}

By considering the six test points of Tab.~\ref{tab:global-bfp} event rate spectra are displayed in Fig.~\ref{fig:Nue} for \nue.
Results without oscillations, with \nue disappearance only and with both \nue disappearance and appearance, generated by \numunue transitions,
are shown.

\begin{figure}[htbp]
  \includegraphics[height=1.3in, width=0.80\textwidth]{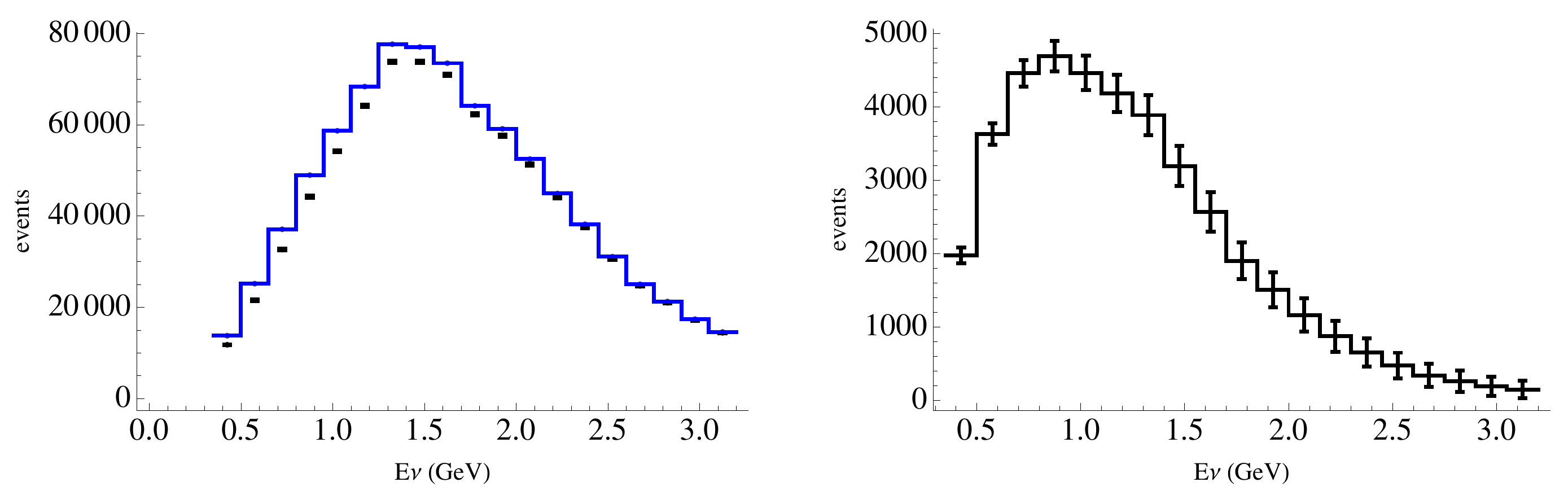}
  \includegraphics[height=1.3in, width=0.80\textwidth]{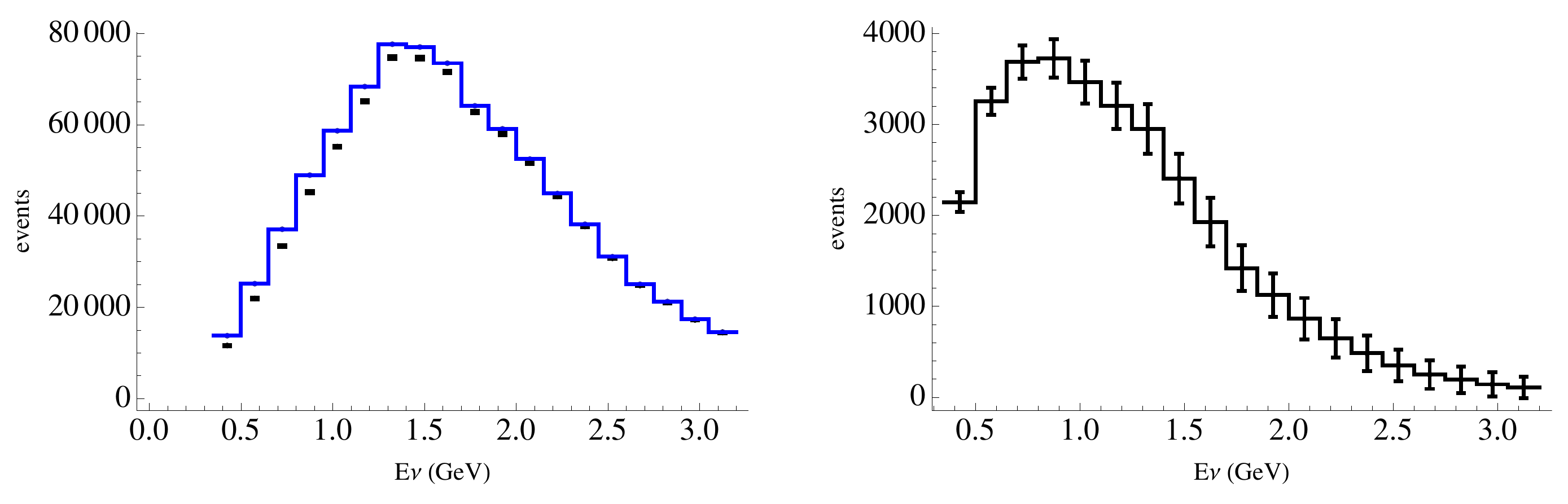}
  \includegraphics[height=1.3in, width=0.80\textwidth]{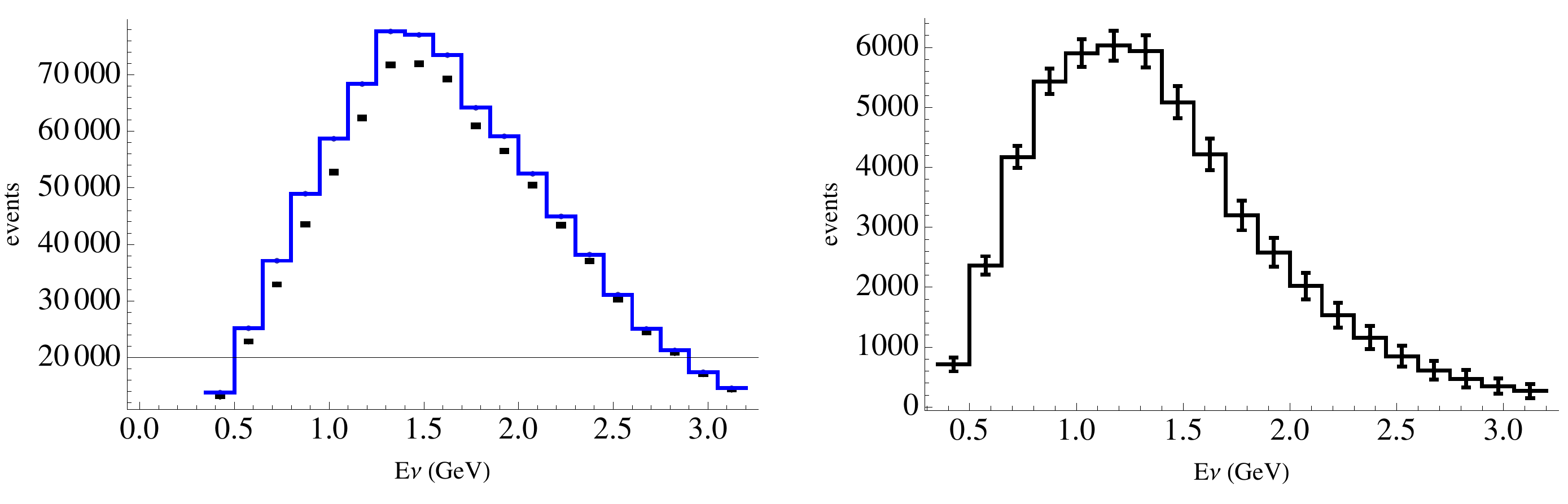}
  \includegraphics[height=1.3in, width=0.80\textwidth]{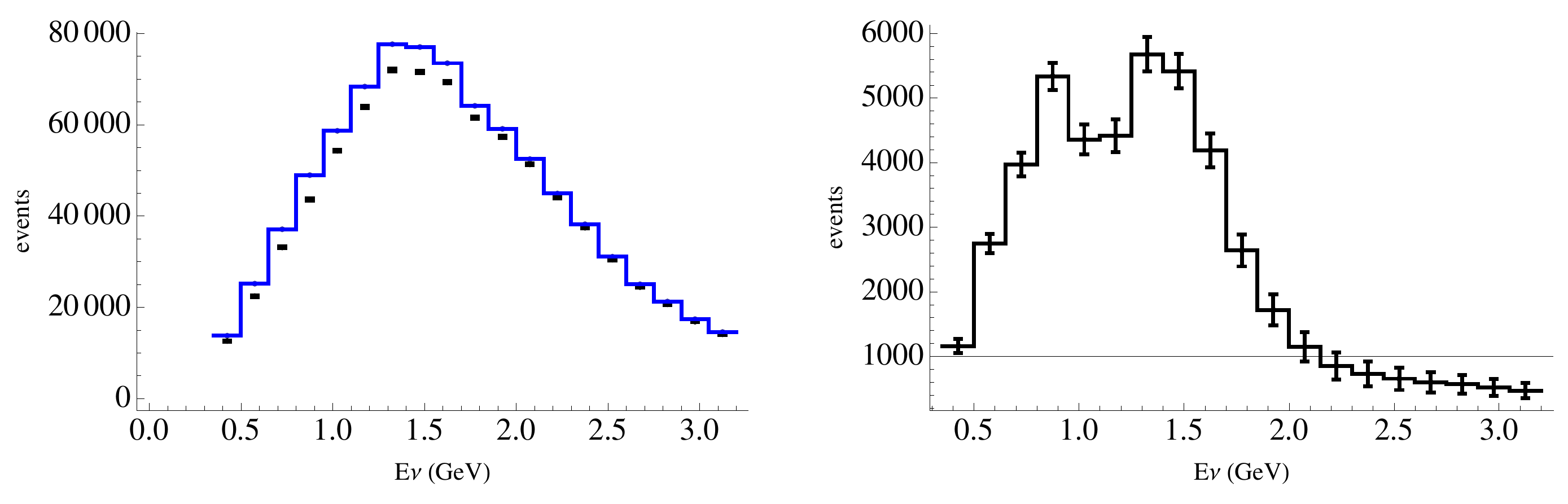}
  \includegraphics[height=1.3in, width=0.80\textwidth]{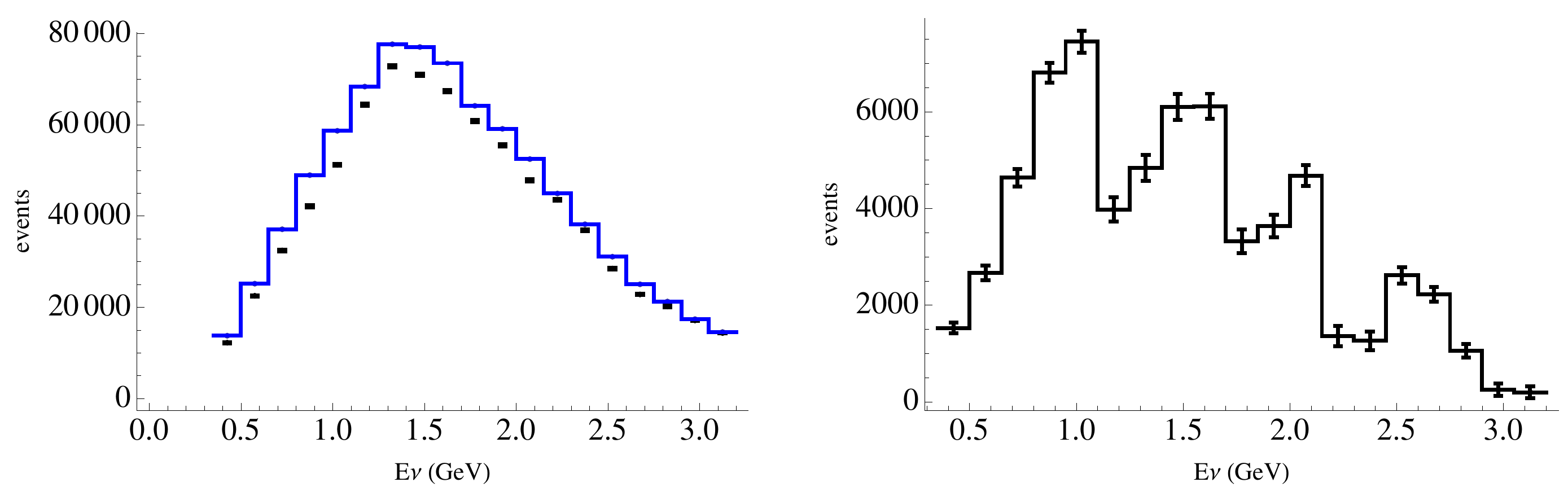}
  \includegraphics[height=1.3in, width=0.80\textwidth]{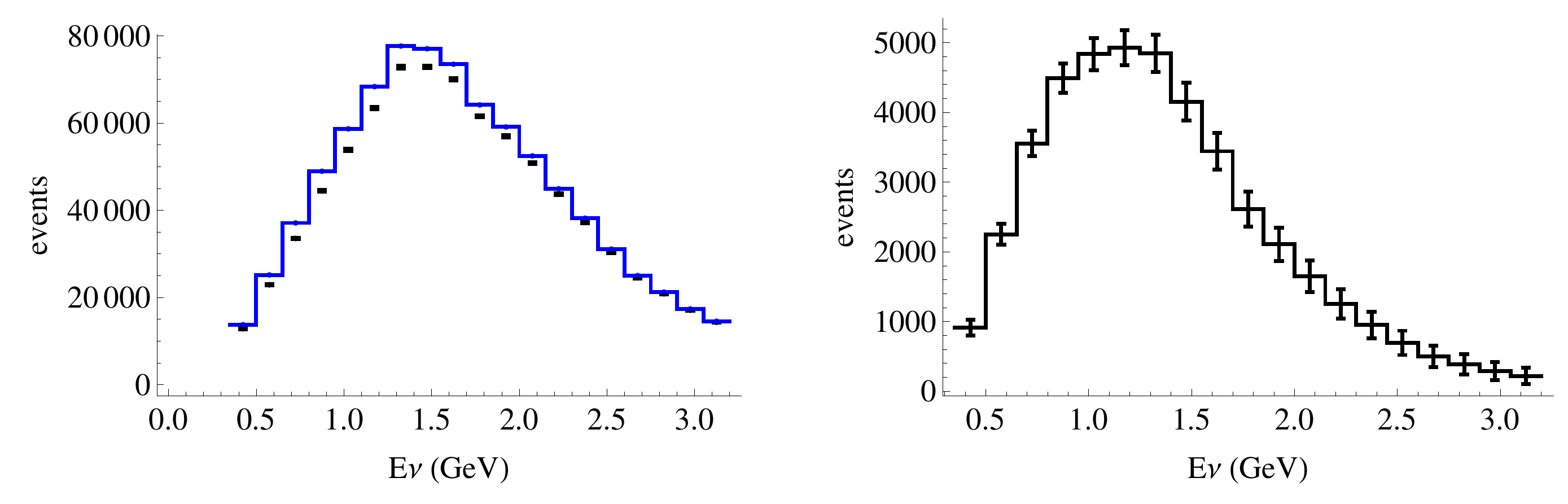}
    \caption{\numu event rates (left) where blue lines correspond to no-oscillation 
    and black points with statistical error bars (not visible) include disappearance. 
     Difference between rates estimated with and without oscillation (right). The 
     six rows refer to the test points discussed in the text.}
     \label{fig:FirstNuMu}
\end{figure}

The number of expected \numu events are displayed in Fig.~\ref{fig:FirstNuMu}, while
event rates in the energy range $0.2\; GeV < E_\nu < 2\; GeV$ are reported
in Tab.~\ref{Tab:Numu} and Tab.~\ref{Tab:Nue-full}, for \numu and \nue, respectively.
 The integral effect is not negligible (up to 6\%)
and the distinctive spectral signature is clearly detectable.

\begin{table}[h]
\begin{tabular}{ccc}
Fit point  & No-Osc. & Disappearance \\
1) & 605792  & 569060 \\
2) & 605792   & 576102\\
3) & 605792 & 560108\\
4) & 605792 & 564136\\
5) & 605792 & 554629\\
6) & 605792 & 567584\\
\end{tabular}

\caption{\numu ~events computed for $0.2\; < E_\nu < 2\; \rm{GeV}$ and two years' run at 30 kW, for the 6 test points described in the text.}
\label{Tab:Numu}
\end{table}

\begin{table}[h]
\begin{tabular}{cccc}
Fit point & No-Osc. & Disapp. & Disapp. + App.\\
1)          & 1424    & 1349   &2065   \\
2)          & 1424    & 1342   &1512   \\
3)          & 1424    & 1306   &1788   \\
4)          & 1424    & 1326   &2018   \\
5)          & 1424    &1349   &1977   \\
6)          & 1424    &1283   &1835   \\

\end{tabular}
\caption{\nue events with $0.3\; < E_\nu < 2\; \rm{GeV}$ computed for two years' run at 30 kW, for the 6 test points described in the text.}
\label{Tab:Nue-full}
\end{table}

\clearpage

\subsection{Antineutrino Rates at the Far detector\label{sec:arates}}
\begin{figure}[htbp]
    \includegraphics[width=0.32\textwidth]{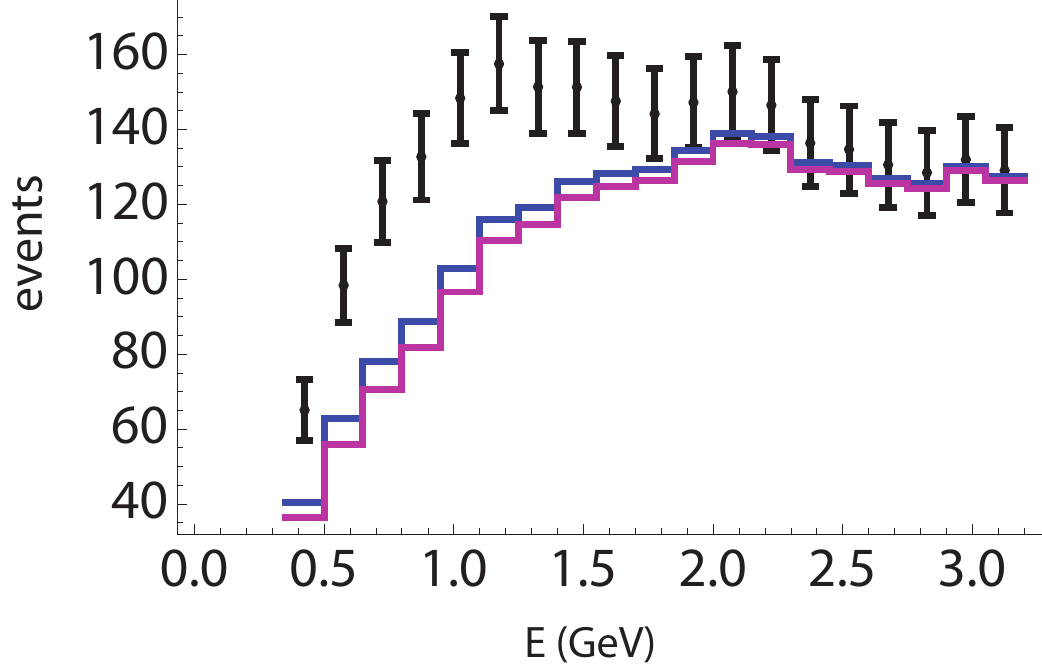}
    \includegraphics[width=0.62\textwidth]{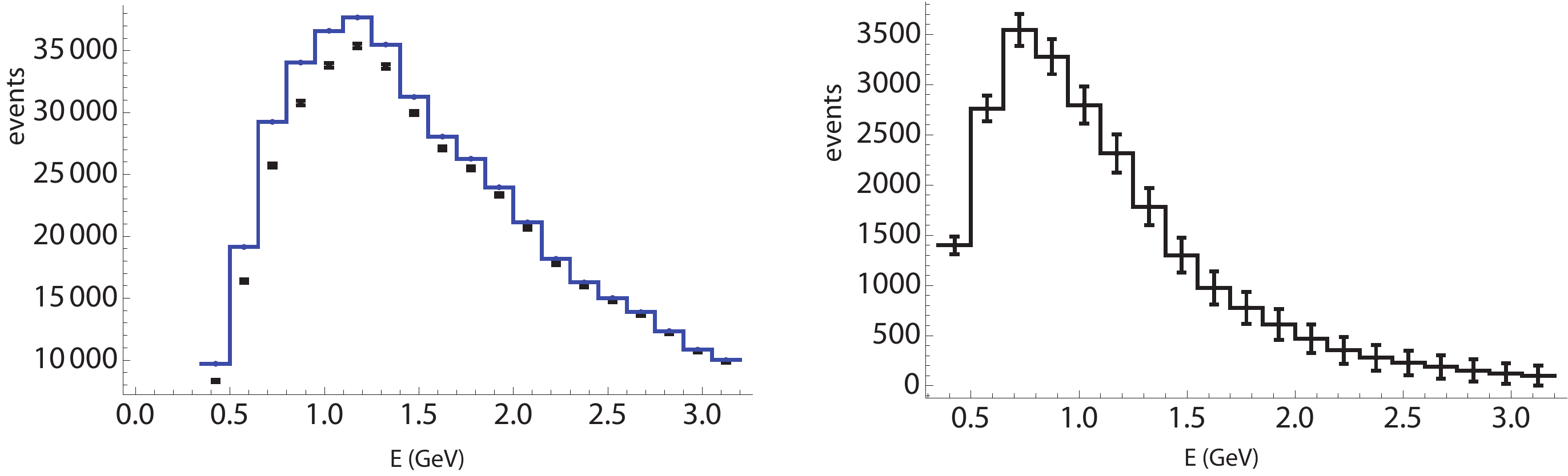}
   \caption{Left panel: \nue+\nubare events in the antineutrino beam, computed 
   for the first test point of Table~1, blue line corresponds to non-oscillated events, 
   magenta line to disappearance only and points with errors correspond to oscillated 
   events. Central panel: \numu+\nubarmu events, blue line corresponds to non-oscillated 
   events, points with errors (not visible) to oscillated events. Right panel:
 \numu+\nubarmu events are shown as expected minus measured.}
\label{Fig:antineutrino-plots}
\end{figure}

If CPT holds, data taken with antineutrino beams should provide
identical \nue and \numu disappearance rates.
If CP is violated, as predicted by all the ``3+2'' best fit points,
\nubare appearance rate results to be rather high and the signal
could not be missed.
As an example we display in Table~\ref{Tab:antineutrino-rates} event rates as
predicted by the ``3+2'' best fit and in Figure~\ref{Fig:antineutrino-plots}
the corresponding signal plots.

Section~\ref{sec:spect2} quantitatively discusses the possibility of
measuring both \numu and \nubarmu disappearance rates with the NESSiE
Spectrometer in the antineutrino run.

\begin{table}[htbp]
\begin{center}
\begin{tabular}{|cccc|}
	\hline
	{\bf events}  & No-Osc. & Disappearance & Disapp. + App. \\
  \nue (0.3-2 GeV)  & 1134 & 1078 & 1475 \\	  
  \numu(0.3-2 GeV)  & 312439 & 290791&  \\	  
	\hline
\end{tabular}
\caption{Event numbers in antineutrino mode, best fit (1), 3 years' of data taking}
\label{Tab:antineutrino-rates}
\end{center}
\end{table}

\subsubsection{``3+1'' neutrino oscillations model}
 
In the so called ``3+1'' model the flavor neutrino basis includes three active neutrinos
$\nu_{e}$, $\nu_{\mu}$, $\nu_{\tau}$ and a sterile neutrino $\nu_{s}$. The effective flavor transition and survival probabilities
in short-baseline (SBL) experiments are given using the standard notation by

\begin{align}
\null & \null P_{\boss{\nu}{\alpha}\to\boss{\nu}{\beta}}^{\text{SBL}} =
\sin^{2} 2\vartheta_{\alpha\beta}\sin^{2}\left( \frac{\Delta{m}^2_{41} L}{4E} \right)
\qquad (\alpha\neq\beta) \,,
\label{trans} \\
\null & \null
P_{\boss{\nu}{\alpha}\to\boss{\nu}{\alpha}}^{\text{SBL}} = 1 - \sin^{2} 2\vartheta_{\alpha\alpha}
\sin^{2}\left( \frac{\Delta{m}^2_{41} L}{4E} \right) \,,
\label{survi}
\end{align}

\noindent for $\alpha,\beta=e,\mu,\tau,s$, with

\begin{align}
\null & \null
\sin^{2} 2\vartheta_{\alpha\beta} = 4 |U_{\alpha4}|^2 |U_{\beta4}|^2 \,,
\label{transsin}
\\
\null & \null
\sin^{2} 2\vartheta_{\alpha\alpha} = 4 |U_{\alpha4}|^2 \left( 1 - |U_{\alpha4}|^2 \right) \,.
\label{survisin}
\end{align}

The key features of this model are:
\begin{enumerate}
\item All effective SBL oscillation probabilities depend only on the absolute value of the largest squared-mass difference
$\Delta{m}^2_{41}$.
\item All oscillation channels are open, each one with its own oscillation amplitude.
\item The oscillation amplitudes depend only on the absolute values
of the elements in the fourth column of the mixing matrix,
i.e. on three real numbers with sum less than unity,
since the unitarity of the mixing matrix implies $\sum_{\alpha} |U_{\alpha4}|^2 = 1$
\item CP violation cannot be observed in SBL oscillation experiments, even if the mixing matrix contains CP-violation phases.
In other words, neutrinos and antineutrinos have the same effective SBL oscillation probabilities.
\end{enumerate}

The global fit of all data in ``3+1'' scheme yields the best-fit values of the oscillation parameters listed in Tab.~\ref{bef}.

\begin{table}[h!]
\begin{center}
\begin{tabular}{cc}&3+1\\
\hline
$\chi^2_{\text{min}}$&$100.2$\\
$\text{NDF}$&$104$\\
$\text{GoF}$&$59\%$\\
\hline
$\Delta{m}^2_{41}\,[\text{eV}^2]$&$0.89$\\
$|U_{e4}|^2$&$0.025$\\
$|U_{\mu4}|^2$&$0.023$\\
\hline
$\Delta\chi^{2}_{\text{PG}}$&$24.1$\\
$\text{NDF}_{\text{PG}}$&$2$\\
$\text{PGoF}$&$6\times10^{-6}$\\
\hline
\end{tabular}
\end{center}
\caption{ \label{bef} Values of $\chi^{2}$, number of degrees of freedom (NDF), goodness-of-fit (GoF) and
best-fit values of the mixing parameters obtained in 3+1 fits of short-baseline oscillation data.
The last three lines give the results of the  parameter goodness-of-fit test, $\Delta\chi^{2}_{\text{PG}}$,
number of degrees of freedom ($\text{NDF}_{\text{PG}}$) and parameter goodness-of-fit (PGoF).}
\end{table}

\begin{figure}[htbp]
\includegraphics*[bb=12 18 576 566, width=\linewidth]{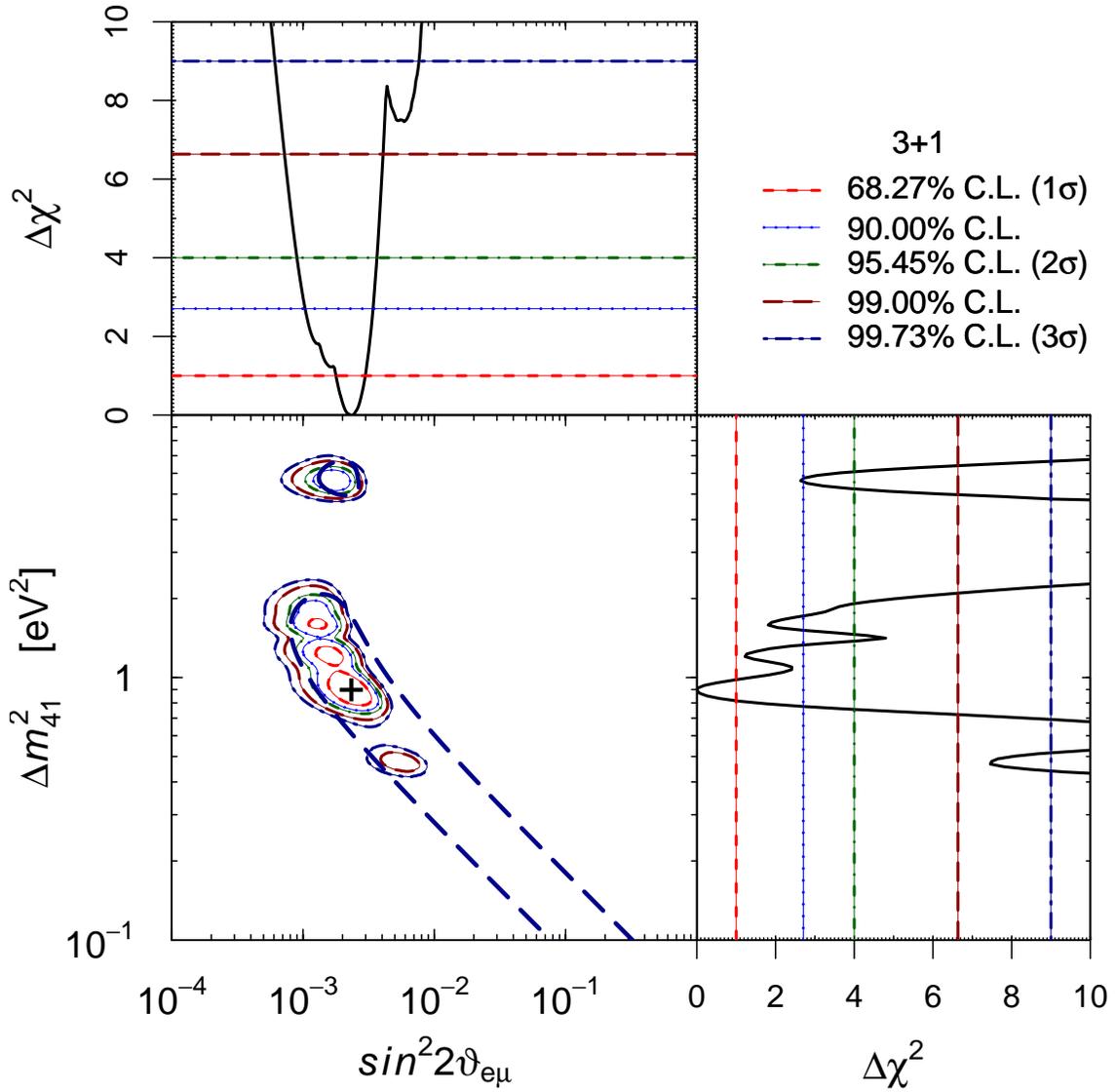}
\caption{ \label{3p1-sem}
Allowed regions in the $\sin^{2}2\vartheta_{e\mu}$--$\Delta{m}^2_{41}$ plane and marginal $\Delta\chi^{2}$'s for
$\sin^{2}2\vartheta_{e\mu}$ and $\Delta{m}^2_{41}$ obtained from the global fit of all the considered data in 3+1 schemes.
The best-fit point corresponding to $\chi^2_{\text{min}}$ is indicated by a cross. The isolated dark-blue dash-dotted 
contours enclose the regions allowed at $3\sigma$ by the analysis of appearance data (the $\bar\nu_{\mu}\to\bar\nu_{e}$ data of the 
LSND, KARMEN  and MiniBooNE  experiments and the $\nu_{\mu}\to\nu_{e}$ data of the NOMAD~\cite{nomad}  and MiniBooNE  experiments).}
\end{figure}

\begin{figure}[htbp]
\includegraphics*[bb=12 18 559 566, width=\linewidth]{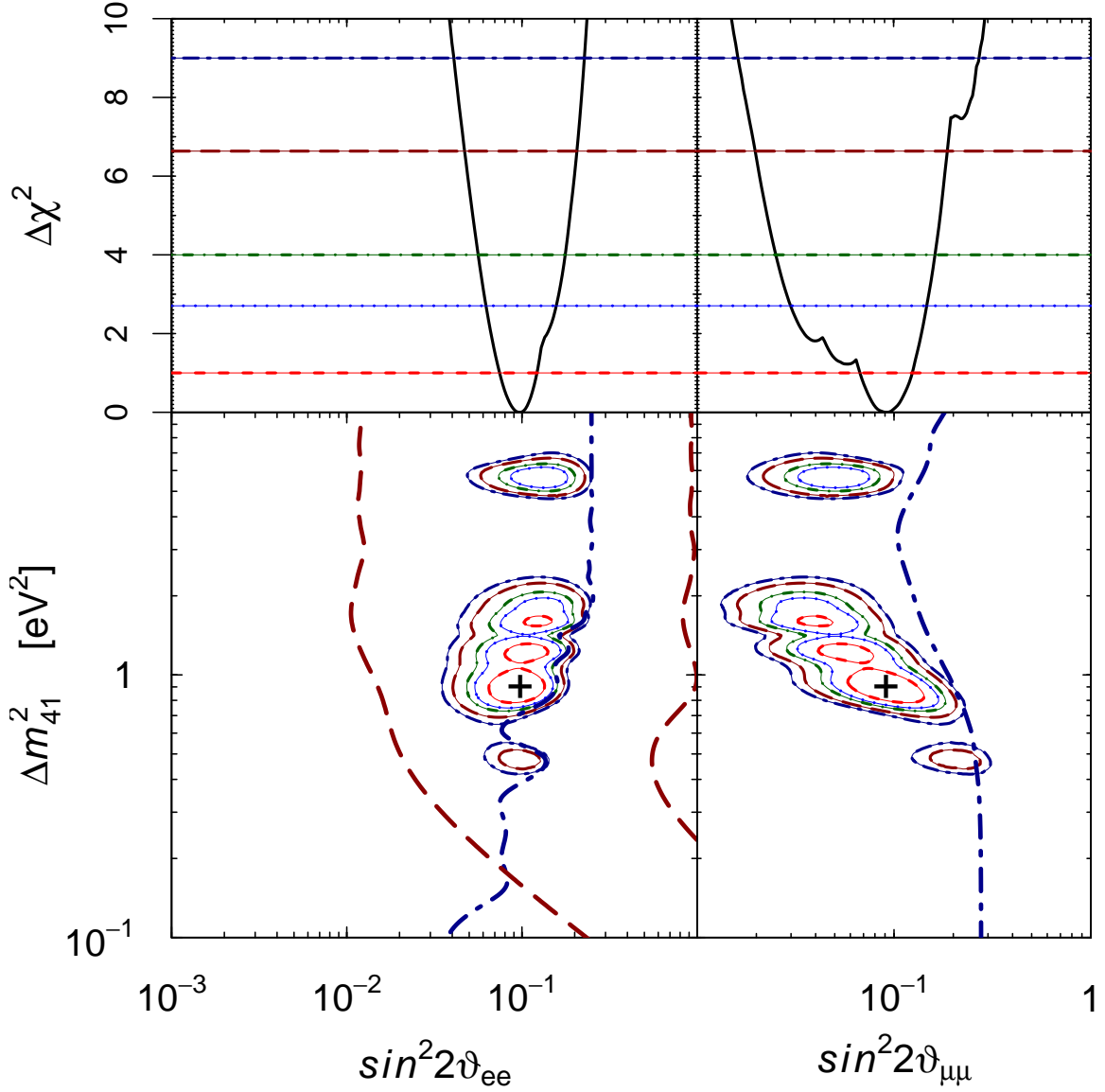}
\caption{ \label{3p1-see-smm}
Allowed regions in the $\sin^{2}2\vartheta_{ee}$--$\Delta{m}^2_{41}$ and
$\sin^{2}2\vartheta_{\mu\mu}$--$\Delta{m}^2_{41}$ planes and marginal $\Delta\chi^{2}$'s for
$\sin^{2}2\vartheta_{ee}$ and $\sin^{2}2\vartheta_{\mu\mu}$
obtained from the global fit of all the considered data in 3+1 schemes.
The best-fit point corresponding to $\chi^2_{\text{min}}$ is indicated by a cross.
The line types and color have the same meaning as in Fig.~\ref{3p1-sem}.
The isolated dark-blue dash-dotted lines are the $3\sigma$ exclusion curves
obtained from reactor neutrino data in the left plot
and from CDHS and atmospheric neutrino data in the right plot.
The isolated dark-red long-dashed lines delimit the region allowed at 99\% C.L.
by the Gallium anomaly.}
\end{figure}

Figures~\ref{3p1-sem} and \ref{3p1-see-smm} show the allowed regions in the
$\sin^{2}2\vartheta_{e\mu}$--$\Delta{m}^2_{41}$, $\sin^{2}2\vartheta_{ee}$--$\Delta{m}^2_{41}$ and
$\sin^{2}2\vartheta_{\mu\mu}$--$\Delta{m}^2_{41}$ planes, respectively, together with 
the marginal $\Delta\chi^{2}$'s for $\Delta{m}^2_{41}$, $\sin^{2}2\vartheta_{e\mu}$, $\sin^{2}2\vartheta_{ee}$
and $\sin^{2}2\vartheta_{\mu\mu}$. 

The proposed NESSiE experiment aims at exploring these regions.

\clearpage

\subsection{3+1 and CPT violation model \label{sec:3+1}}

The only implementation among 3+1 models able to fit global data
is the 3+1 and CPT violation model of Giunti-Laveder~\cite{Giunti:2010zu, Giunti:2010zs}\footnote{Indeed 
also``Non Standard Neutrino Interactions'' or quantum decoherence have been
proposed~\cite{Akhmedov:2011zz}.}.
The model was inspired by the analysis of the electron neutrino data of the Gallium radioactive source experiments
and the electron antineutrino data of the Bugey~\cite{bugey} and Chooz~\cite{Chooz-final} reactor experiments
in terms of neutrino oscillations allowing for a CPT-violating difference of the squared-masses and mixings of
neutrinos and antineutrinos.

It was found that the discrepancy between the disappearance of electron neutrinos
indicated by the data of the Gallium radioactive source experiments
and the limits on the disappearance of electron antineutrinos
given by the data of reactor experiments reveal a positive
CPT-violating asymmetry of the effective neutrino and antineutrino mixing angles.
If there is a violation of the CPT symmetry,
it is possible that the effective parameters governing neutrino and antineutrino oscillations
are different.
From a phenomenological point of view, it is interesting to
consider the neutrino and antineutrino sectors independently,
especially in view of the experimental tests considered in this proposal.

The parameters of the model are reported in Tab.~\ref{Tab:trepuno}.

\begin{table}[htbp]
\begin{center}
 \begin{tabular}{|ccc|ccc|}
	  \hline
  $\Delta m^2_{41}$ & $|U_{e4}|$ & $|U_{\mu 4}|$ & 
   $\overline{\Delta m}^2_{41}$ & $|\overline{U}_{e4}|$ & $|\overline{U}_{\mu 4}|$ \\
\hline
   1.92 & 0.275 & 0.0   &     0.47 & 0.068 & 0.886 \\
 \hline
 \end{tabular}
\end{center}
\caption{Best fit parameters of the 3+1 and CPT violation model}\label{Tab:trepuno} 
\end{table}

\begin{figure}[htbp]
   \includegraphics[width=0.90\textwidth]{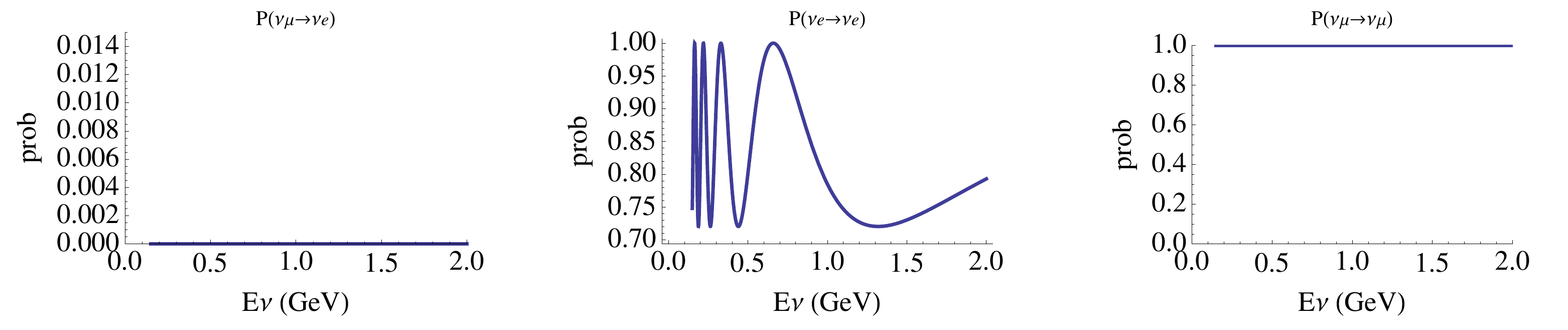}\\
   \includegraphics[width=0.90\textwidth]{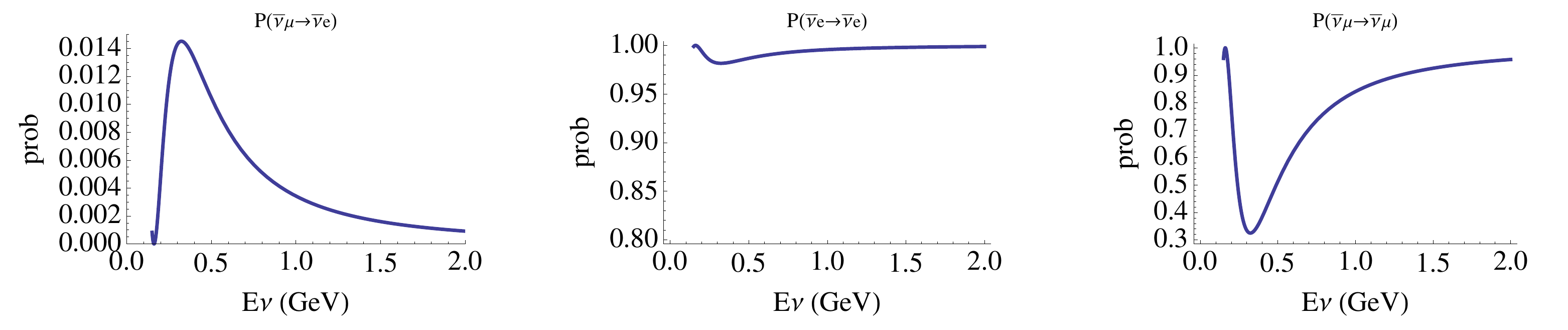}
   \caption{Oscillation probabilities computed for the  3+1 and CPT violation model.
  \pnumunumu and \pnumunue are not displayed because
they are predicted to be null by the model.
}
    \label{fig:3+1Prob}
\end{figure}

In this scenario, neutrinos undergo \nuenue transitions only,
see Fig.~\ref{fig:3+1Prob}, while
antineutrinos have a much richer phenomenology~\cite{Giunti:2010jt, Giunti:2010uj}.

\begin{figure}[htbp]
   \includegraphics[width=0.45\textwidth]{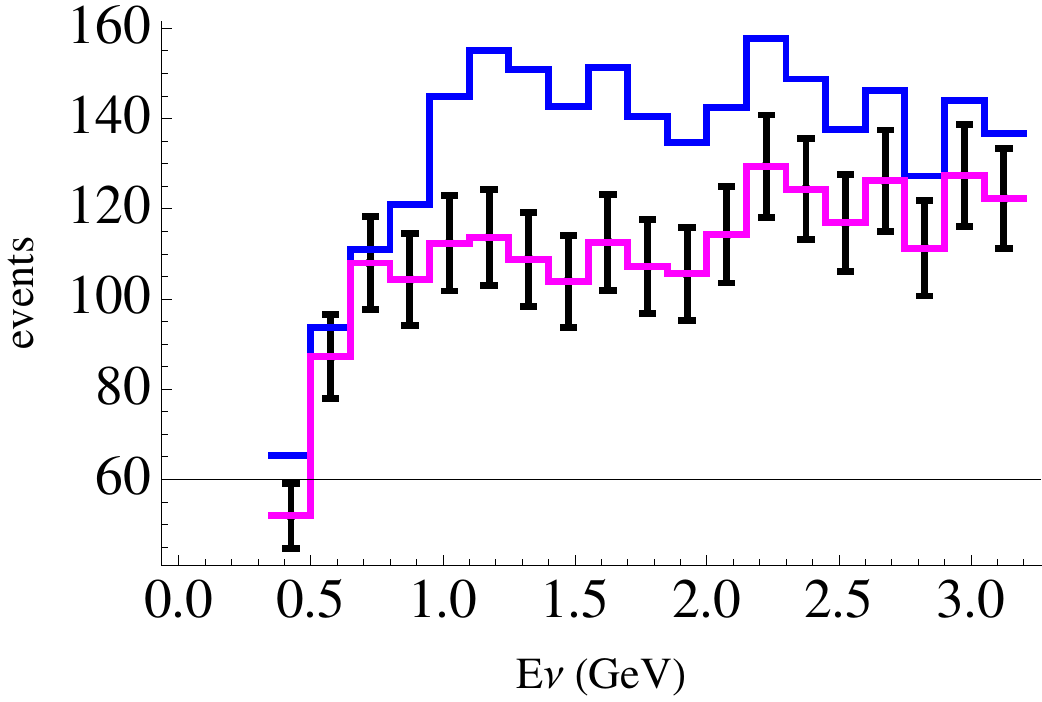}
   \includegraphics[width=0.42\textwidth]{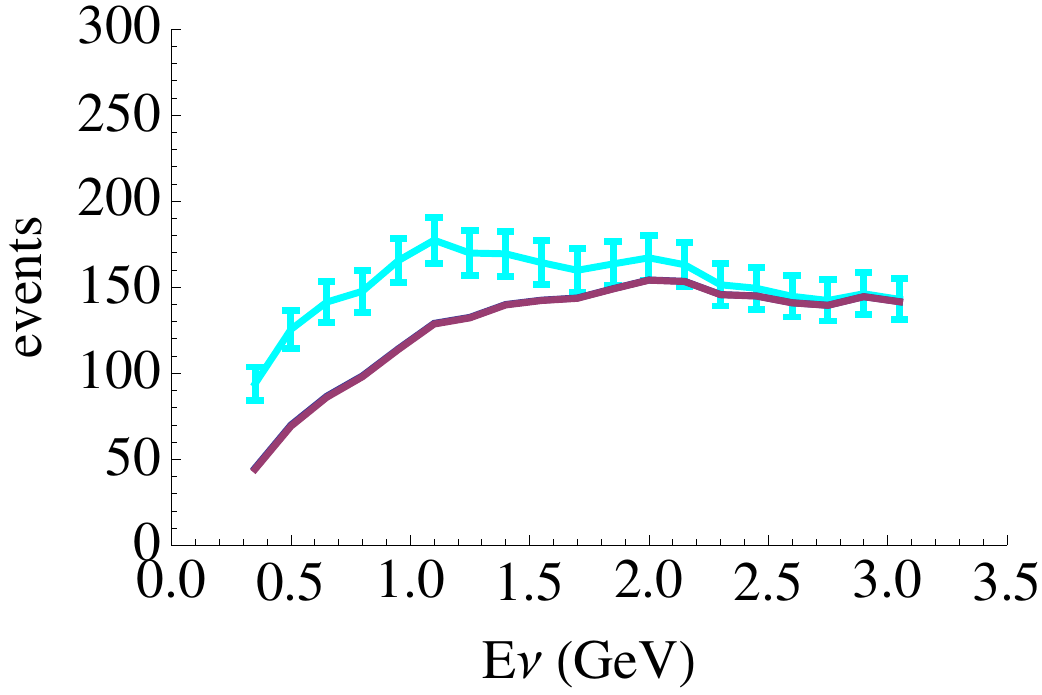}
   \includegraphics[width=0.85\textwidth]{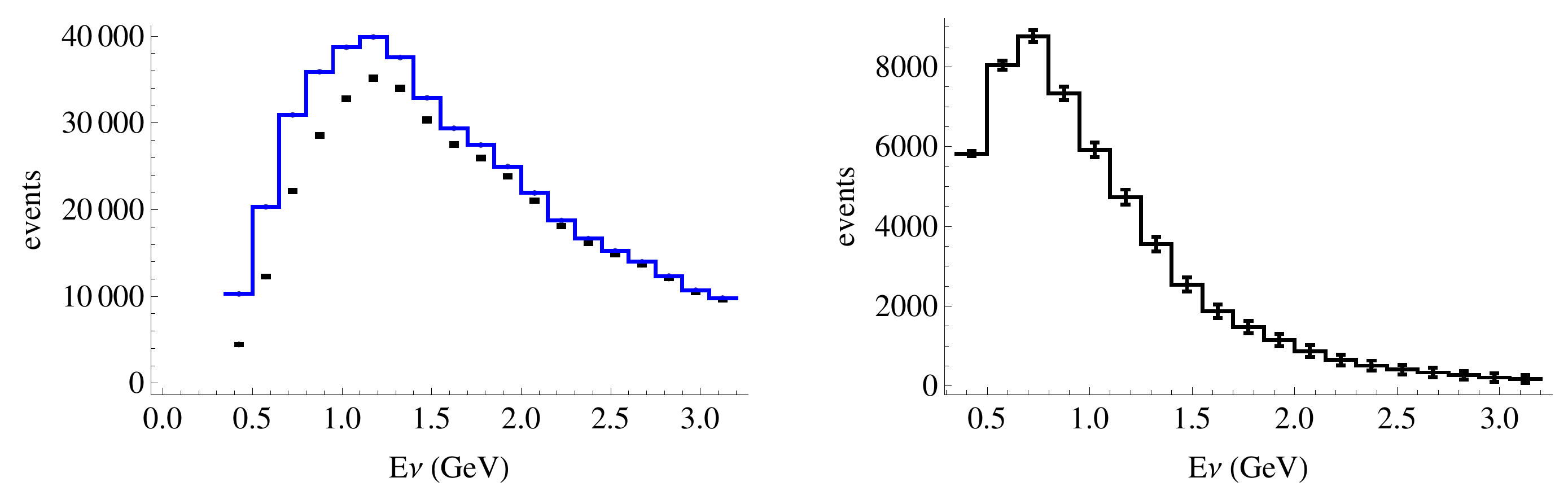}
   \caption{Top panel: \nue  event rates in the negative focussing beam (left) and \nubare event rates in the positive 
   focussing beam (right) computed assuming the 3+1 and CPT violation model. Non-oscillated rates are displayed as a blue histogram.
   Bottom panel:  \nubarmu  event rates in the  positive focussing beam:
 on the right absolute event rates computed with (point with errors) and
without (blue histogram) oscillations  following the 3+1 and CPT violation model;
on the left the difference between expected and measured events is shown.} 
   \label{fig:3+1}
\end{figure}

The \nue spectra computed for two years' run at 30 kW
 are reported in Fig.~\ref{fig:3+1}.
According to 3+1 and CPT violation model \nue disappearance should be clearly detectable 
1127  events detected against
a prediction of 1424 for $0.3 < E_\nu < 2$ GeV (Tab.~\ref{tab:3+1res}).

\begin{table}[htbp]
\begin{center}
\begin{tabular}{|ccccc|}
	\hline
{\bf events} & & No-Osc. & Disapp. & Disapp. + App. \\
 $\nu$ mode &\nue (0.3-2 GeV)  & 1424 & 1127 & 1127 \\	  
$ \overline{\nu}$ mode & \nue (0.3-2 GeV)  & 1260 & 1254 & 1699 \\	  
$ \overline{\nu}$ mode & \numu(0.3-2 GeV)  & 329328 & 277335&  \\	  
	\hline
\end{tabular}
\end{center}
\caption{Events expectation for the 3+1 and CPT violation model}\label{tab:3+1res} 
\end{table}

 
\clearpage 
 
\section{The Neutrino Beam}\label{sec:beam}
The NESSiE detector is planned for exposure to the CERN-PS neutrino beam-line, 
originally used by the BEBC/PS180 collaboration~\cite{BEBC} and re-considered later
by the I216/P311~\cite{I216_99} proposal. The baseline setup 
used by BEBC/PS180 consisted in a 80 cm long, 6 mm diameter 
beryllium-oxide target followed by a single pulsed magnetic horn operated
at 120 kA. The PS can deliver 3 $\cdot$ 10$^{13}$ protons per cycle
at 19.2 GeV kinetic energy in the form of 8 bunches of about 60 ns 
in a window of 2.1 $\mu s$,  integrating about 1.25 $\cdot$ 10$^{20}$ protons on target (p.o.t.)/year
under reasonable assumptions\footnote{180 days' run per year allocating one third of the protons are assumed, 
with present performances.}.
%
The existing decay tunnel, which has a cross section of 3.5 $\times$ 2.8 m$^2$ for 
the first 25~m of length and 5.0 $\times$ 2.8 m$^2$ for the remaining 20~m, 
is followed by a 4 m thick iron shield and 65~m of earth.
With respect to the original configuration, 
the target and the horn must be redesigned and reconstructed due to 
the present level of radioactivity while the proton beam line magnets and supplies 
could be recovered (see Sect.~\ref{sect:rea}).

\subsection{Beam simulation}
The BEBC/PS180 fluxes were reproduced by 
I216/P311 in their Letter of Intent~\cite{I216_97} using a simulation based 
on GEANT3 and GFLUKA.
The comparison was updated using a simulation adapted from the 
one used for the CNGS beam for their proposal~\cite{I216_99}. 
For this memorandum a GEANT4 and FLUKA~\cite{fluka} based simulation
has been developed to profit of a modern programming framework 
and to investigate possible improvements in the beam performance.
 
The generation of proton-target interactions is done with FLUKA-2008 while
GEANT4 is used for tracking in the magnetic field and the materials and
for the treatment of meson decays. The simulation program, thoroughly 
described in~\cite{mySPLarticle}, has been modified to take finite-distance effects into account, which are
particularly important due to the short baselines involved 
(127 and 885 $m$ measured from the target to the beginning of the LAr detector). 
Neutrinos crossing the LAr and Spectrometer volumes are directly scored
using a full simulation avoiding any weighting approach\footnote{This technique
is relatively CPU-consuming but nevertheless affordable.}. A sample of $10^7$
simulated p.o.t. was produced and has been used in the following.

\begin{figure}
\centering
\includegraphics[scale=0.3]{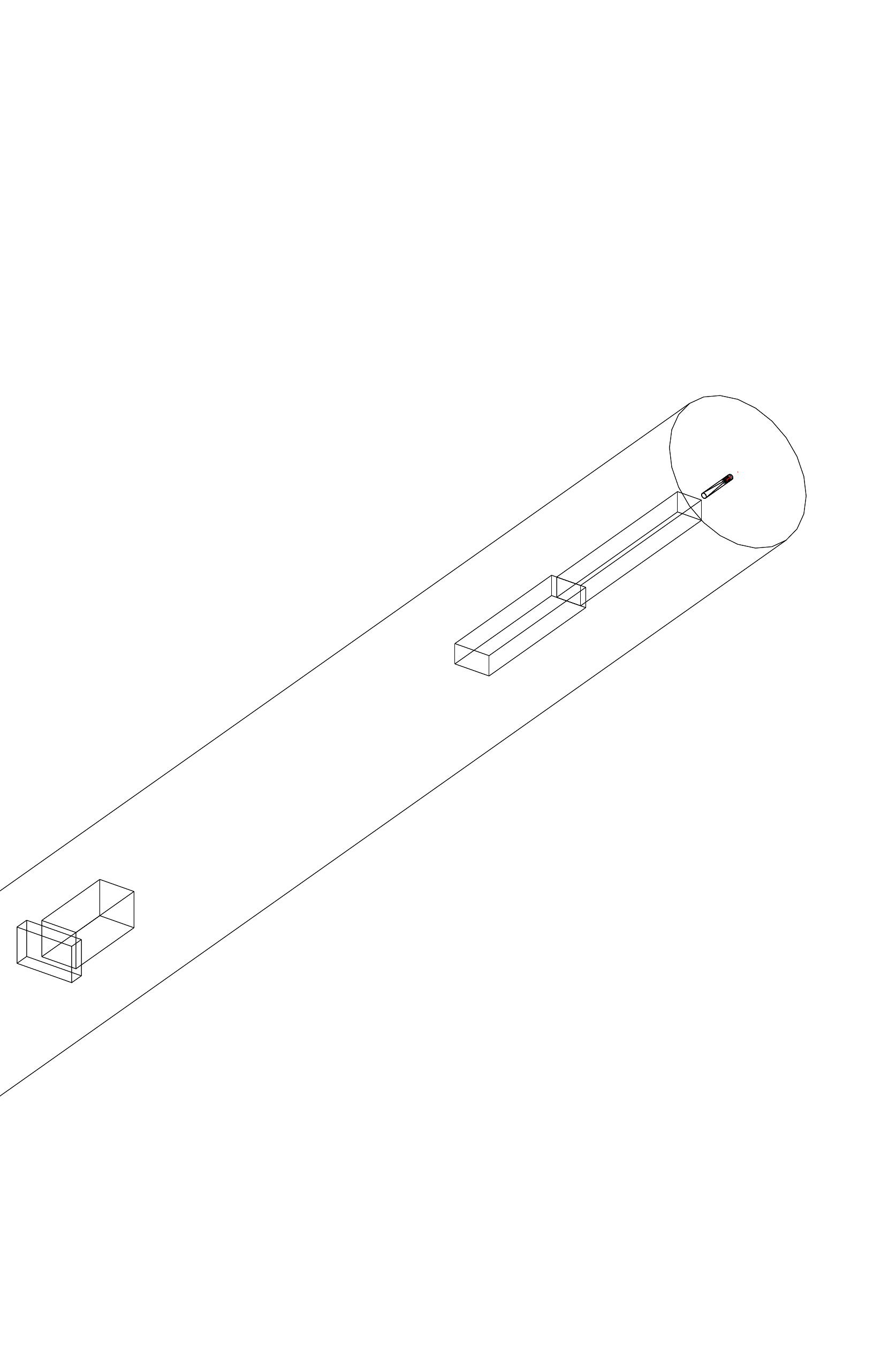} 
\caption{Layout of the focussing system and Near detector station in the GEANT4 simulation.}
\label{fig:layout}
\end{figure}

In order to benchmark the GEANT4 simulation the existing setup used
for the BEBC/PS180 experiment has been reproduced~\cite{roberta}. 
The layout of the considered volumes is shown in Fig.~\ref{fig:layout}.

\begin{figure}
\centering
\includegraphics[scale=0.6]{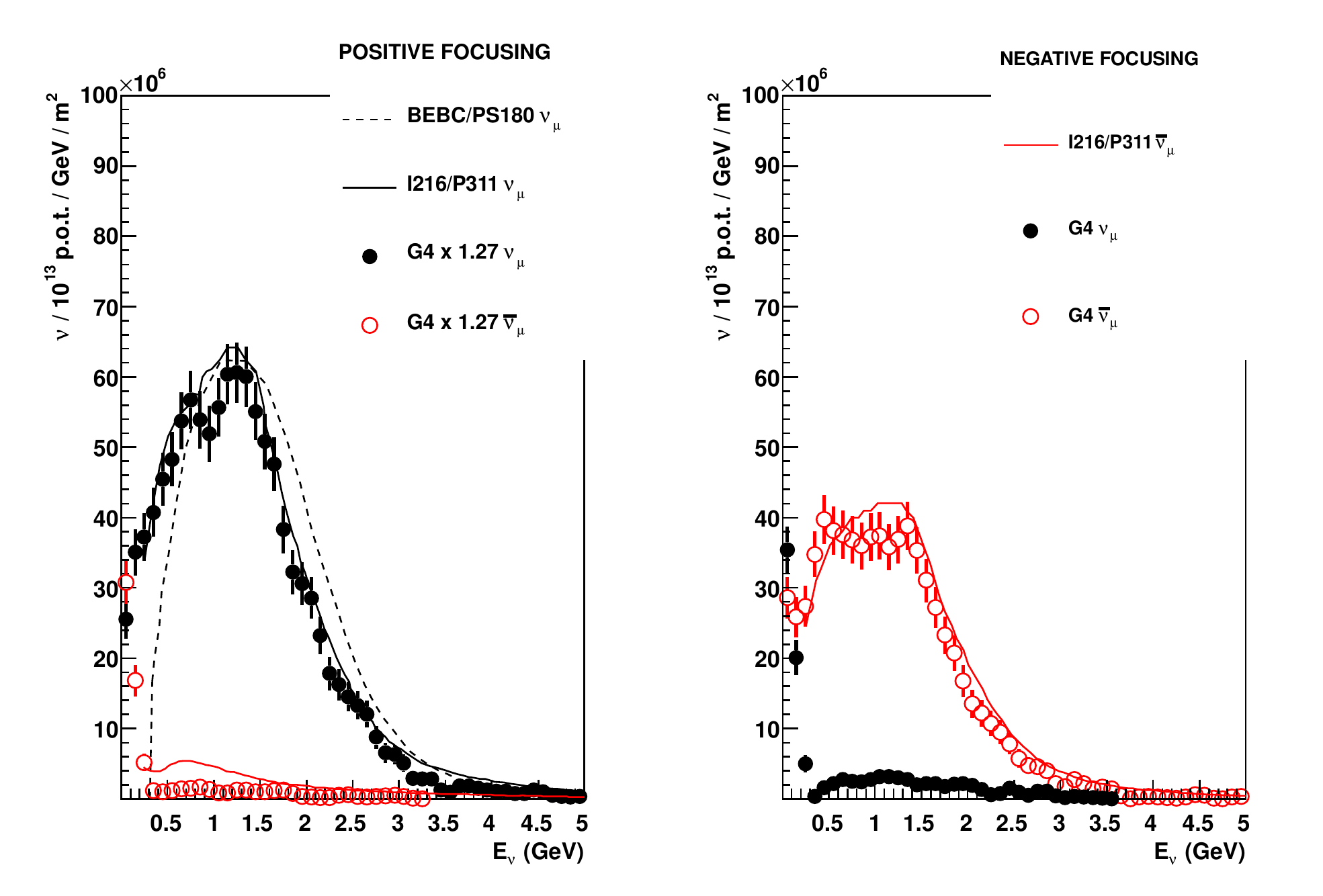} 
\caption{Comparison of the present simulation (dots) with those of BEBC 
(dashed line) and I216/P311 (continuous line) for positive-focussing mode (left)
and negative-focussing mode (right). G4 stands for the {\em new} simulation obtained with GEANT4 which 
introduces a 1.27 reduction factor with respect to the {\em old} simulation.
Fluxes are referred a distance of 825 $m$ from 
the target (BEBC site). The target starts at 35 $cm$ upstream of the BEBC/PS180 horn,
which is operated at 120 kA.}
\label{fig:prev_comp}
\end{figure}

%
%
The spectrum shape of the $\nu_\mu$ is in good agreement with that calculated by the
I216/P311 experiment whereas the obtained normalization is instead  27\% lower (Fig. \ref{fig:prev_comp}). 
Investigations are ongoing to understand this difference, in particular the implementation 
of the geometry and the hadro-production models are under checking.
%
The flux reduction resulting from negative-focussing, which 
amounts to about 40\%, is well reproduced.

\begin{figure}[htbp]
\centering
\includegraphics[scale=0.6]{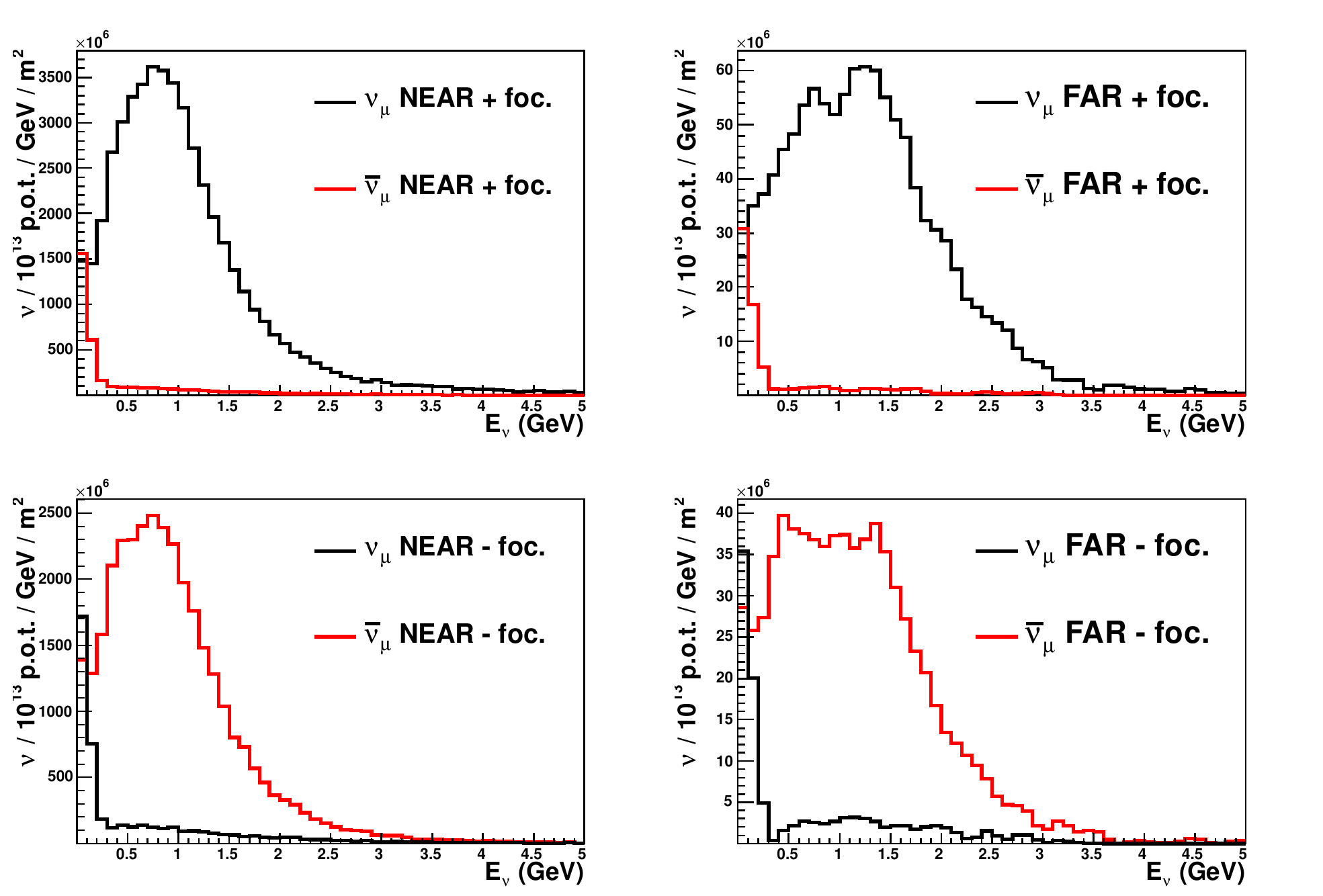} 
\caption{$\nu_\mu$ fluxes (black) and $\bar{\nu}_\mu$ fluxes (red) at the Near 
(left column) and Far (right column) Spectrometers in positive-focussing 
(upper row) and negative-focussing (lower row) modes.}
\label{fig:fluxes}
\end{figure}

Neutrino fluxes at the Far and Near Spectrometers in positive and negative-focussing 
are shown in Fig.~\ref{fig:fluxes}. Accounting for the Spectrometer geometries 
and locations we obtain a ratio $\nu$/p.o.t. of about $10^{-2}$ in the Near station 
and a further reduction of a factor of about 20 in the Far location. 
The fractions of ($\nu_\mu$, $\bar{\nu}_\mu$, $\nu_e$, $\bar{\nu}_e$) in the Near
Spectrometer are (92.3, 6.6, 0.87, 0.26)\%
in positive-focussing mode and  
(12.2, 86.6, 0.49, 0.75)\% 
in negative-focussing mode.
In the Far detector numbers are very close to the previous ones. The electron neutrino contamination is in 
agreeement with the estimates of I216/P311. 

\begin{figure}[htbp]
\centering
\includegraphics[scale=0.6]{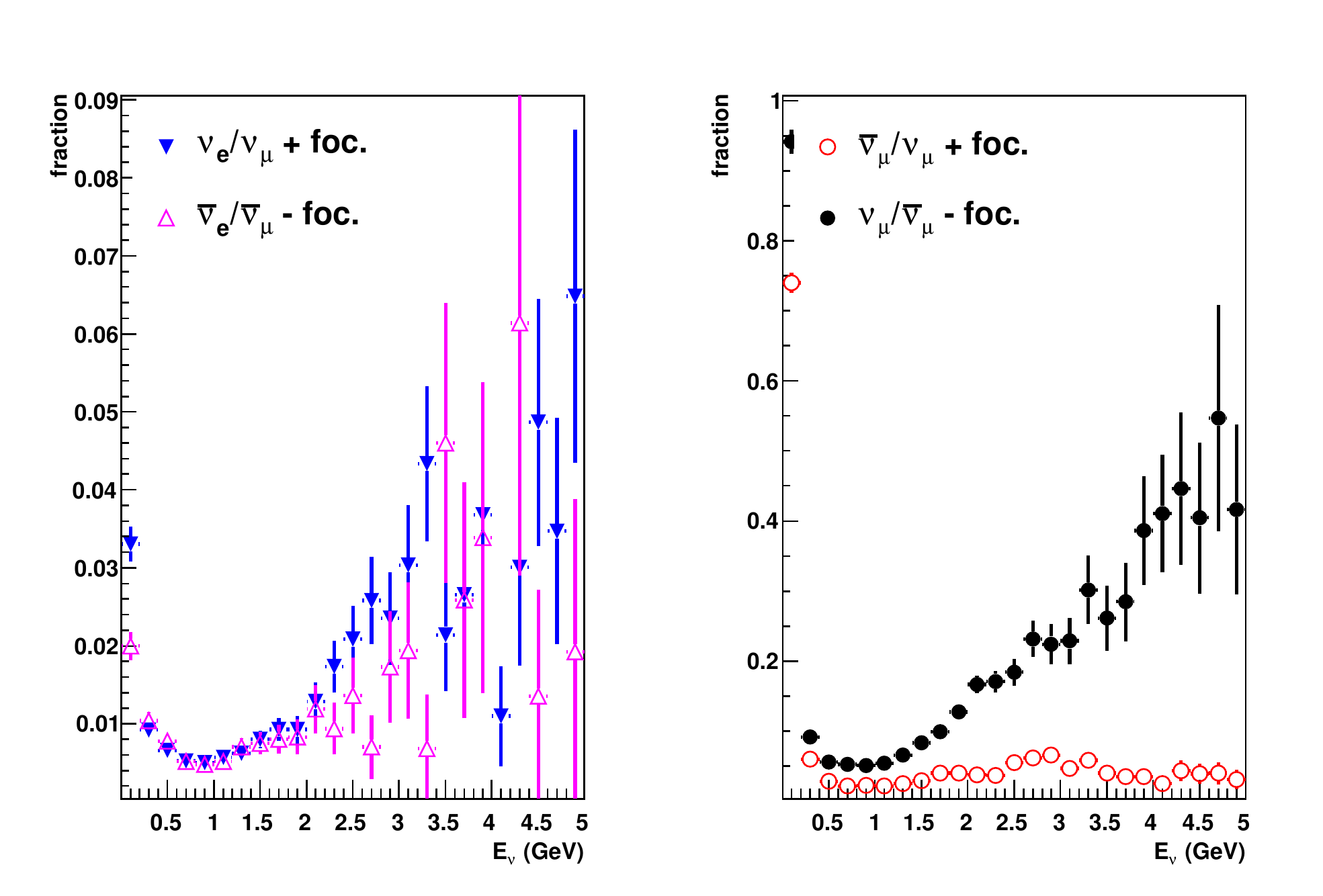} 
\caption{
Left: $\nu_e/\nu_\mu$ ratio in positive-focussing mode
and $\bar{\nu}_e/\bar{\nu}_\mu$ ratio in negative-focussing mode
in the Near Spectrometer as a function of the neutrino energy. 
Right: $\bar{\nu}_\mu/\nu_\mu$ ratio in positive-focussing mode 
and $\nu_\mu/\bar{\nu}_\mu$ ratio in negative-focussing mode in the Near 
Spectrometer as a function of the neutrino energy.}
\label{fig:contamination}
\end{figure}

The energy dependence of the electron and wrong-sign muon contamination 
is shown in Fig.~\ref{fig:contamination}.
It can be noticed that the $\nu_e$ contamination in the energy region between 500 $MeV$
and 1 $GeV$ is very low, approaching a minimum of 0.4\%, to the benefit of the $\nu_e$ 
appearance search. The contamination of $\bar{\nu}_\mu$ 
in the $\nu_\mu$ beam is quite significant especially at high energy where the
spectrometer charge separation becomes thus very important.

\begin{figure}
\centering
\includegraphics[scale=0.4]{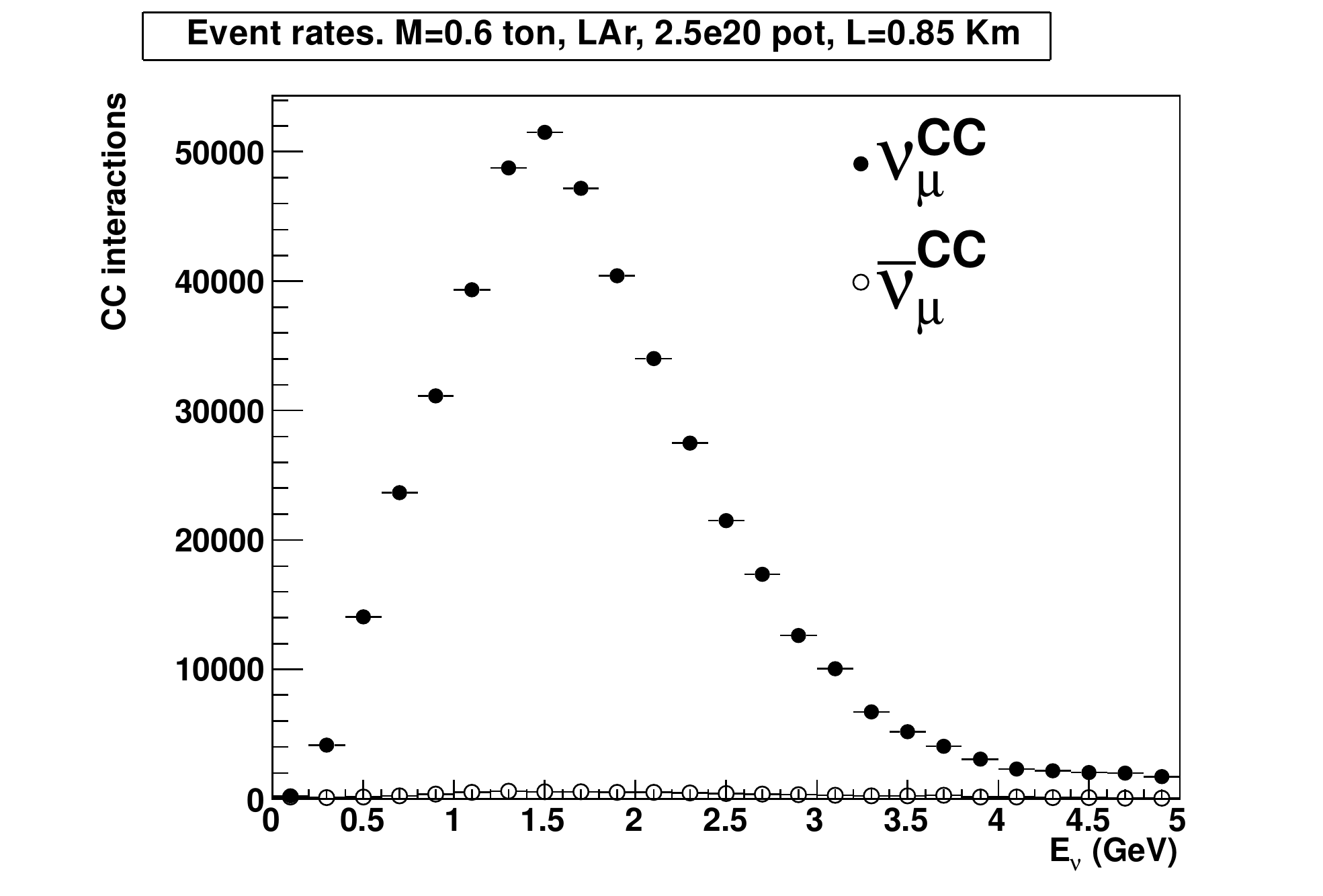}%
\includegraphics[scale=0.4]{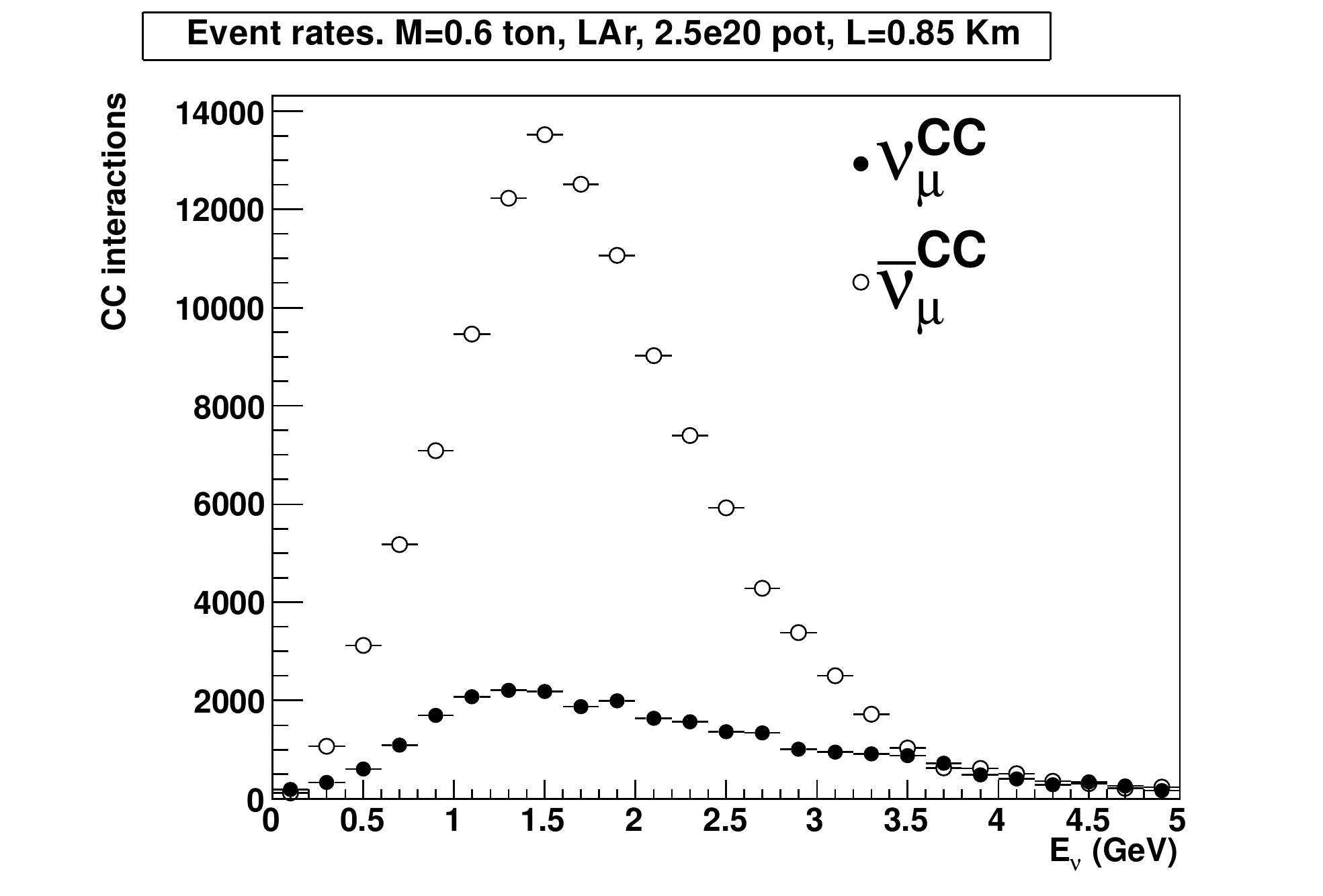}
\caption{Comparison of $\nu_\mu^{CC}$ and $\bar{\nu}_\mu^{CC}$ event rates for 
the positive (left) and negative-focussing (right) beams.}
\label{fig:rates}
\end{figure}

In particular, after folding the spectrum with the CC
cross section (Fig.~\ref{fig:rates}), one can see that the 
negative-focussing beam at high energy
yields a comparable mixture of $\mu^-$ and $\mu^+$.
The Spectrometer charge separation allows studying the behavior
of both CP states at the same time.

\begin{figure}
\centering
\includegraphics[scale=0.6]{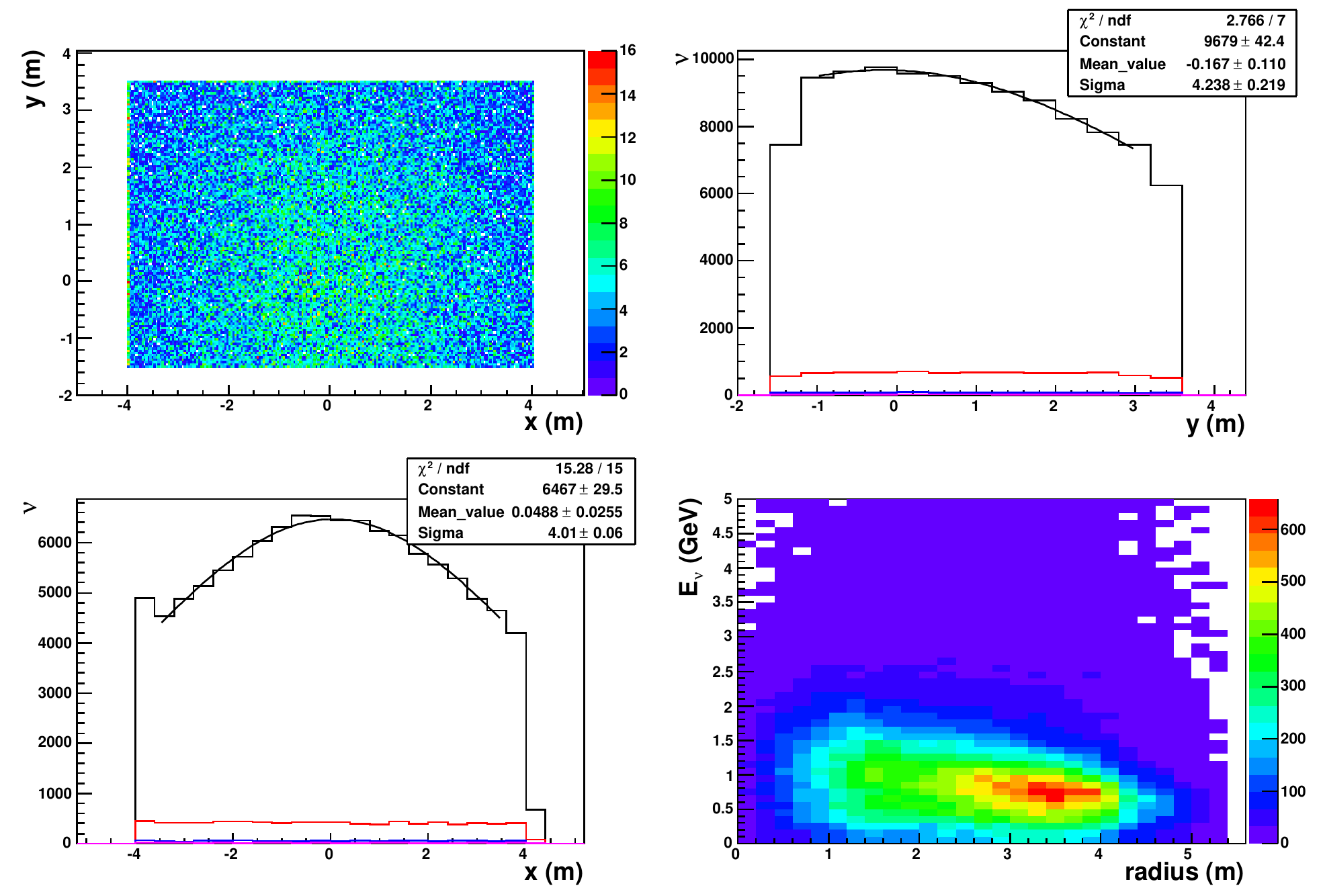} 
\caption{Up left. Distribution of the neutrino impact points
in the transverse plane, $X$ vs $Y$. Up right: $X$ projection. 
Bottom left: $Y$ projection. The black histogram represents the 
$\nu_\mu$, the red histogram the $\bar{\nu}_\mu$. Bottom right: 
$E_{\nu_\mu}$ vs $\sqrt{X^2+Y^2}$. Results refer to positive-focussing mode.}
\label{fig:spread}
\end{figure}

The distribution of the impact points of the $\nu_\mu$
in the Near Spectrometer is shown in the upper left plot of Fig.~\ref{fig:spread}.
The center of the beam is displaced in the bottom direction of 1 m and
the radial beam profile can be well fitted with a Gaussian having a $\sigma$ of about 4 m
(Fig.~\ref{fig:spread} upper right and lower left plots)

\begin{figure}
\centering
\includegraphics[scale=0.4]{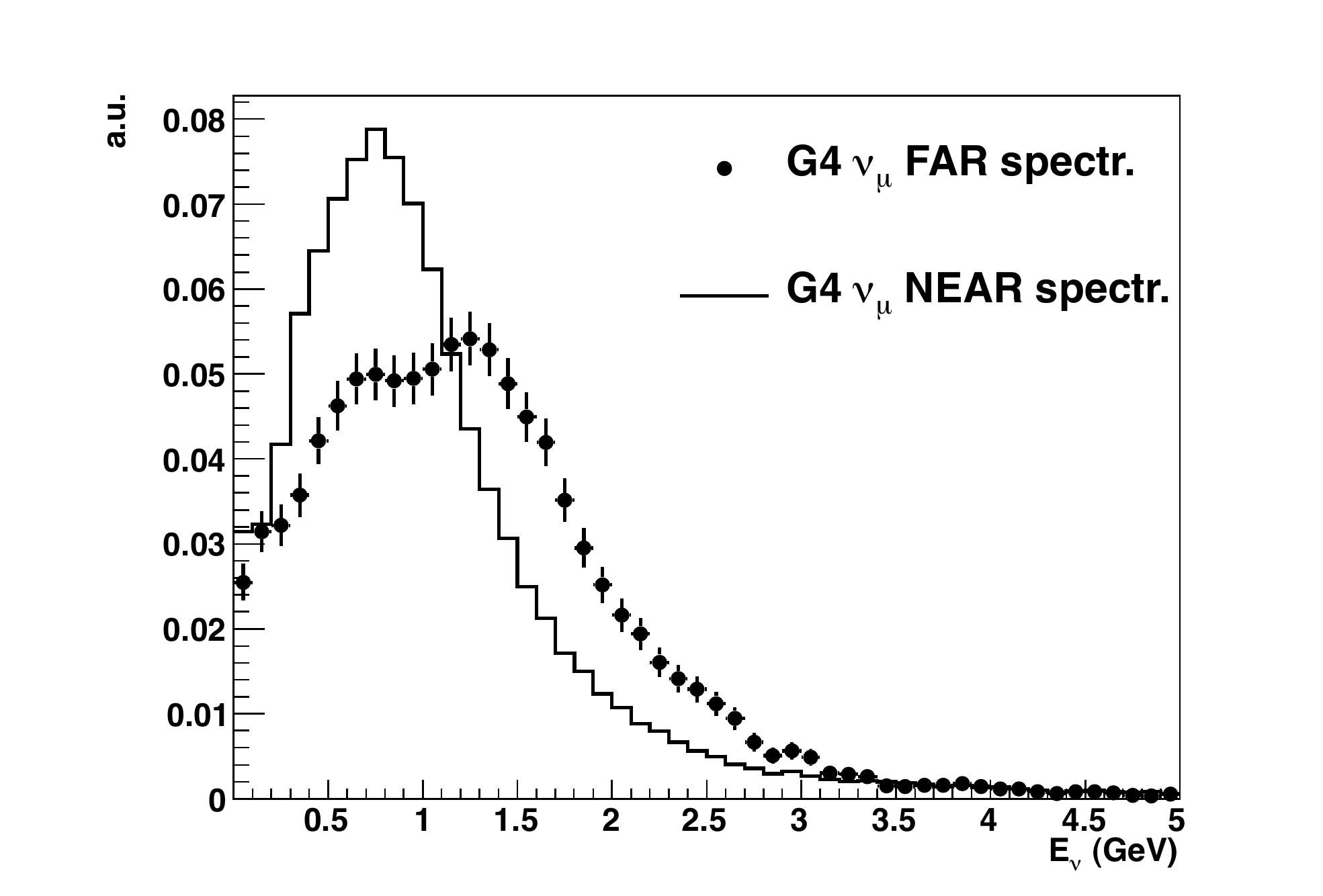}%
\includegraphics[scale=0.4]{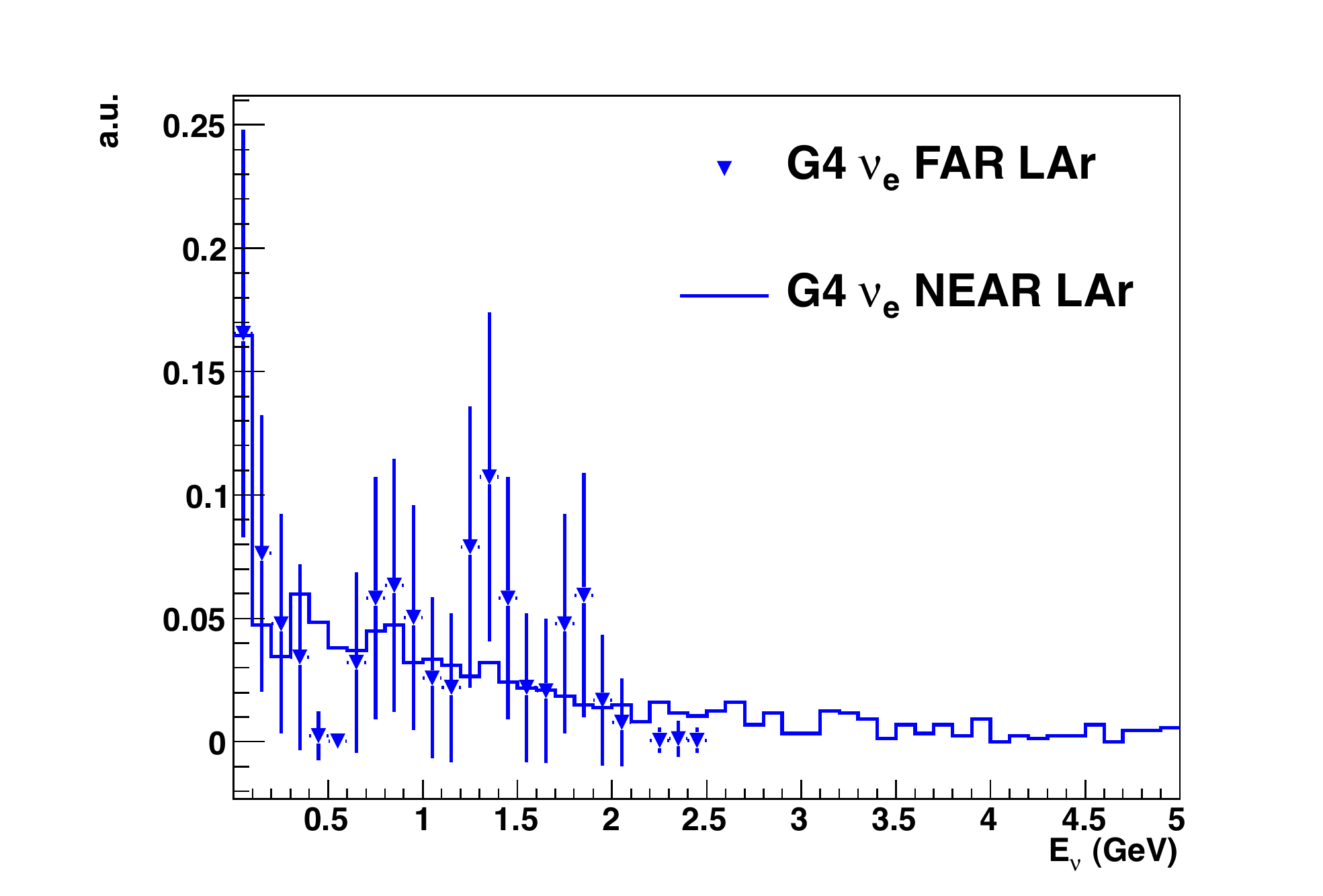} 
\caption{Comparison of the $\nu_\mu$ (left) and $\nu_e$ (right) 
fluxes in the Near and Far LAr detector in positive-focussing mode. G4 stands for the {\em new} simulation obtained with GEANT4 
which introduces a 1.27 reduction factor with respect to the {\em old} simulation.}
\label{fig:nuNF}
\end{figure}

The spectrum of $\nu_\mu$ at the Near detector peaks at lower energies
due to the fact that many interacting neutrinos are off-axis and then
tend to have a lower mean energy (Fig \ref{fig:nuNF}, left). 
Nevertheless, restricting to the central region, i.e. the region that subtends 
an angle similar to that of the Far detector,
the shapes tend to get closer. This selection allows smaller corrections
in the Near/Far ratio thus decreasing the systematics. The correlation between the
energy and the radius of the neutrino impact point is shown the lower right 
plot of Fig.~\ref{fig:spread}. The shape of the $\nu_e$ spectrum on the other
hand is more similar in the Near and Far locations (Fig \ref{fig:nuNF}, right)
due to the predominantly 3-body decay origin.

\subsection{New horn designs}

By keeping the basic geometry of the beam (size
of the target and horn hall, decay tunnel,
shielding) unchanged, studies of several modified horns and targets 
 in addition to the BEBC original setup are undergoing,
an optimized design of new and possibly improved beam optics being pursued.
%
Our aim is to move towards a high intensity $\nu_\mu$ flux in order to reduce the statistical errors,
peaking at $\langle E_{\nu_\mu}\rangle \sim$ 1 GeV to match  the $\Delta m^2 \sim 1$ eV$^2$ region.
For the reduction of systematics in the $\nu_\mu \to \nu_e$ appearance channel
the high energy tail above $\sim$ 2.5 GeV should be reduced to suppress $\pi^0$ production
and the intrinsic $\nu_e$ contamination should be kept small.

\begin{figure}[htbp]
\centering
\includegraphics[scale=0.6]{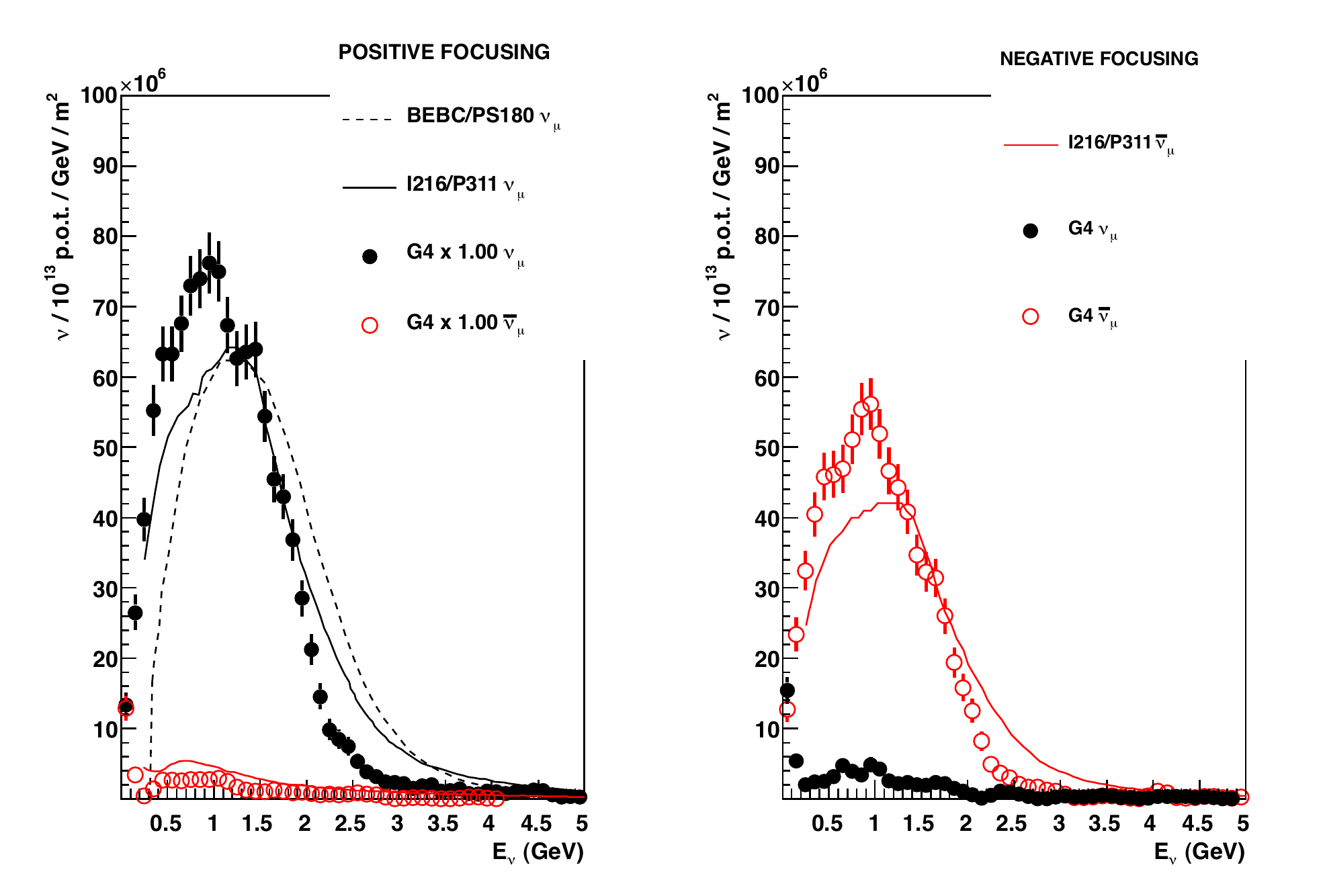} 
\caption{New horn fluxes compared to BEBC and I216/P311. The meaning of the symbols is as in Fig.~\ref{fig:prev_comp},
in particular G4 stands for the {\em new} simulation obtained with GEANT4 which introduces a 1.27 reduction factor
with respect to the {\em old} simulation.}
\label{fig:newhornflux}
\end{figure}

Preliminary studies performed with bi-parabolic horns \emph{\`a la} NuMI~\cite{numi} indicate
room for improvement. In particular in the last decades the feasibility
of horns pulsed at high currents of order 300 kA has been demonstrated.
The fluxes obtained with a new bi-parabolic horn pulsed at 246 kA, are shown in 
Fig.~\ref{fig:newhornflux}. An overall flux gain is obtained, particularly at low energy. 
This would be particularly useful to extend the sensitivity towards smaller $\Delta m^2$.

Since the interval of the interesting $\Delta m^2$ regions is large
it would be desirable to get a configuration capable of scanning different
neutrino energy regions thus adapting to the physics indications coming from 
the data taking.
Despite the fact that the overall features of the neutrino beam 
are dictated by the proton energy, an effective way for varying 
the mean energy consists in using a tunable target-horn distance.

\begin{figure}[htbp]
\centering
\includegraphics[scale=0.5]{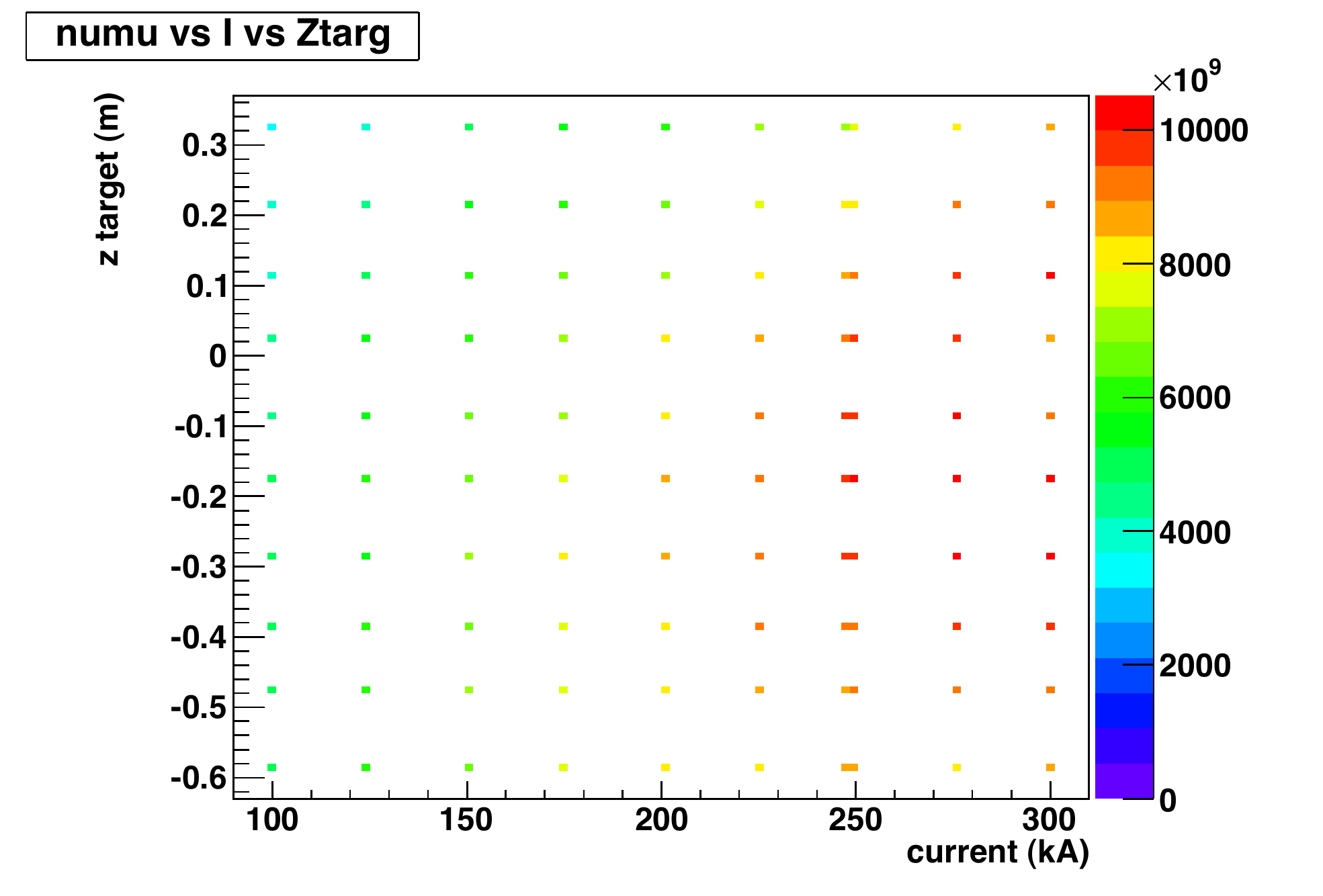}
\includegraphics[scale=0.5]{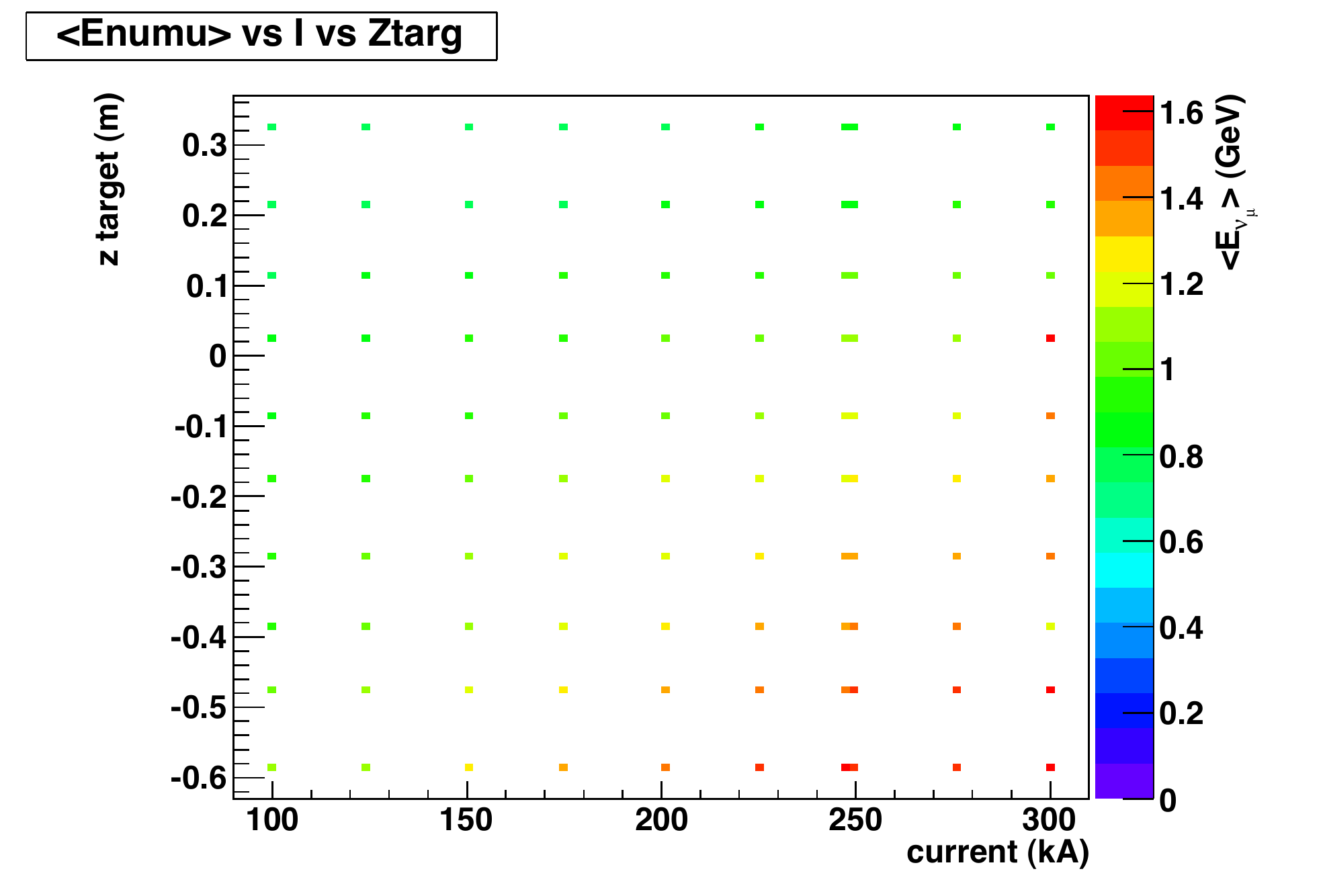} 
\includegraphics[scale=0.5]{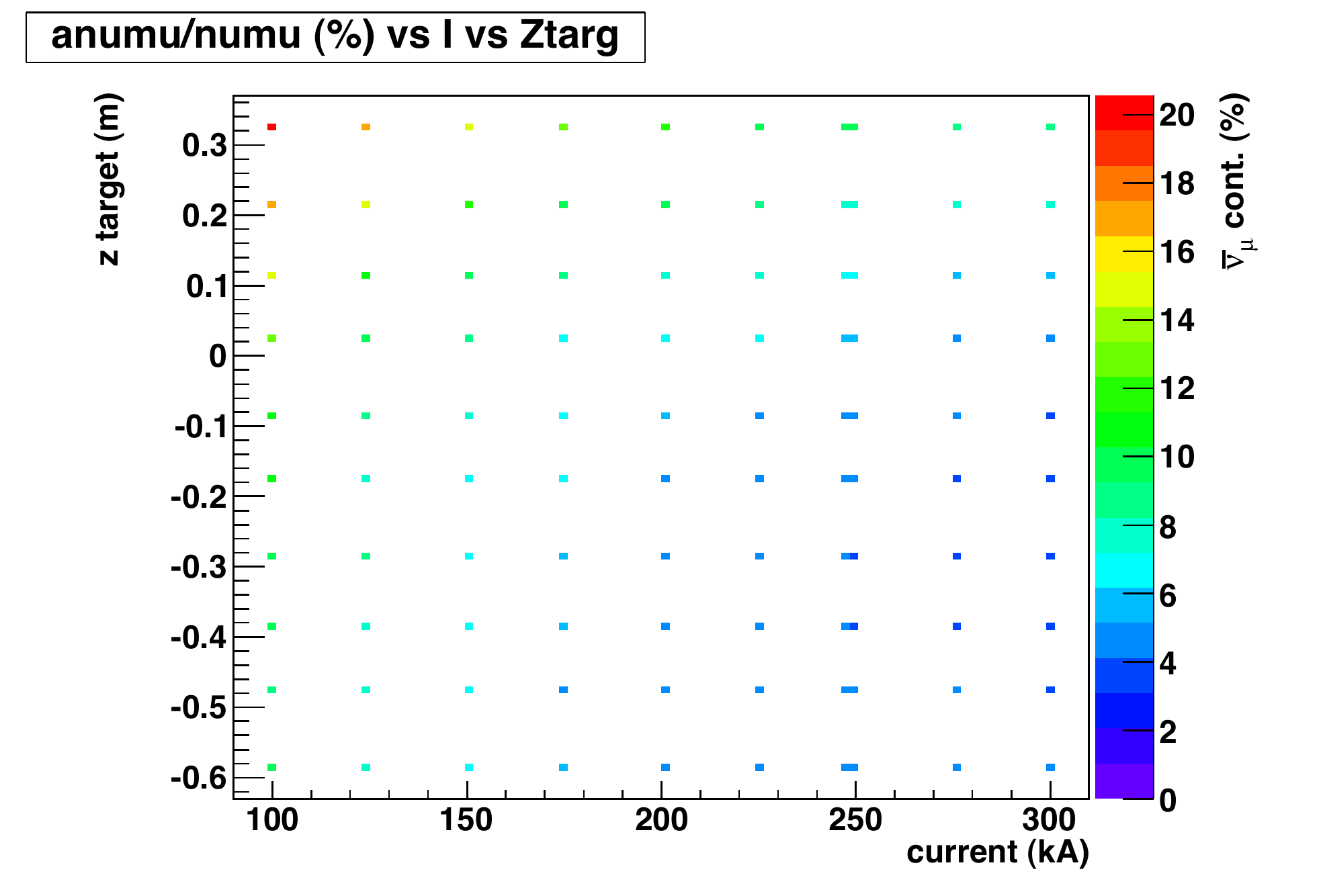} 
\caption{Dependence of the integral $\nu_\mu$ flux (top), the $\nu_\mu$ mean energy (middle) 
and the $\bar{\nu_\mu}$ contamination (bottom) from ($I$,~$z_{targ}$) with a test bi-parabolic horn.
Results are shown for positive-focussing at the Far location. }
\label{fig:trends}
\end{figure}

\begin{figure}[htbp]
\centering
\includegraphics[scale=0.6]{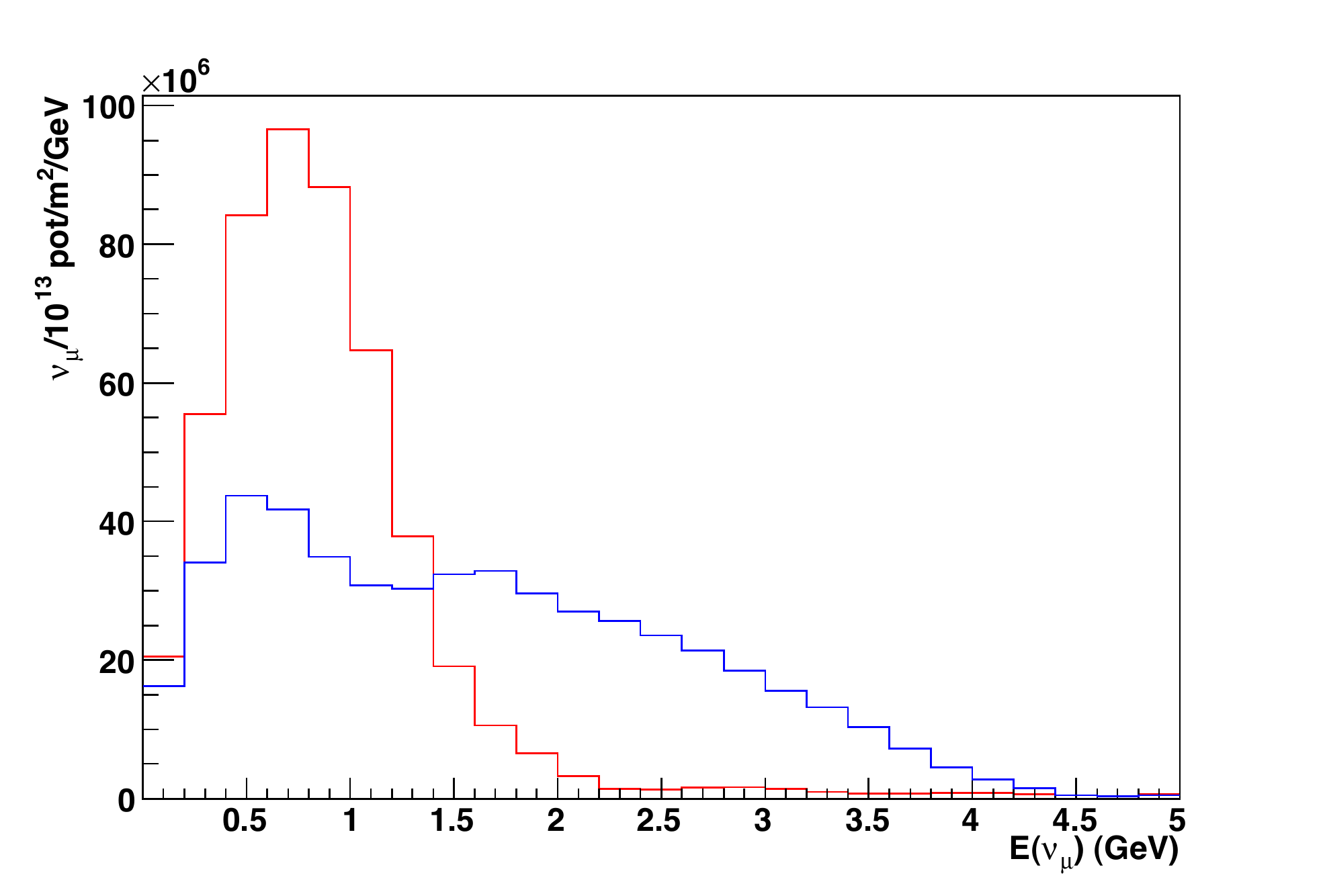} 
\caption{Two extreme low (red) and high-energy (blue) configuration fluxes arising 
from the scan in the ($I$, $z_{targ}$) plane. The current is 300 kA 
and the position of the beginning of the target with respect to 
the beginning of the horn is +32 and -58 cm respectively. The fluxes
refer to a distance of 825 m.}
\label{fig:Extremes}
\end{figure}

While keeping the shape of the original horn unchanged, a scan in the scatter plane 
(current ($i$) vs target longitudinal position ($z_{targ}$)) 
was performed to study the effects of the fluxes in terms 
of $\nu_\mu$ normalization, energy distribution and contaminations. 
Results are shown in Fig.~\ref{fig:trends}
for positive-focussing at the Far location. 
Plots show how the integral $\nu_\mu$ flux is increased
going to higher currents; pushing the target upstream (downstream)
corresponds to decrease (increase) the $\bar{\nu}_\mu$ contamination 
and to probe the high(low)-energy region.
%
%
Two configurations yielding very different spectra are
shown in Fig.~\ref{fig:Extremes}. It is interesting to notice
that these fluxes are obtained using the same current of 300 kA 
just varying the position for the target.

\subsection{Beam reactivation\label{sect:rea}}
Preliminary evaluations for a renovated TT7 PS neutrino
beam line have been already completed at CERN~\cite{PSrefur}. 
The TT7 transfer line, the target chamber and the decay tunnel are
in good shape and available for the installation of the proton beam line, 
target and magnetic horn. The main dipoles, quadrupoles, correction 
dipoles and possibly the transformer for the magnetic horn can be recuperated, 
reducing significantly the cost and time schedule.
The target and secondary beam focussing design can
profit of the CNGS experience as well as monitoring 
systems, primary beam steering and target alignment~\cite{Edda}. 

The study dating back to 1999~\cite{I216_99}
estimated the time required to be approximately 2 years 
for a total cost of about 4.2 MSF detailed as follows:
\begin{itemize}
\item power converters of TT1 and TT7 magnets: 
consolidation and lower level electronics (1.1 MSF);
\item civil engineering, mostly new housing for converters (0.5 MSF);
\item removal of 400 m$^3$ radioactive waste material in TT7, provisions and installations for radiation protections, access control ($<$ 0.5 MSF);
\item beam line installation, vacuum chamber, general mechanics (0.4 MSF);
\item beam monitoring instrumentation (0.4 MSF);
\item new target and horn (0.4 MSF);
\item new pillars and platform in the Near experimental hall (50 KSF).
\end{itemize}
%


%
%
%
%

\clearpage

\section{Spectrometer Design Studies}\label{sec:spect1}
The main purpose of a spectrometer placed downstream of the target section is to provide  
charge and momentum reconstruction of muons escaping from LAr detection.  
This choice would provide a double benefit with respect to a LAr detector running in stand-alone mode. 
Firstly a precise muon momentum reconstruction would allow a good kinematical closure of CC events 
occurred in LAr in particular in the high energy tail where the muon momentum resolution is poorer. 
Secondly muon charge separation would allow us to disentangle $\nu_\mu$ and $\bar{\nu}_\mu$ 
disappearance channels in particular in the negative-focussing option where wrong-sign contamination 
is larger. This would allow tackling CP and CPT violating scenarios in an unambiguous way. 

In addition, a proper mass-granularity combination would allow a coarse reconstruction of $\nu_\mu^{CC}$ events 
occurring within the Spectrometer itself and provide a way to use the Spectrometer in stand-alone mode, too. 
 
We characterize the detector by evaluating the  physics performances in terms of CC disappearance
and   the sensitiveness to ($\Delta m^2 , \sin^2 2\theta$) values predicted by the models 
described in Sect.~\ref{sec:physics}. We recall that with a suitable detector  the 
old CDHS \numu disappearance limit will be tested in just one day of data taking.

Moreover we  assumed some ``a priori" constraints by defining a realistic, conservative, relatively inexpensive apparatus.

\subsection{Magnetic field in Iron}

A magnetized iron spectrometer was chosen as baseline option. 
Its design was such to match the required sensitivity of the experiment with respect to the 
{\em sterile} neutrino models of Sect.~\ref{sec:physics}.
The basic parameters to be tuned are the transverse width\footnote{Here and in the following the  
transverse directions are defined by the horizontal $x$ and the vertical $y$ axis while the $z$ axis  
runs in the beam direction.}, the longitudinal dimensions, the iron slab thickness and the tracking 
detector resolution. 
The transverse and longitudinal dimensions constrain the detector acceptance and hence the detectable number of events 
which are directly related to the $|U_{\mu4}|$ and $|\overline{U}_{\mu4}|$ mixing parameters (see in particular
Table~\ref{tab:global-bfp} of Sect.~\ref{sec:physics}). 
On the other hand the target segmentation sets bounds on muon momentum and charge reconstruction and are 
directly related to $\Delta m^2_{41}$ and $\overline{\Delta m}^2_{41}$. 
 
The transverse width has to be large enough in order to maximize the detection of muons escaping 
from LAr. Monte Carlo simulation has shown that a transverse size 
matching the LAr acceptance ($\sim 8 \times 5$ m$^2$ in the Far site (FD) provides a good performance while still keeping the detector 
size at a feasible level (doubling the dimensions would improve the sensitivity to $|U_{\mu4}|$ by 5\%). 

\begin{figure}[htbp] 
\begin{center} 
\includegraphics[height=4.0in]{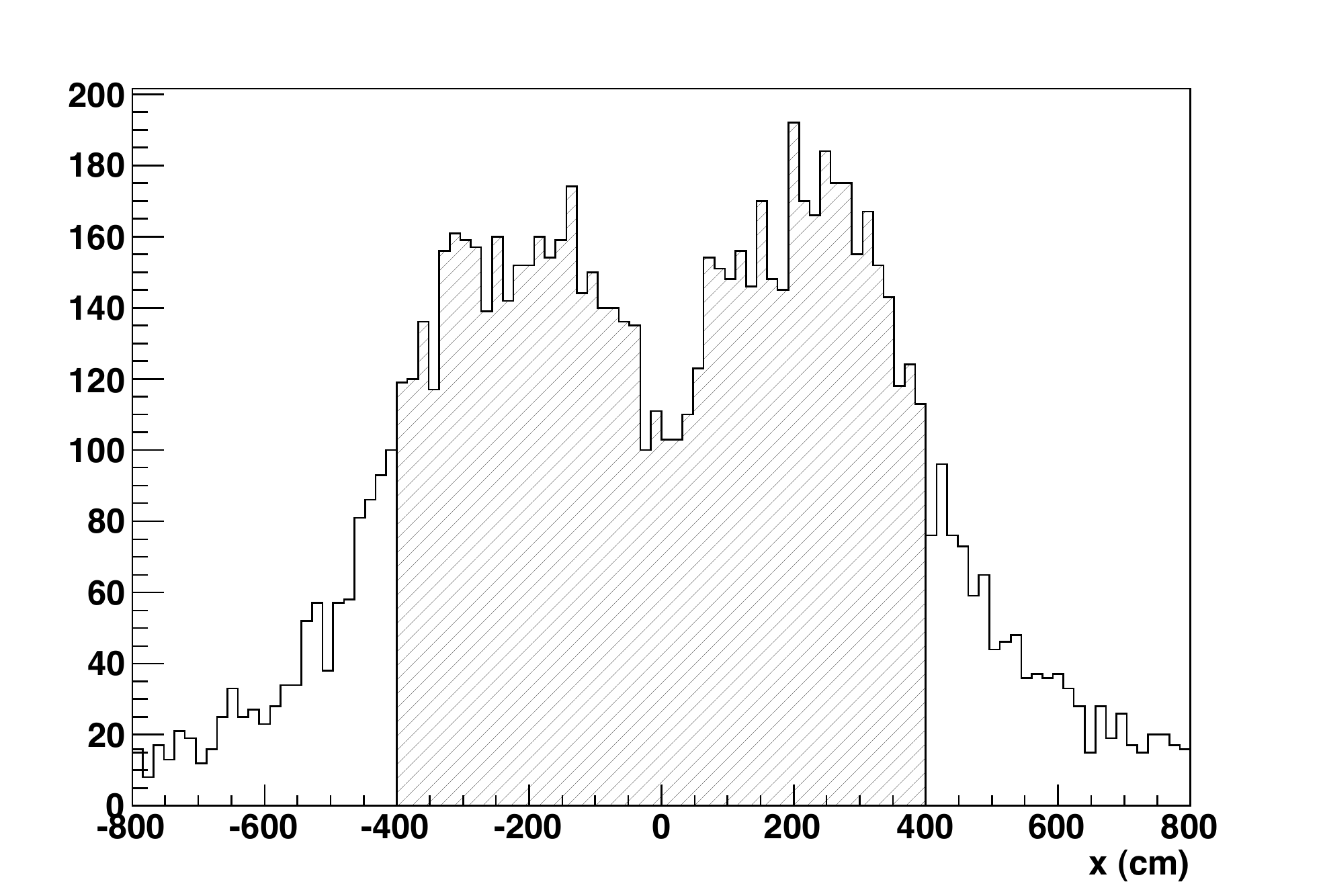} 
\caption{Distribution of muon impact points along the $x$ axis, at the Far site (FD). The dip at the center of the distribution and the sharp tails on the sides 
are due to the convolution with the LAr detector acceptance. The shaded area corresponds to the 78\% of the total one.} \label{fig:fig1} 
\end{center} 
\end{figure}

Fig.~\ref{fig:fig1} shows the muon impact point distribution along the $x$ axis at FD. 

\begin{figure}[htbp] 
\begin{center} 
\includegraphics[height=4.0in]{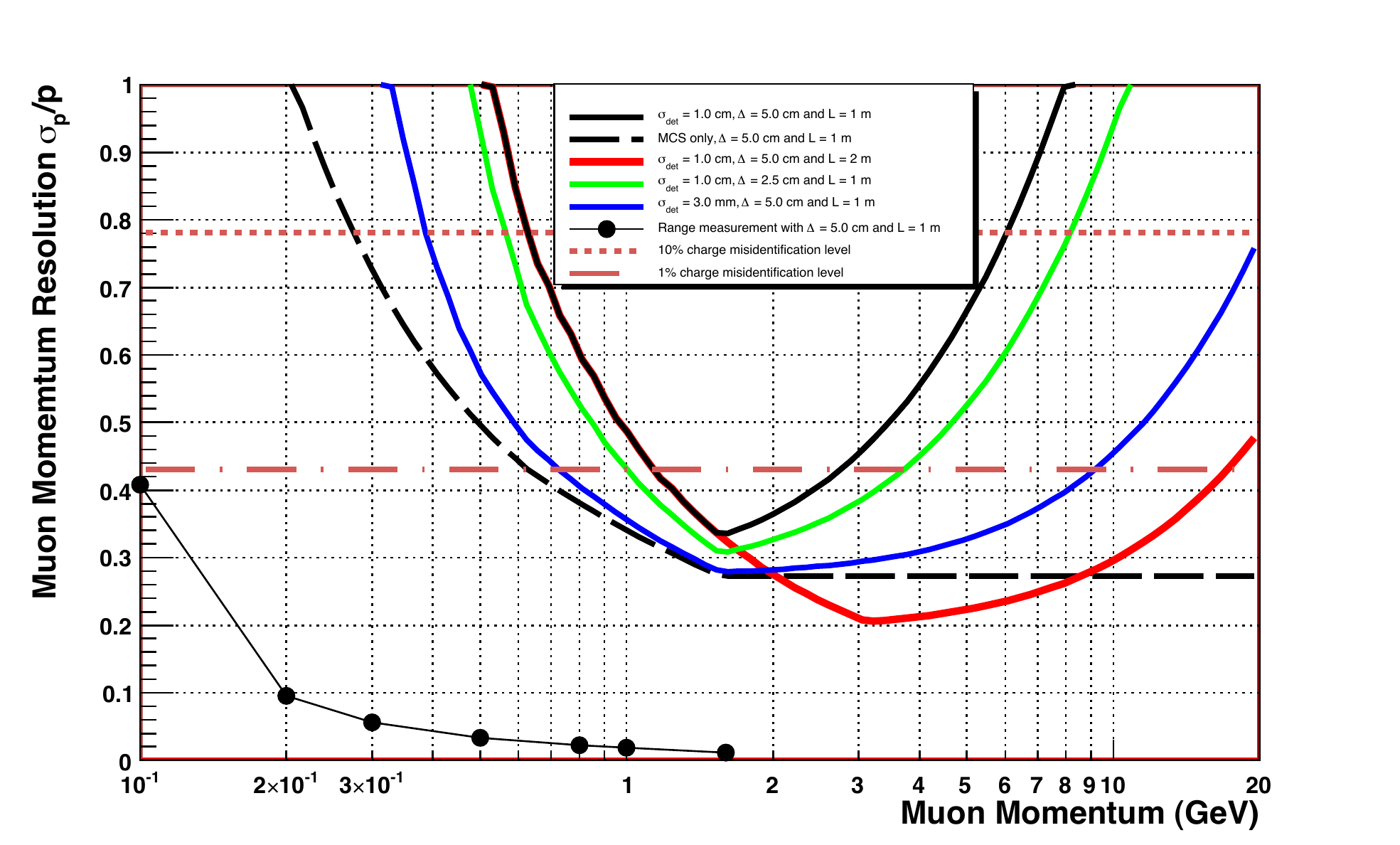} 
\caption{Muon momentum resolution as calculated for several Spectrometer configurations, with a magnetic field $B=1.5\ T$.} 
\label{fig:MomResCalc} 
\end{center} 
\end{figure}

Due to the relatively low energy spectrum of the PS neutrino beam the  Spectrometer longitudinal size
has to be large enough in order to allow energy measurement by {\em range} in a wide energy interval (see later on in this Section).
Simulations have shown that 2 m of Iron contain $\sim$90\% of muons in 
positive-focussing ($\sim$85\% in negative-focussing).  
5 cm thick iron slabs would allow a spectrum bin size of $\sim$60$\div$100 MeV/$\sqrt{12} \simeq 20\div 30$ MeV 
for perpendicularly impinging muons. Momentum reconstruction for passing-through muons ($E_{\mu} > 3$ GeV/c) 
can be performed exploiting the track bending in the magnetic field. A detector resolution of $\sim$1 cm provides 
a $\sigma_p/p$ resolution ranging from 20\% at 3 GeV/c up to 30\% at 10 GeV/c (see Fig.~\ref{fig:MomResCalc}).

\begin{figure}[htbp] 
\begin{center} 
\includegraphics[height=3.5in]{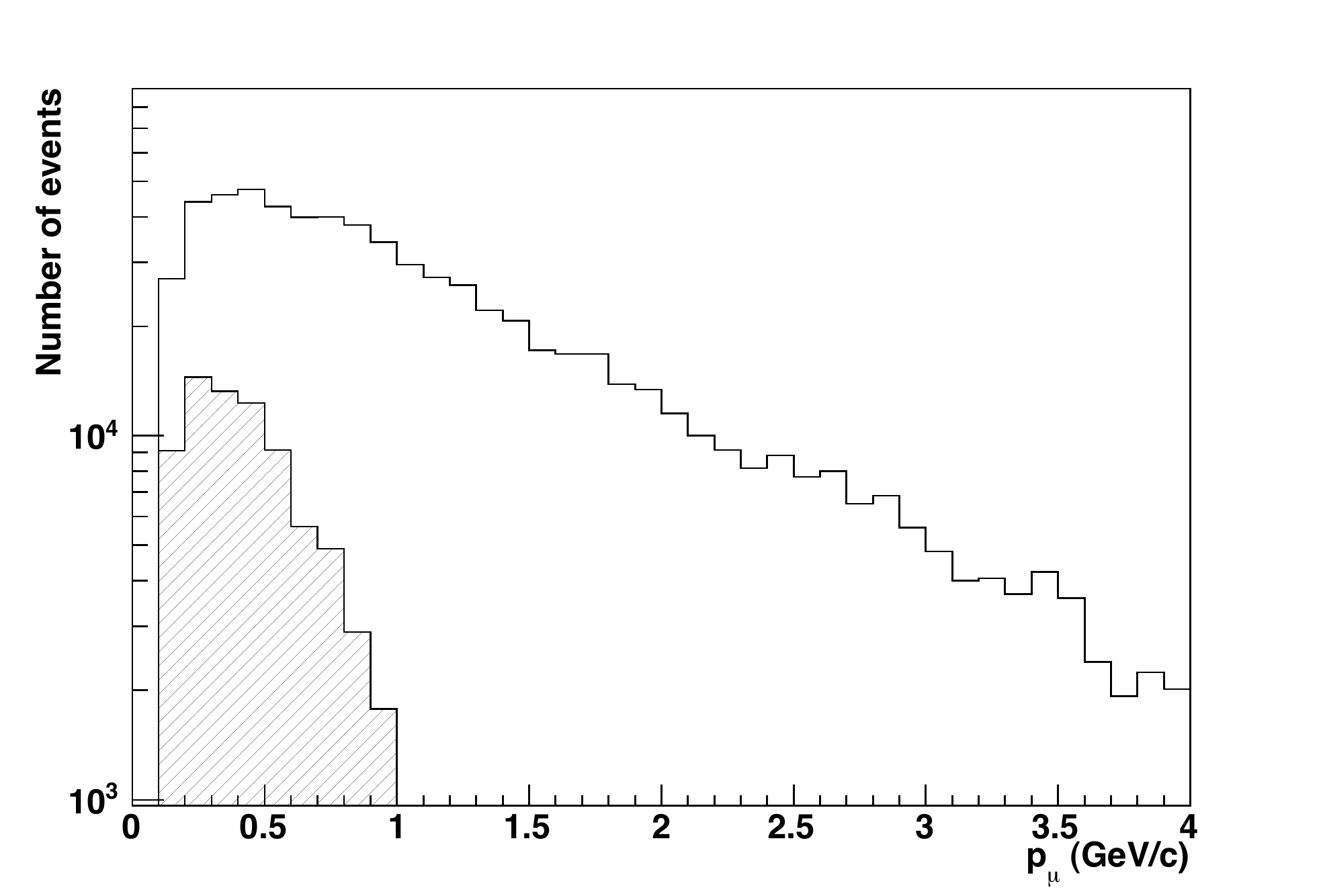} 
\caption{Energy spectrum of $\mu^{+}$ intercepted by the Spectrometer in the FD in negative-focussing for 3.75 p.o.t. The shaded region is the subsample of muons 
from events such that 1.27 $\times$ 1 eV$^2 \times L/E_{\bar{\nu}} > \pi/2$.} \label{fig:fig2} 
\end{center} 
\end{figure}

Muon charge assignment is performed measuring the direction of track curvature.  
The experimental sensitivity to $\Delta m^2$ scales linearly with the neutrino energy, $E_{\nu}$, and only with $\sqrt[4]{N_{events}}$. 
Therefore the possibility to explore 
CPT-violating $\nu_\mu$ disappearance at $\Delta m^2$ values below $\sim$1 eV$^2$ in a baseline of $\sim 850$ m 
relies on the capability to identify the muon charge with good accuracy down to $E_\nu$ of about few hundreds of MeV 
and even lower if we consider the muon residual energy reaching the Spectrometer (see Fig.~\ref{fig:fig2}). 
However in this momentum region one has to cope with the severe limits imposed  
by Multiple Scattering in iron which completely dominates the charge identification capability. 

Since for stopping muons the path-length $L$ in the $B\cdot L$ product is fixed by range $B$ has 
to be as large as possible ($B \simeq 1.5$ T in iron). 
Once the magnetic field is fixed the momentum resolution $\sigma_p/p$ and the charge mis-identification $\eta$ 
(defined as the fraction of muon tracks whose charge is wrongly assigned)
are limited at low energies by Multiple Coulomb Scattering (MCS). 
In order to understand qualitatively this behavior let us define $\eta$ as the fraction of muon tracks whose 
charge is wrongly assigned. $\eta$   is related - in the Gaussian approximation of the Moliere distribution- to the momentum resolution by 
\begin{equation} 
\eta = \frac{1}{2}\textrm{erfc}\left[{\frac{1}{\sqrt{2}(\sigma_p/p)}}\right] 
\label{eq:1} 
\end{equation} 
where {\em erfc} is the complementary error function. 
The capability to separate charge-conjugated oscillation patterns depends on the available statistics and on the value of $\eta$: 
the lower the number of events the lower $\eta$ has to be in order to assess a separation at a given significance 
level. For instance neglecting systematic and statistical uncertainties on $\eta$ a $\nu/\bar{\nu}$ separation 
at 10\% level would require $\eta \simeq 30\%$ in energy bins with 10000 events and $\eta \simeq 5\%$ with 1000 events. 
 
The momentum resolution (or equivalently $\eta$) is the combination of two terms, the first one related to MCS and the other to measurement errors:
\begin{equation} 
\frac{\sigma_p}{p} = \sqrt{\left(\frac{\sigma_p}{p}\right)_{MCS}^2 + \left(\frac{\sigma_p}{p}\right)_{res}^2 } 
\label{eq:2} 
\end{equation} 
The two terms can be calculated e.g. for the case of uniform detector spacing as in~\cite{scott1949,gluckstern1963,innes1993}.
The first term is almost independent of the number of measurement points along the trajectory and 
it is expressed as 
\begin{equation} 
\left(\frac{\sigma_p}{p}\right)_{MCS} \simeq 27\% \left(\frac{1.5\textrm{T}}{B}\right)\left(\frac{1\textrm{m}}{L}\right)^{1/2} 
\label{eq:3} 
\end{equation} 
where L is the muon path-length in iron and is actually limited by the muon range and the spectrometer size at low and high momenta respectively.
This term sets the lower irreducible limit that can be obtained by a measurement of this kind (note that $\eta \simeq 5\%$ 
roughly corresponds to $\sigma_p/p \simeq 50\%$). 
 
The second term depends on the detector resolution and on the number of measurements (or equivalently on the iron slab thickness,
$\Delta$) 
and, in principle, can be decreased by changing the detector sampling and/or space resolution.
In this case, for $L >> 5 \Delta$,
\begin{equation} 
\left(\frac{\sigma_p}{p}\right)_{res} 
\simeq 
13\% 
\left( \frac{p}{1\textrm{GeV/c}} \right) 
\left(\frac{\Delta}{5\textrm{cm}}\right)^{1/2}
\left(\frac{1\textrm{m}}{L}\right)^{5/2} 
\left(\frac{1.5\textrm{T}}{B}\right)
\left(\frac{\sigma_{det}}{1\textrm{cm}}\right)
\label{eq:4} 
\end{equation} 
with the same considerations on $L$ as for equation~\ref{eq:3}.

\begin{figure}[htbp] 
\begin{center} 
\includegraphics[height=4.0in]{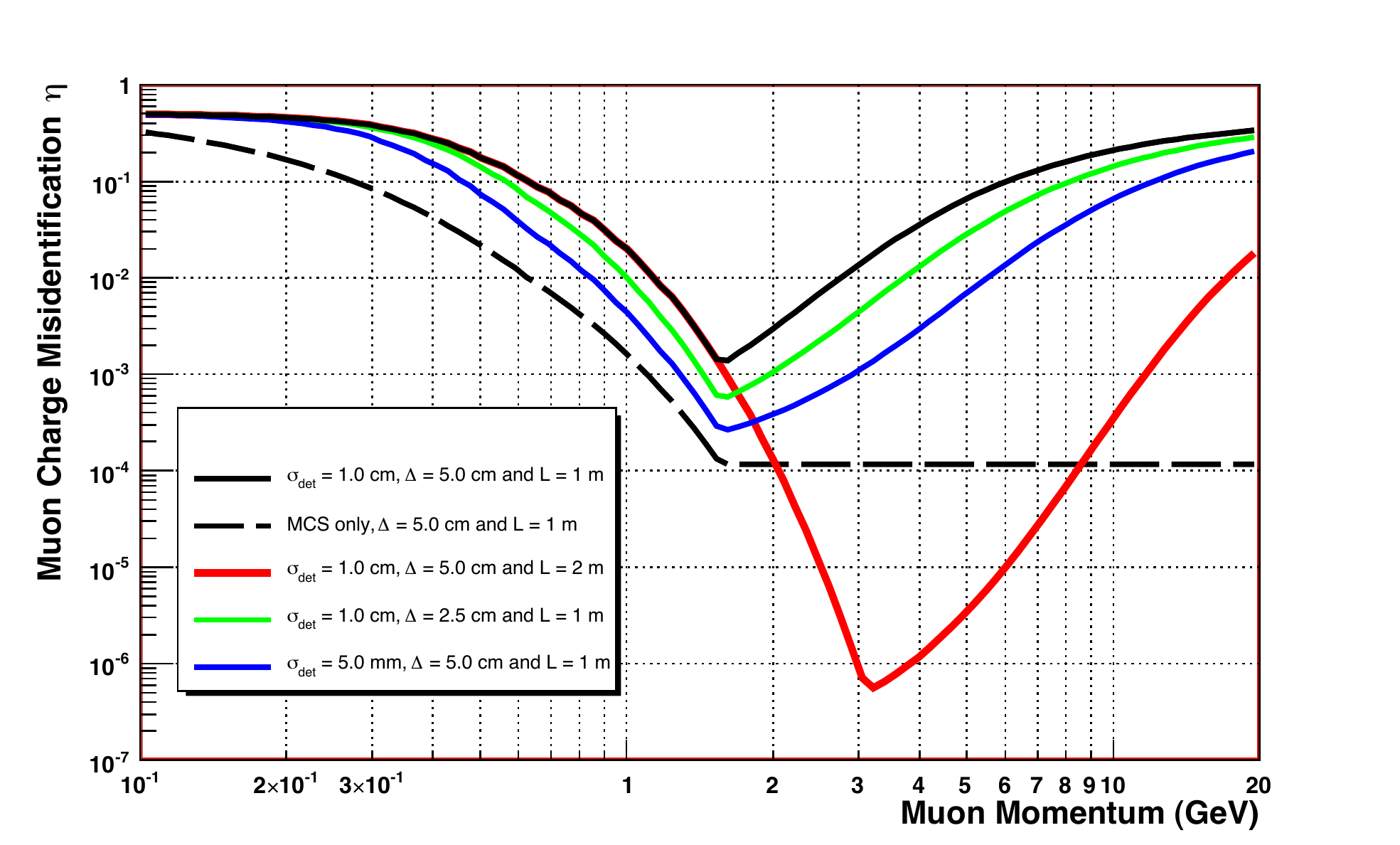} 
\caption{Muon charge mis-identification $\eta$ as calculated for several Spectrometer configurations, with a magnetic field $B=1.5\ T$.} 
\label{fig:MisQCalc} 
\end{center} 
\end{figure}

Curves reported in Fig.~\ref{fig:MomResCalc} and Fig.~\ref{fig:MisQCalc} show the momentum resolution and the $\eta$ dependence on 
the muon momentum for various choices of $\Delta$ and $\sigma_{det}$  (for orthogonally impinging muons).
It is apparent that at best a mis-identification $\eta$ below 5\% can be obtained only 
for muons with momenta above $\simeq$ 500 MeV/c. 

An option to improve charge identification capability below 500 MeV is to equip the Spectrometer with a magnetic field in air just  
in front of the first iron slab. This possibility  is discussed in the next Section. 
 
From now on we will assume as baseline option for the FD site a dipolar magnetic Spectrometer made of two arms separated by 1 $m$ air gap. 
Each arm consists of 21 iron slabs 585$\times$875 cm$^2$ wide and 5 cm thick interleaved with 2 cm gaps hosting 20 Resistive Plate Chambers
(RPC) detectors 
with $\sim$1 cm tracking resolution. The Near detector (ND)  Spectrometer is a downsized replica of the FD one,
the only difference being in fact the transverse dimensions  (351.5$\times$625 cm$^2$). 

\vskip 10pt

\begin{figure}[htbp]
\vspace{-0.2cm}
\centering
 {\centering\resizebox*{15cm}{!}{\includegraphics{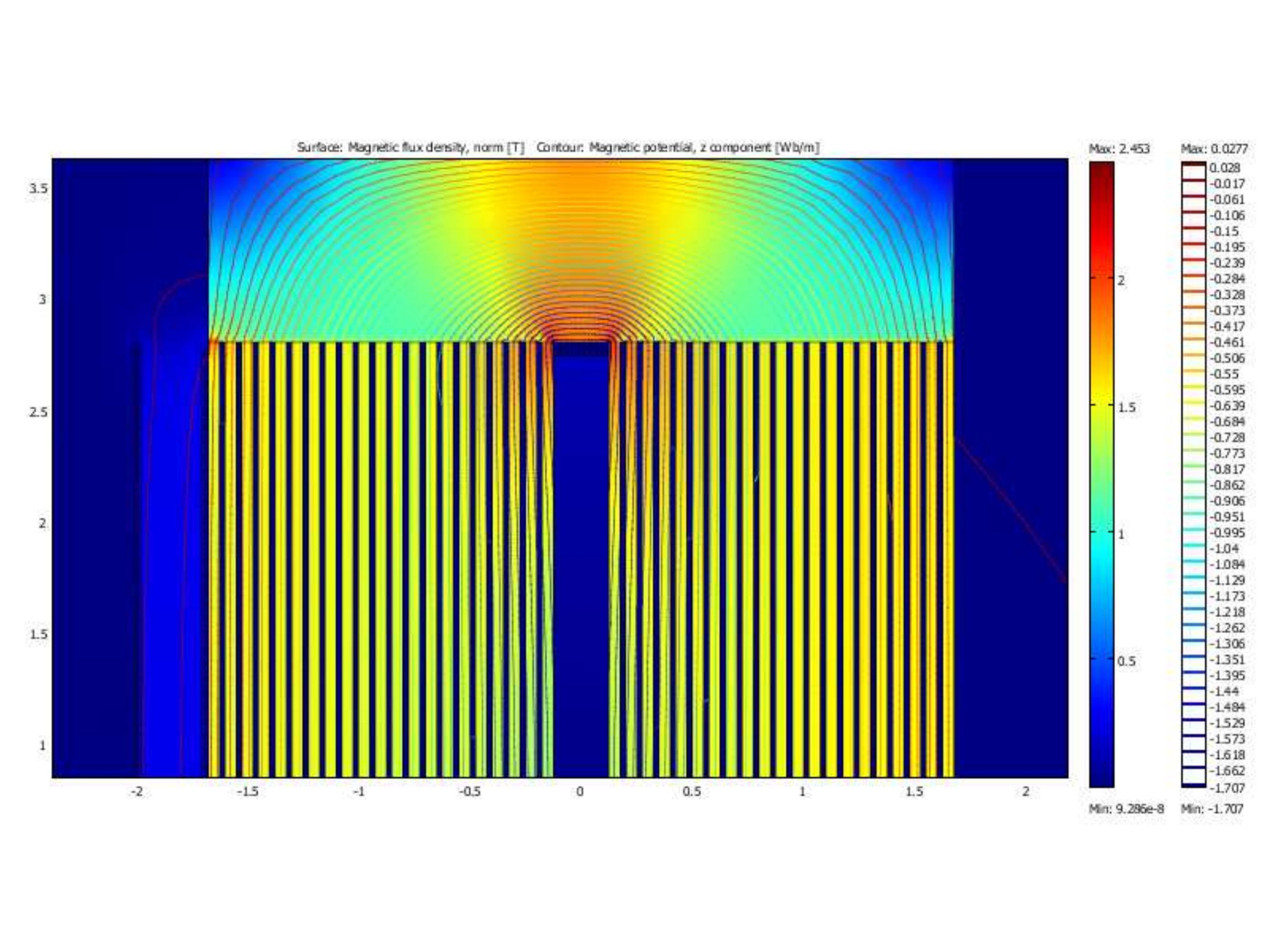}}}
\begin{quote}
\vspace{-0.5cm}
\caption{Simulated magnetic field distribution in the Iron Spectrometer (yellow regions) and in the air volume (light blue region on the left side, see Sect.~\ref{sec:air_field}). The horizontal axis is in meters; the color bar is in Tesla. The iron field is generated by two  coils wrapped around the top and bottom iron flux return (only the upper coil is shown in the figure).}
\label{fig:CM1}
\vspace{-0.5cm}
\end{quote}
\end{figure}

\begin{figure}[htbp]
\vspace{-0.2cm}
\centering
 {\centering\resizebox*{12cm}{!}{\includegraphics{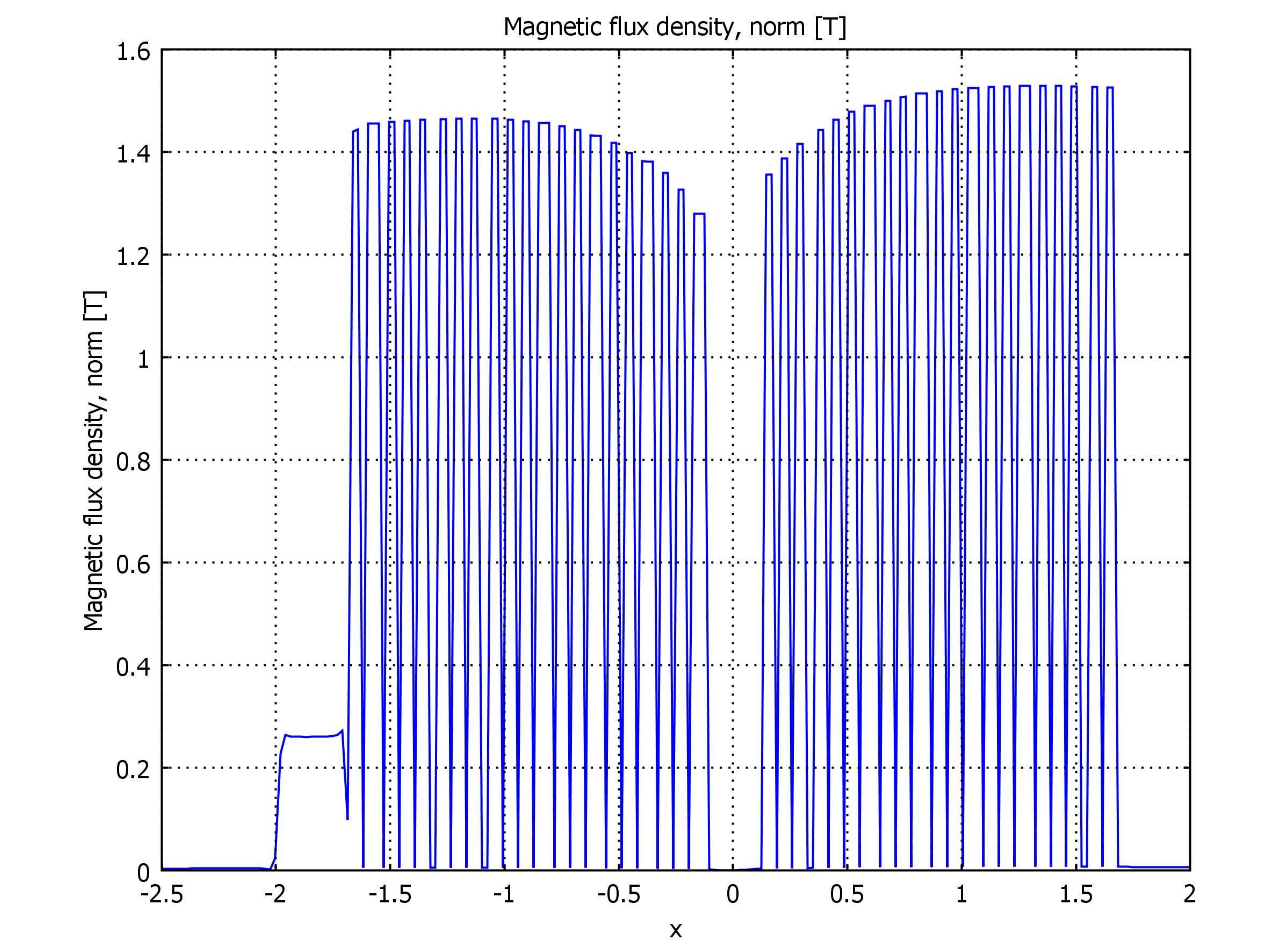}}}
\begin{quote}
\caption{Magnetic field distribution in the 21+21 iron slabs and in the air volume upstream of the first iron slab. The vertical axis indicates the magnetic field in Tesla. }
\label{fig:CM2}
\vspace{-0.5cm}
\end{quote}
\end{figure}

A calculation of the magnetic field map was performed with the COMSOL$^{ª}$ code~\cite{comsol}.  Fig.~\ref{fig:CM1} shows the distribution of the
magnetic field in the 21+21 iron layers (for convenience we show in the figure  the entire magnetic system, i.e. in iron and in air,
the latter to be discussed later on in the Section). The fringe field is $70\div 100~G$ at $10~cm$ distance from the iron slab edge 
(Fig.~\ref{fig:CM2}).
  
The simulation was performed  assuming two symmetric coils wrapped around the top and bottom flux return path 
 (following the same concept applied in the OPERA magnet~\cite{bopera}). The turns (36) are made of aluminum bars $27\times 19~mm^{2}$ with inner water
 cooling. The current density is $3~A/mm^{2}$ for a total current of $1255~A$. The total resistance is $39~mOhm$ per coil, the voltage is $49 ~V$ and the 
 power $61~ kW$. The conductor cross section can be increased to reduce the power dissipation.

\subsection{Magnetic field in Air} \label{sec:air_field}

\begin{figure}[htbp]
\begin{center}
\includegraphics[height=1.8in]{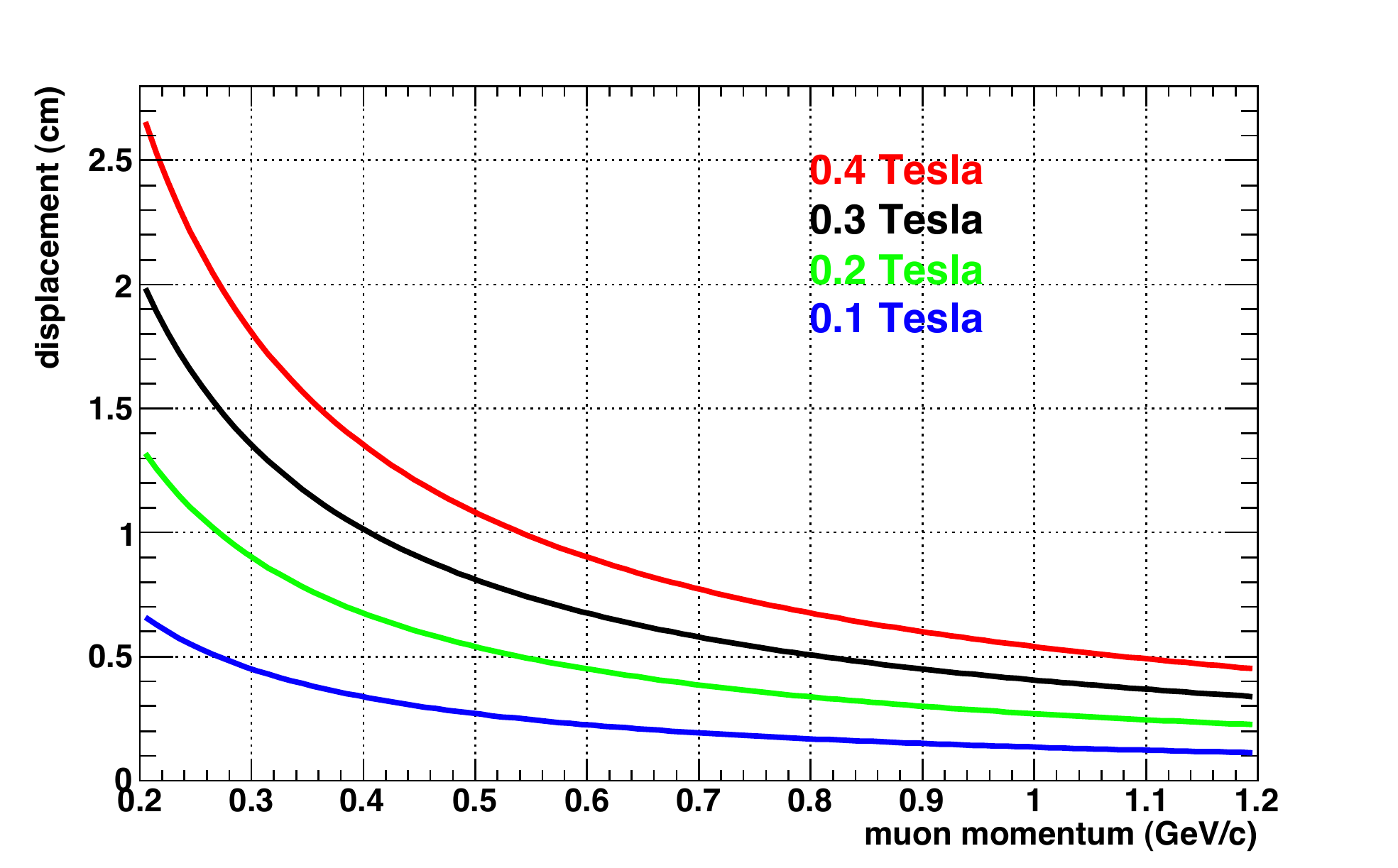}
\includegraphics[height=1.8in]{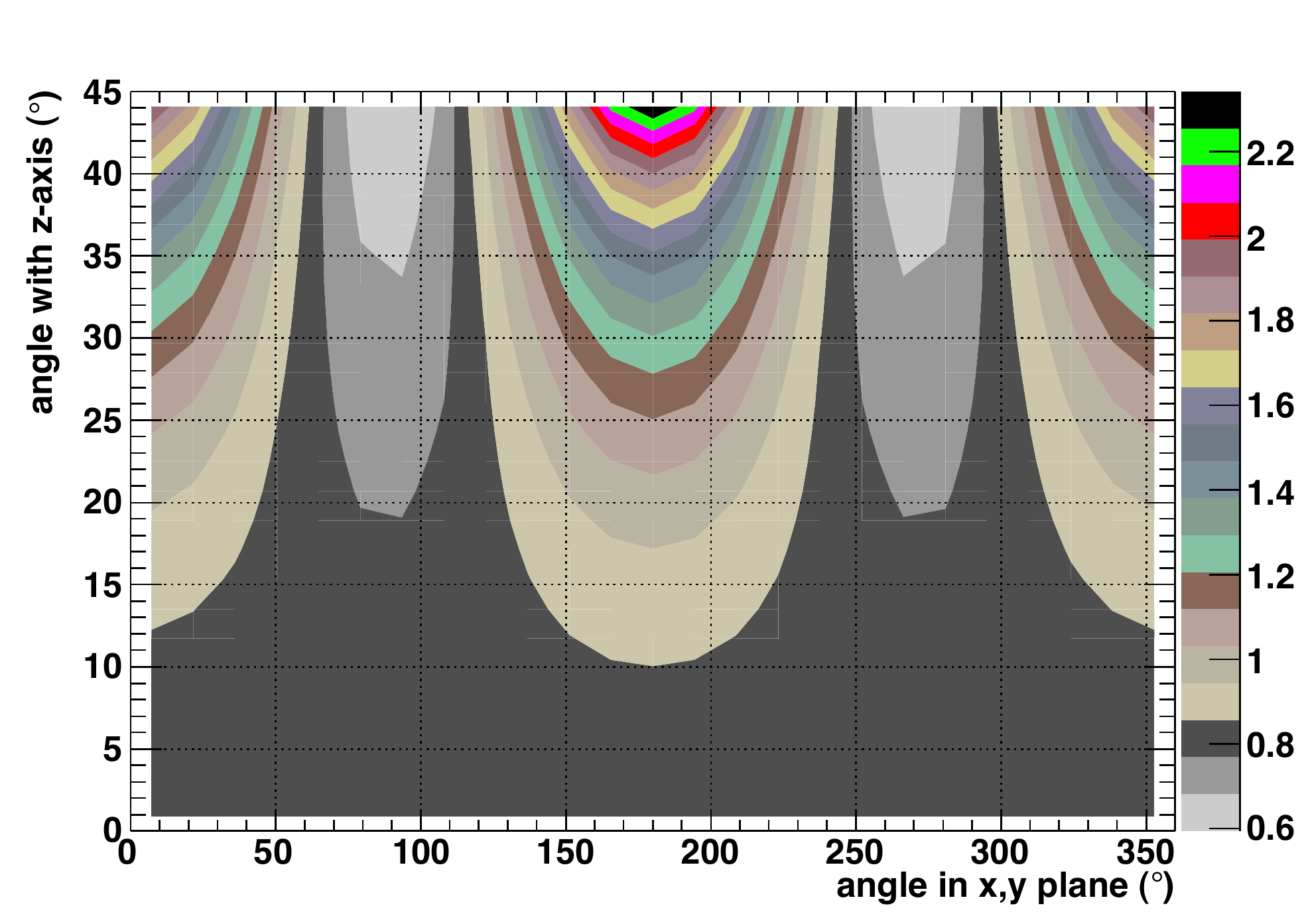}
\caption{Left: Spatial displacement of muon  trajectory in the bending plane as a function of 
 their momentum  after crossing an air magnetic field volume of
 $30\ cm$ thickness. Different field values in the 0.1-0.4 T range are considered. The incoming muon direction is perpendicular to
the detector planes. Right:  the displacement of $0.5\ GeV$ muon tracks in 
the $x,z$ view ($B=0.3\ T$) as a function of the initial muon angle (the
color bar scale is in $cm$).} \label{fig:displacement}
\end{center}
\end{figure}

For low momentum muons the effect of Multiple Scattering in iron is comparable
to the magnetic bending and therefore the charge mis-identification increases  (see Fig.~ \ref{fig:MisQCalc})
For muon momenta  $ < 1\ GeV/c$ the charge measurement can be performed by means 
of a magnetic field in air. In Fig.~\ref{fig:displacement} the displacement 
expected in the bending plane is shown for muons crossing a magnetized air volume of $30\ cm$ depth. A uniform magnetic field oriented along the  $y$ axis (the bending plane is the $z,x$ one) is assumed.
In the left plot the shift is shown as a function of the muon momentum for some values of the magnetic field in the $0.1-0.4$ $T$ range. 
In the right panel the spatial displacement in the bending plane estimated for muons of $0.5\ GeV$ in a magnetic field of $0.3\ T$ 
as a function of the incoming angles is plotted.  \par

\begin{figure}[htbp]
\vspace{-0.2cm}
\centering
 {\centering\resizebox*{15cm}{!}{\includegraphics{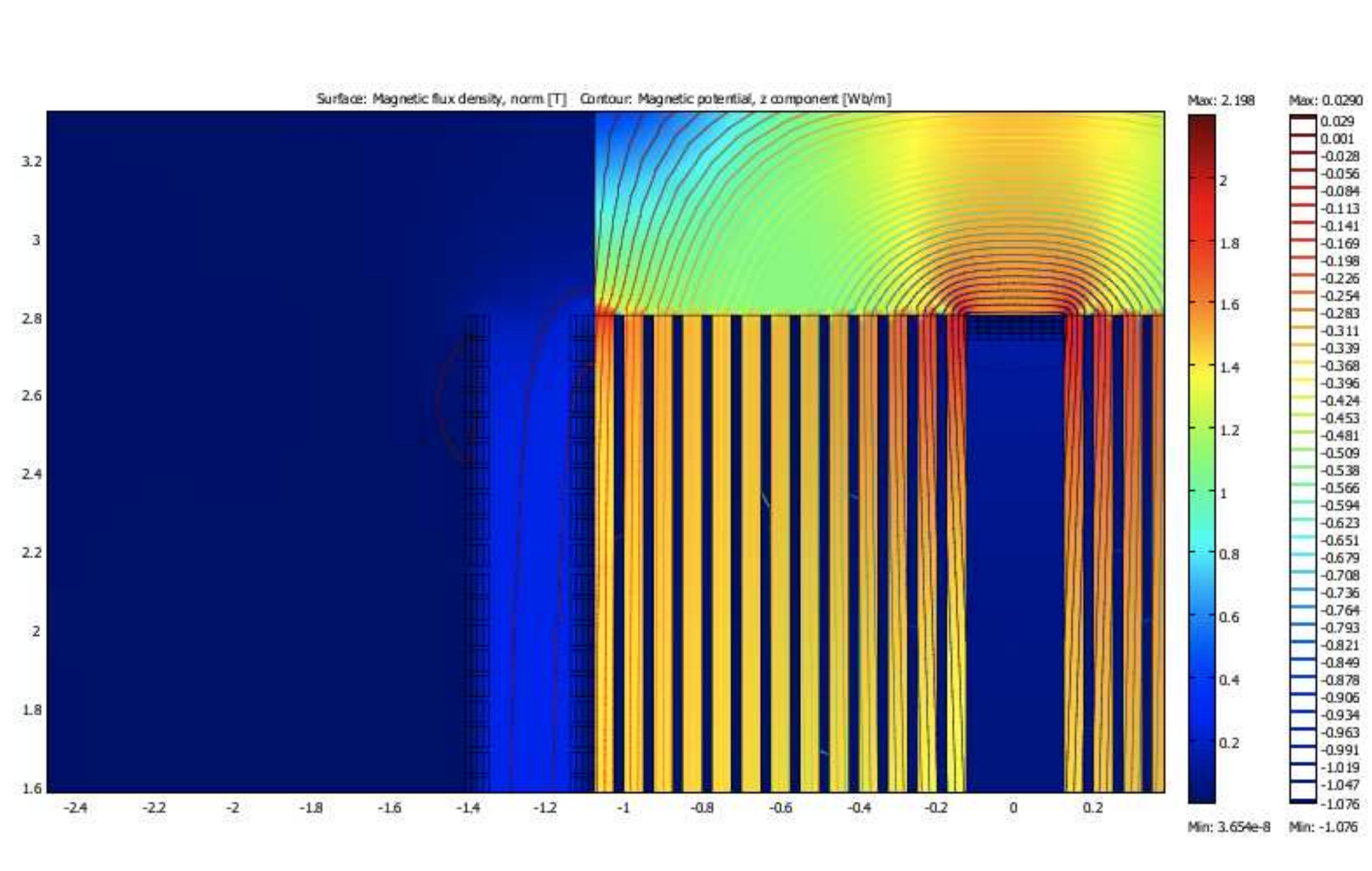}}}
\begin{quote}
\vspace{-0.5cm}
\caption{Simulated magnetic field distribution in the Iron Spectrometer (yellow regions) and in the air volume (light blue region on the left side). The horizontal axis is in meters. The color bar scale is in Tesla. The vertical pairs of coils used to generate the field in air are  shown (on the left of the picture).}
\label{fig:CM3}
\vspace{-0.5cm}
\end{quote}
\end{figure}

A simulation of the magnetic field in air was realized  based on a coil wound on a large conductor ($54~mm\times~19~mm$) with 170 turns distributed along  the spectrometer height.   A uniform magnetic field of $0.25 ~T$  along the $y$ axis is obtained with a coil current density of $8 ~A/mm^{2}$ (Fig.~\ref{fig:CM1}, $7000~A$ current). The fringe field  at $10 ~cm$ distance from the edge (Fig~\ref{fig:CM3}) is $70\div 200~G$. 
The total resistance is  $0.1~\Omega$, the voltage is $700~ V$ and the power $4.9 ~ MW$. This set of parameters shall be intended as preliminary and needs to be optimized for reducing the electrical power and for splitting the single large coil into a number of smaller (in height) and more manageable coils.

As discussed in Sec.~\ref{sec:hpt} charge measurement for low momentum muons can  be performed by designing a precision tracker detector of $\sim 1\ mm$ spatial resolution installed  in the $z,x$ plane. The reconstruction in the $z,y$ view would be not
required  since the momentum could be measured by the muon range in the Iron Spectrometer.

\clearpage

\section{Monte Carlo Detector Simulation and Reconstruction}\label{sec:MC}

The present proposal has been extensively developed  using full-detail programming 
and up-to-date packages to obtain precise
understanding of acceptances, resolutions and physics output. Although not all  possible options 
have been studied, a rather exhaustive list of different magnet configurations and detector designs 
has been adopted as benchmark for further studies once the 
detector structure will be finalized.

\subsection{Simulation}

The aim of the simulation of the apparatus is to help the design studies reported here to 
understand the main
features of the proposed experiment. The simulated detector consists of a ND 
and FD part, both made of an Liquid Argon target followed 
by a magnetic Spectrometer. The relative position and the dimensions of the LAr target have been 
kept fixed with respect to the Spectrometer for all the studies described below.

Muon neutrino and antineutrino fluxes for positive and negative beam polarity were
assumed as those reported in the LAr proposal~\cite{Icarus-PS} since the beam 
analysis was still in progress.
In the simulation the beam has 1 $mrad$ tilt with respect to the horizontal.
For the time being no angular dependence of the neutrino fluxes has been considered. 
For the future we plan to take profit of the full simulation 
of the secondary beam line reported in the previous Sect.~\ref{sec:beam} in order to improve the 
simulation of the neutrino beam at both the Near and Far detector positions.

Neutrino interactions are generated in the Argon target using GENIE~\cite{ref_GEN} 
with standard parameters and including all interaction 
processes (QEL, RES, DIS, NC). In additions, neutrino interactions have been generated 
in iron to explore the capability of the Spectrometer to 
reconstruct self-contained events taking advantage of its additional mass. 

The propagation in the detector is implemented with either GEANT3 or FLUKA
and the geometry of the detectors is described 
using the ROOT geometry package. The main features of the geometry 
implemented in the simulation are briefly described below.

The ND LAr target has dimensions $352.8 \times 362.5 \times 910$ $cm^3$ 
and a total mass of 162~$t$. The target is surrounded by about 1 $m$
 thick light material standing for the dead material of the cryostat.
The basic Spectrometer is an instrumented dipolar magnet made of two magnetized 
iron walls producing a field of 1.5 T intensity in the tracking region; 
field lines are vertical and have opposite directions in the two walls whereas
track bending occurs on the horizontal plane.
The thickness of the iron planes is at present envisaged to be 5 cm.
Planes of bakelite RPC's are interleaved with the iron slabs of each wall
to measure the range of stopping particles and to track penetrating muons. 
The Spectrometer is equipped with 20 planes of 3 rows, each consisting of 2 
RPC's. At present, additional high precision Drift Tube detectors are not simulated. 

The FD target consists of 2 LAr volumes each with dimensions 
$352.8 \times 362.5 \times 1820$~$cm^3$. The total LAr mass is 648~$t$. 
The FD Spectrometer is assumed to be 
similar to the one for the ND (20 planes each with 5 rows and 3 columns of RPC's).

The response of the Liquid Argon detector has been sketched by sampling the position of the charged particle 
with 0.5 cm resolution. For the RPC's we assume digital read-out using 2.5~$cm$ strip width and a position resolution of about 1~$cm$.
 
\subsection{Reconstruction}

A framework based on standard tools (ROOT, C++)  has been developed for the reconstruction in both the Near and Far detectors. 

The reference frame is defined to have the Z-axis along the beam direction, Y perpendicular to the floor pointing upwards and X 
completing a right-handed frame. Event reconstruction is performed separately in the two projected views XZ (bending plane) 
and YZ (vertical plane). 

A simple track model is adopted to describe the shape of the trajectory of tracks travelling through the detector. 
The model is based on the standard choice of parameters used in {\em forward} geometry (i.e.  intercepts, slopes, particle momentum, 
particle charge, ...  ). The reconstruction strategy is optimized to follow a single long track (the muon escaping from the neutrino-interaction 
region) along the Z-axis. 

The reconstruction is performed in the usual two steps: Pattern Recognition (Track finding) and Track Fitting.

The task of the Pattern Recognition is to group hits into tracks. Taking into account that most 
of neutrino interactions generated in the LAr target have just one track reaching the Spectrometer (the muon) and assuming 
a read-out capable of avoiding event overlap, we postponed the development of a dedicated algorithm
For the time being all the hits of the events are associated to a single track.

The Track Fitting has to compute the best possible estimate of the track parameters according to the track model. 
A parabolic fit is performed in the XZ plane (bending) whereas a linear fit is used for YZ plane (vertical). Particle charge and momentum 
are determined from the track sagitta measured in the bending plane; the track fit is corrected for material interactions (Multiple Scattering and energy loss). 
Each spectrometer arm provides an independent measurement of charge/momentum.  A better estimation of momentum is 
obtained by range for muons stopping inside the Spectrometer. The implementation of a track fitting algorithm based on a {\em Kalman} filter is eventually
foreseen.

\section{Physics performances}\label{sec:spect2}

In this section we address the physics performance of the FD and ND Spectrometers with and without the B field in air option.

We studied the performance of the Spectrometer using the Monte Carlo simulation detailed in the previous section.
In particular we report on the negative-focussing option which is the most promising for our purposes.
We have generated $10^6$ $\nu_{\mu}$ (NC+CC) events and $10^6$ $\bar{\nu}_{\mu}$ (NC+CC) events according to spectra shown in the right panel of 
Fig.~\ref{fig:rates}.
Neutrino vertices were generated within the LAr volume and each outgoing particle from the vertex was propagated in the LAr $+$ Spectrometer geometry.
Each event was classified into one of the following category:

\begin{enumerate}
	\item {\bf Fully contained events}. These are events in which the neutrino vertex is contained the LAr fiducial volume and the muon track stops in the LAr fiducial volume.
	We considered events with a muon track escaping from the vertex at least 20 cm long. 
	\item {\bf Partially contained in LAr not reaching the Spectrometer}. These are events with the neutrino vertex within the LAr fiducial volume and the muon track which escapes from the LAr fiducial volume which do not intercept the Spectrometer. In this case we require a muon track at least 200 cm long in order to reconstruct the muon momentum with a good momentum resolution by MCS in LAr.
	\item {\bf Partially contained in LAr reaching the Spectrometer}. These are events with the neutrino vertex within the LAr fiducial volume and the muon track which escapes from the LAr fiducial volume which do intercept the Spectrometer. In this case we accept also muon track shorter than 200 cm in LAr since the muon momentum information are recovered with the Spectrometer.
\end{enumerate}

The last sample is further subdivided into events with a muons stopping in the Spectrometer and events passing through the whole Spectrometer 
(or escaping from the side).
In the first case the muon momentum is reconstructed by range, in the second case by magnetic bending. The muon charge sign is always reconstructed 
by magnetic bending,
in the air field or in the magnetized iron according to the muon energy.

\begin{figure}[htbp]
\centering
\includegraphics[width=9cm, angle=-90]{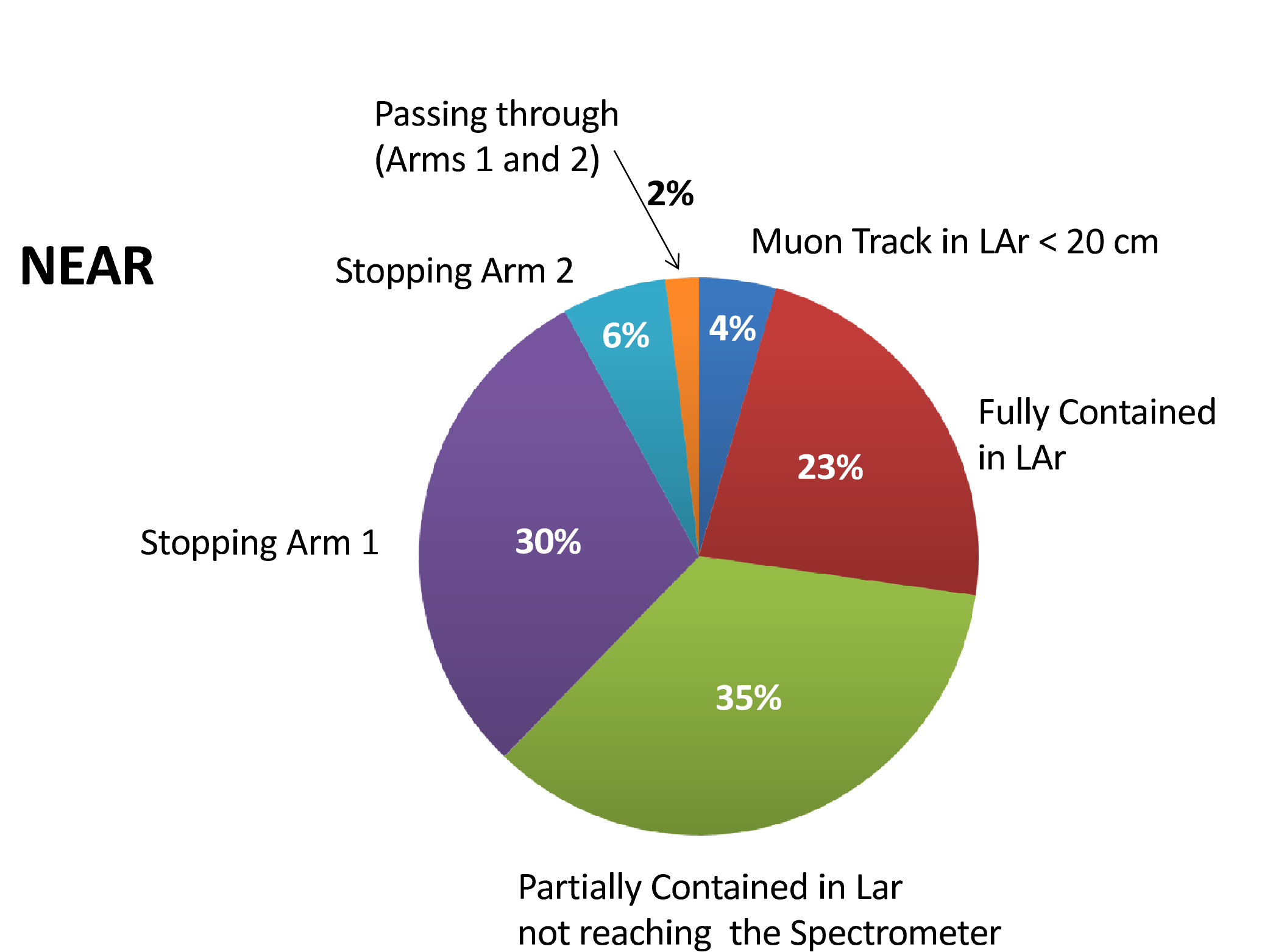}
\caption{Fraction of events collected at the Near site by the LAr and the Spectrometer detetctors. The different topologies are detailed in the text.}
\label{fig:topologies-N}
\end{figure}

\begin{figure}[htbp]
\centering
\includegraphics[width=9cm, angle=-90]{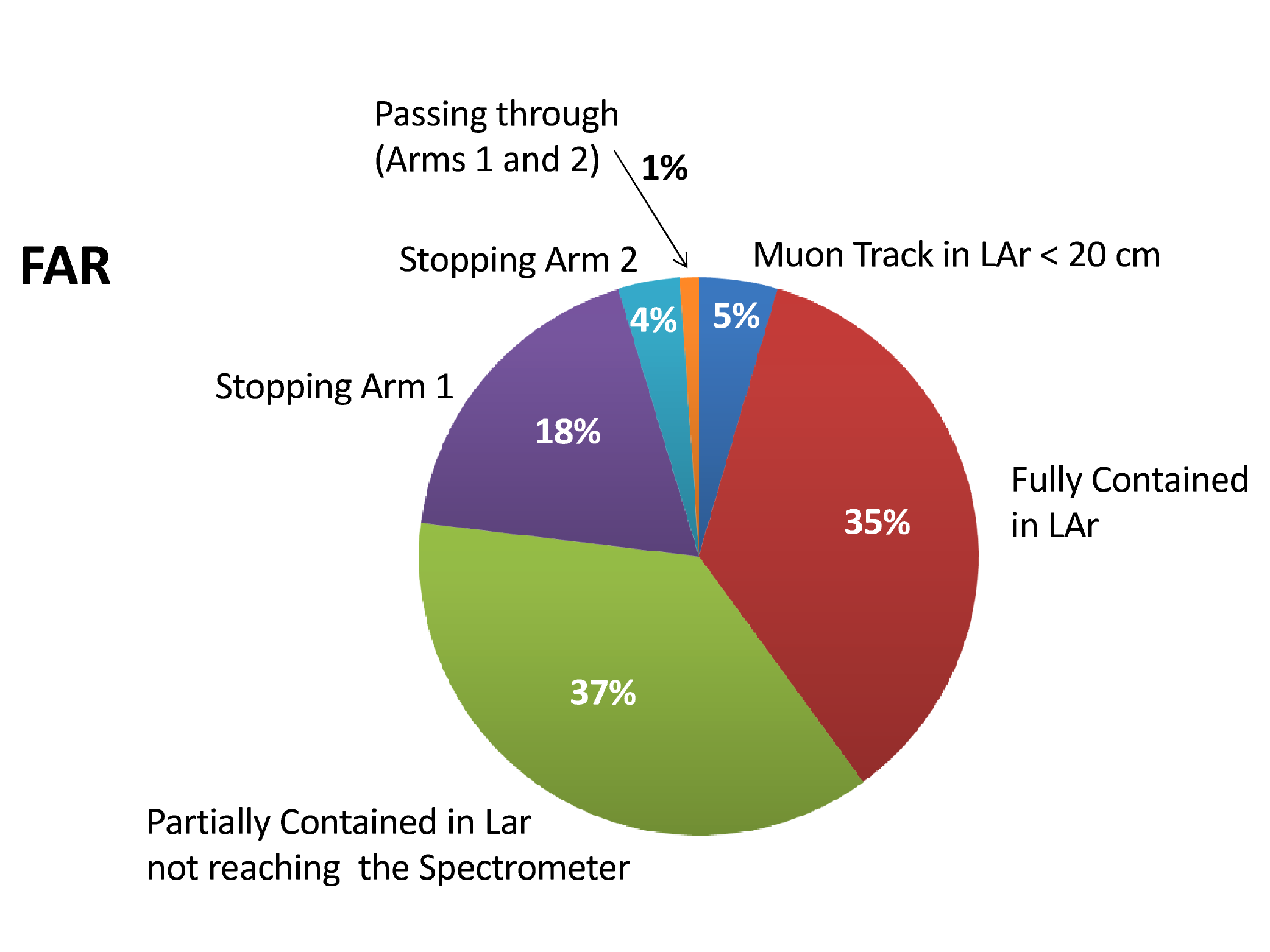}
\caption{Fraction of events collected at the Near site by the LAr and the Spectrometer deetctors. The different topologies are detailed in the text.}
\label{fig:topologies-F}
\end{figure}

Figg.~\ref{fig:topologies-N} and ~\ref{fig:topologies-F} report the fraction of each event topology in the negative-focussing option in the Near and Far detector,
respectively.

\begin{figure}[htbp]
\centering
\includegraphics[width=12cm]{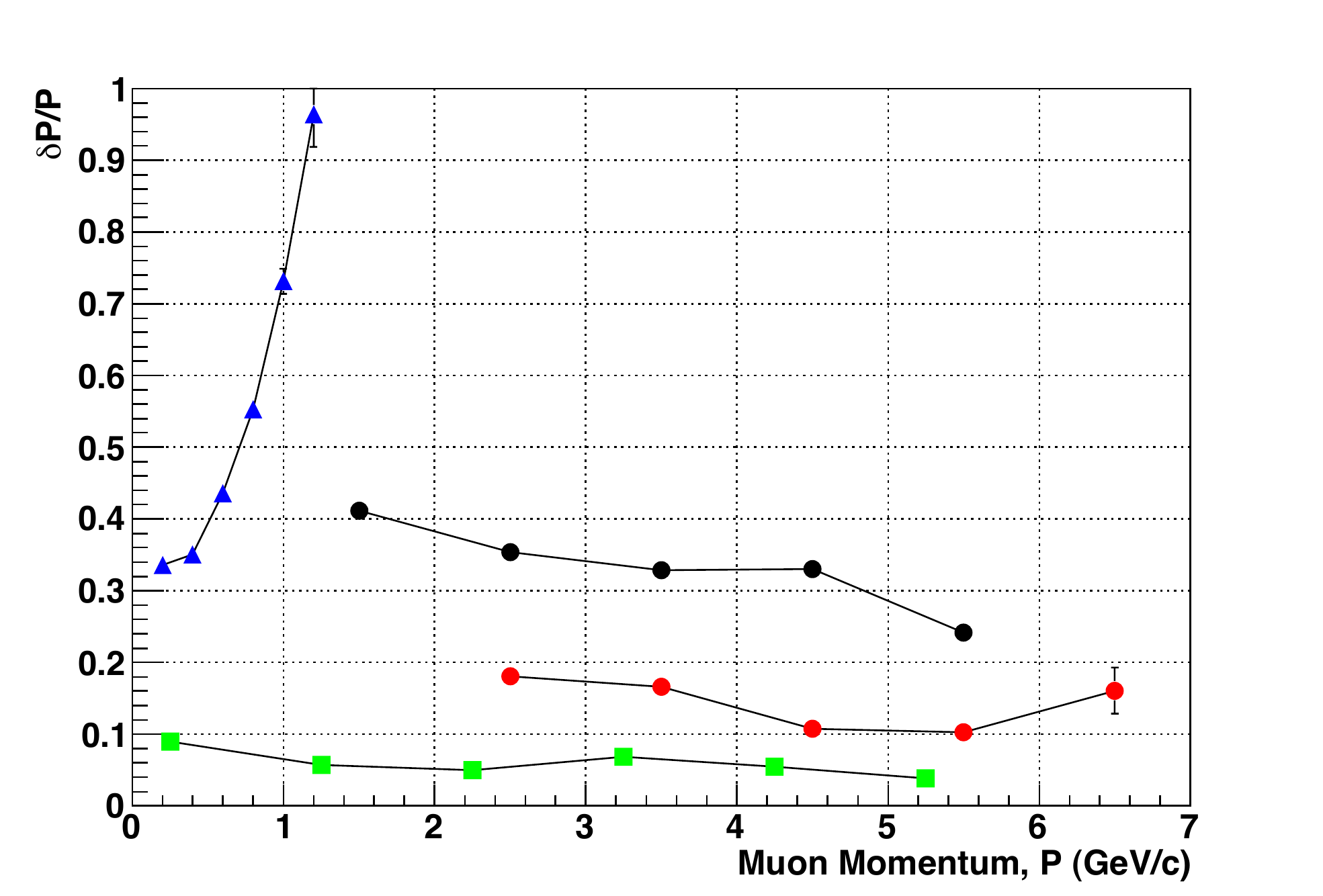}
\caption{The obtained momentum resolution percentage including all the selection, efficiency  and reconstruction procedures as described in the text.
The blue dots correspond to the measure performed by the magnetic field in air, the red (black) dots correspond to what can be achieved by using
the {\em slope} algorithm in the magnetic field in iron with the two (one) arms. The achievable resolution with the dE/dX {\em range} corresponds to the green dots.}
\label{fig:momres}
\end{figure}

In Fig.~\ref{fig:momres} we show the Spectrometer performance in terms of momentum reconstruction. In the figure 
the separate contributions of stopping muon, passing through going muons and low energy muons are displayed.
The total muon momentum at the neutrino vertex is computed adding the contribution of
the energy lost in LAr.

\begin{figure}[htbp]
\centering
\includegraphics[width=12cm]{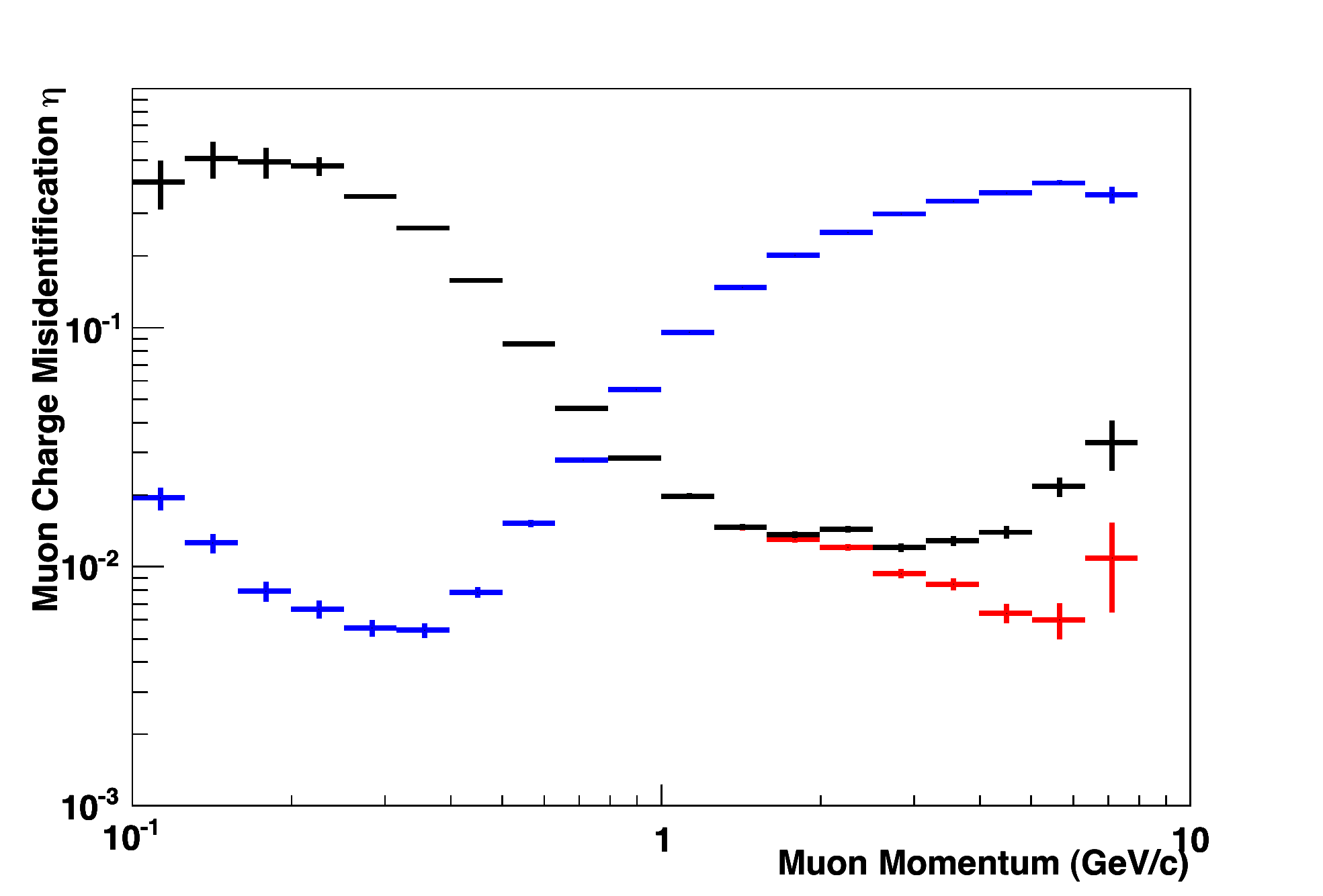}
\caption{The obtained charge mis-identification percentage including all the selection, efficiency  and reconstruction procedures as described in the text.
The blue dots correspond to the measure performed by the magnetic field in air, the red (black) dots correspond to the magnetic field in iron with the two (one) arms.}
\label{fig:charge-rec}
\end{figure}

In Fig.~\ref{fig:charge-rec} we show the Spectrometer performance in terms of charge sign reconstruction. In the figure the separate contribution
of muon bending in the air magnet and in the magnetized iron are shown.

Once the muon momentum is computed, the neutrino energy is estimated according to the event topology. For QE events the two-body kinematics allow a 
precise reconstruction of the incident neutrino energy from the measured momentum and direction of the outgoing muon:

\begin{equation}
E_\nu^{\rm rec}=
\frac{1}{2}\frac{(M_p^2-m^2_\mu)+2E_\mu(M_n-V)-(M_n-V)^2}{-E_\mu+(M_n-V)+p_\mu
\cos{\theta_\mu}}
\end{equation}

\noindent where $M_p$, $M_n$, $m_\mu$, $E_\mu$, $p_\mu$, $\cos{\theta_\mu}$ $V$ are the proton, neutron, muon masses, muon energy, momentum 
and angle with respect to the incoming neutrino direction and the  LAr nuclear potential, set at 27 MeV, respectively.
For non-QE events the neutrino energy is reconstructed adding up the muon energy to the energy of hadronic system, with a gaussian smearing 
$\sigma_E/E = 0.3/\sqrt{E}$ 

\begin{figure}[htbp]
\centering
\includegraphics[width=14cm]{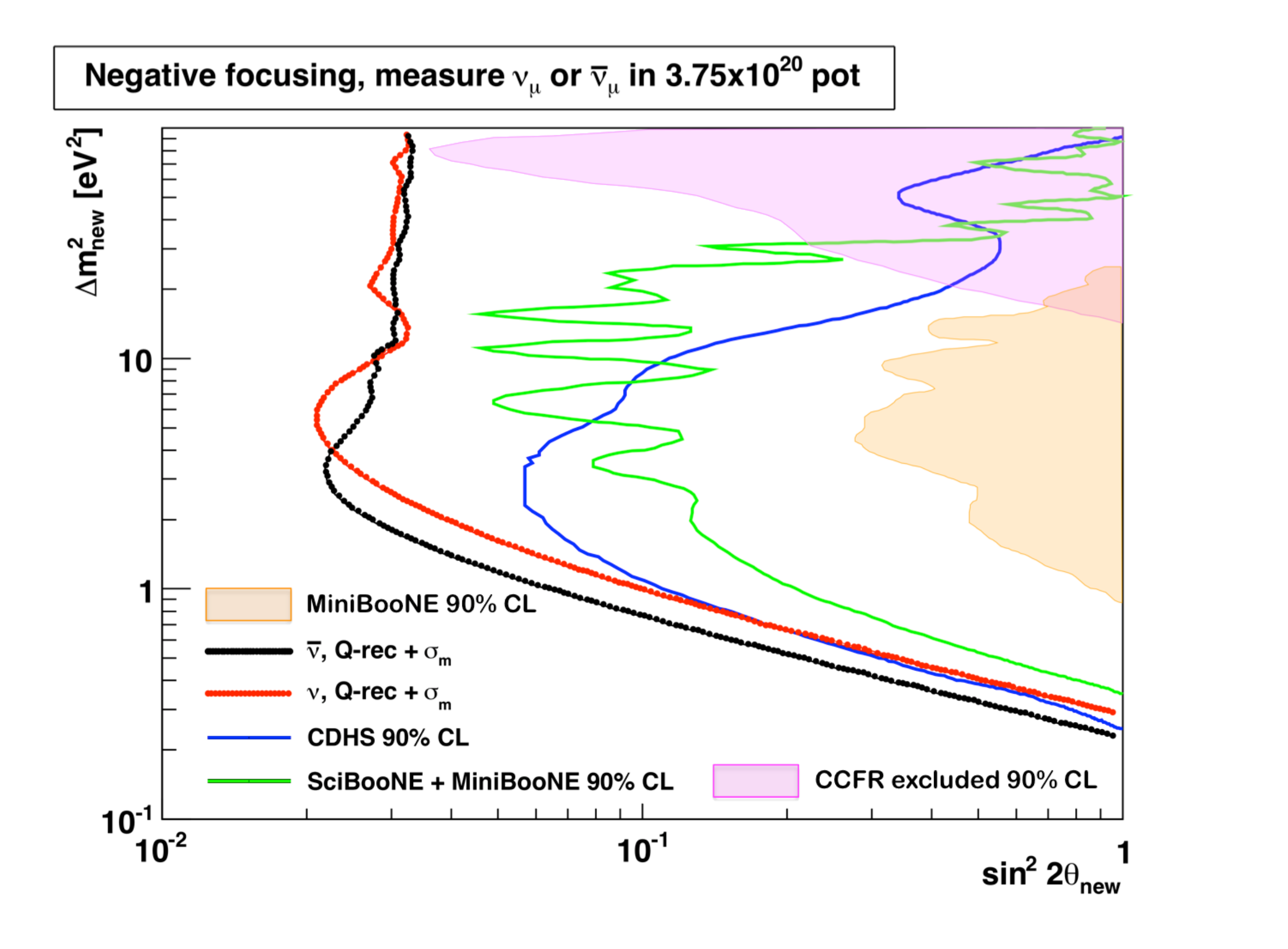}
\caption{The sensitivity plot (t 90\% C.L.) for the negative-focussing option assuming 3.5$\times$10$^6$ pot with and without the air magnet option.
Black (red) line: \nubarmu (\numu )exclusion limit. Note that the contribution of the magnetic field in air to the sensitivity has not being included, yet.
Blue (green) line: old (recent) exclusion limits on \numu from previous (CDHS) and recent~(\cite{sciboone})measurements.The two filled areas 
correspond to the present exclusion limits on the \nubarmu from CCFR~(\cite{ccfr} and MiniBooNE~\cite{MiniBooNE-numu} experiments (at 90\% C.L.). }
\label{fig:sensitivity}
\end{figure}

A two-flavor sensitivity plot was computed assuming $\nu_{\mu}$ disappearance and a $\Delta \chi^2$ approach. We compared the FD measured neutrino 
spectrum with the non-oscillated spectrum derived from the ND data. We assumed a 5\% overall systematic uncertainties on the absolute normalization derived 
from the ND spectrum. Fig. \ref{fig:sensitivity} shows the 90\% C.L. sensitivity curve for the negative-focussing option assuming 3.5$\times$10$^6$ pot with 
the iron magnet (the achievable exclusion limits by including also the magnet in air, are in progress).

\vskip 10pt
The study of the possible physics performances of the Spectrometers with respect to the neutrino interactions inside the iron is in progress.

\section{Mechanical Structure}\label{sec:mech-struc}
    
\begin{figure}[htbp]
\centering
\includegraphics[width=9cm]{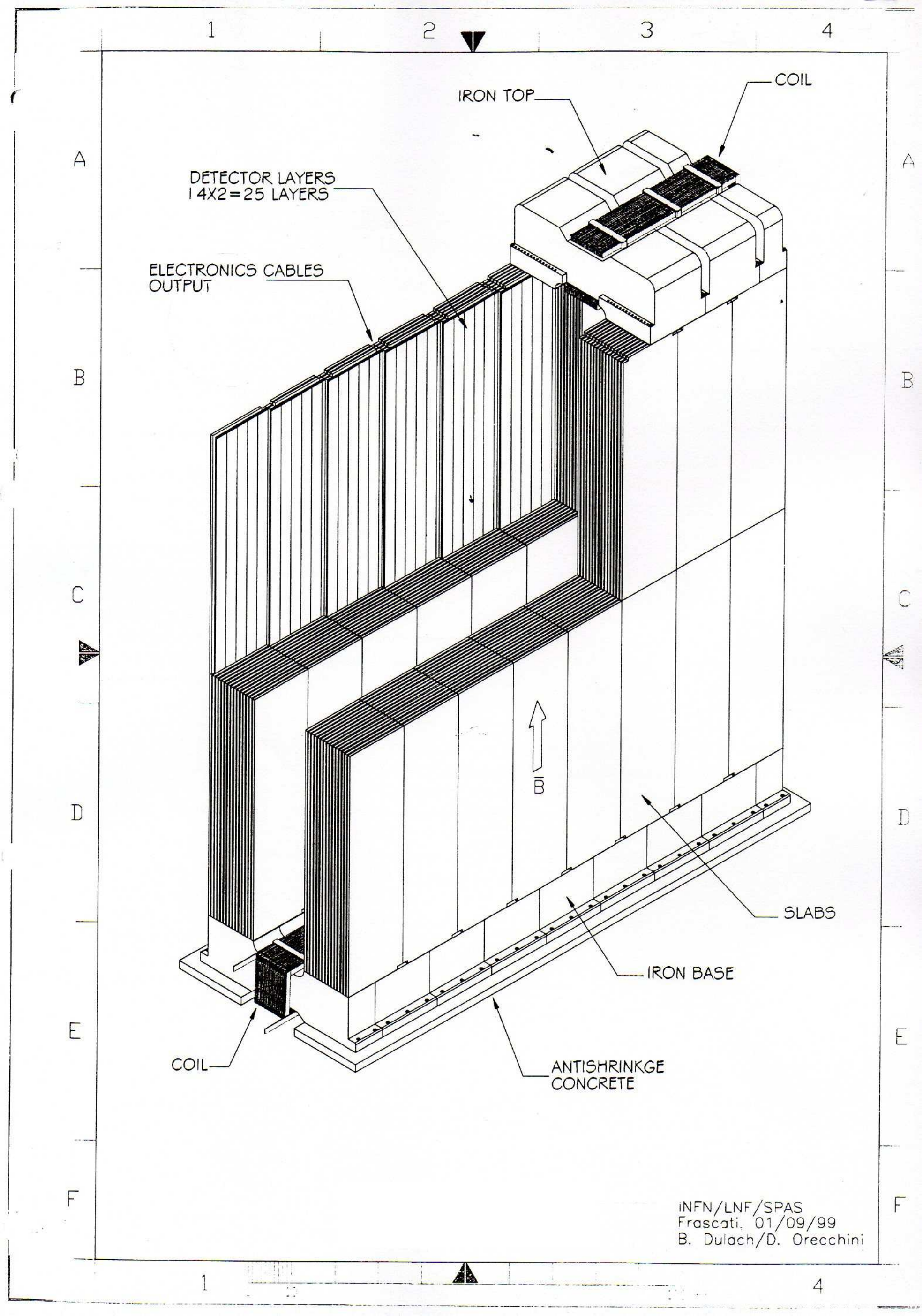}
\caption{OPERA magnet scheme.}
\label{fscheme}
\end{figure}

The two OPERA iron dipole magnets~\cite{bopera} can be taken as an example
for the design of NESSiE Spectrometers.
They are made of two vertical walls with rectangular cross 
section and of the top and bottom flux return yokes, as shown in Fig.~\ref{fscheme}.
Each wall is composed of 12 layers of iron slabs 5 cm thick, separated by
11 gaps, 2 cm thick, hosting RPC detectors.

The iron layers are made of 7 vertical slabs, for a total area of 
$(8.75 \times 8.2) \mbox{m}^2$. 
Including the top and bottom return yokes, the total height of the magnet is 
about 10 m and its length along the beam 2.85 m.
The magnet weight is around 1 $Kton$.

The slabs, the top and the bottom yokes are held together by means of screws,
while more screws (about $1/\mbox{m}^2$) are used to keep slabs straight with
spacers to ensure the thickness uniformity of the gaps hosting the detectors.

The Spectrometers are magnetized by coils located at the top and bottom
return yokes, as shown in Fig.~\ref{fscheme}.

The installation of each magnet was performed according to the following
time sequence:
\begin{itemize}
\item bottom yoke installation;
\item internal support structure construction;
\item iron/RPC layers installation (one plane in each arm at the time in order
to keep the structure balanced);
\item top yoke installation;
\item removal of the internal structure. 
\end{itemize} 

For the NESSIE experiment a very similar setup should be used to profit of the available detectors which have been designed to
obtain the maximum acceptance coverage and the strip signal configuration. The different geometrical requirements for NESSiE
corresponds to an equal width (7 vertical  slabs) and a smaller total height (5 rows of RPC's instead of 7) for the Far Spectrometer.
The ND will instead be constructed with 5 vertical slabs of 3.4 m high.
For the OPERA experiment a total of 336 slabs, (1.25*8.2) $m^2$ wide,
  were employed.  Such slabs can be cut into the shape needed for the present
  baseline NESSiE configuration, but a new production will also be needed.
   The top and bottom returns must be redesigned and the copper coils
  modified accordingly.

\section{Magnet Power Supplies and Slow Controls}\label{sec:mag-SC}
The following description is based on the past experiences with the OPERA iron magnets (see~\cite{bopera} and references therein).
Similar applications are foreseen for the present Proposal, with minor changes due to the magnetic field
in air not present in OPERA and the thicker magnet.

\subsection{Power Supply features}

The magnetomotive force to produce the B field is provided by DC power 
supplies, located on the top of the magnet. 
They are single-quadrant AC$\rightarrow$DC devices providing a maximum 
current of 1700 A and a maximum voltage of 20 V. 
As a single-quadrant power supply can not change continuously the sign of 
the voltage, the sign of the current is reversed by ramping down
the power supply and inverting the load polarity through a motorized breaker.
The power supplies are connected to the driving coil wound in the return 
yokes of the magnet by means of short flexible cables. 

\vskip 10pt
\underline{Ancillary systems}

\begin{itemize}
\item Coil:
The coil is made of $100 \times 20 mm^2$ copper 
(type Cu-ETP UNI 5649-711) bars. The segments are connected through bolts 
after polishing and gold-plating of the contact surface. Each coil has
20 turns in the upper return yoke connected in series to 
20 more turns in the bottom yoke.
The two halves are linked by vertical bars running along the arm. Rexilon 
supports provide spacing and insulation of the bars.
\item Water cooling: 
Water heat exchangers are positioned between these supports and the bars 
while the vertical sections of the coil are surrounded by protective plates 
to avoid accidental contacts. More than 160 junctions have been made for
each coil and the quality of such contacts was tested measuring the overall 
coil resistance during mounting.
Cooling ensures an operating temperature of the RPC detectors  lower than 
20$^0$ C.
\end{itemize}

\underline{Current status of the two OPERA power supplies}

Even if the 1$^{rst}$ power supply stops with a monthly frequency\footnote{No firm conclusions about  the cause 
of the failures has been reached after several years of investigations.}
the 2$^{nd}$ power supply is instead working without troubles.
Altogether the overall downtime of the two OPERA magnets is about 0.1\%.

A similar framework is foreseen for the NESSiE proposal. Both the Far and the Near magnets
will need a single power supply each.

\subsection{Monitored quantities for every magnet}
Here is the list of the foreseen values to be monitored:
\begin{itemize}
\item continuous measurement of the magnetic field strength by Hall probes or
pickup coils at magnet ramp down (not routinely done in OPERA)
\item electrical quantities:
\begin{enumerate}
\item current: check for maximum / minimum range
\item voltage: check for maximum / minimum range
\item ground leakage current: check for maximum range (not automatically 
done in OPERA, via web cam or onsite check)
\end{enumerate}

\item temperatures and cooling:

\begin{enumerate}
\item coil temperatures: check for maximum range
\item cooling water input temperature: check for maximum / minimum range
\item cooling water output temperature: check for maximum / minimum range
\item cooling water pressure, cooling water flux: check for maximum / minimum 
values (not automatically done in OPERA, via web cam or onsite check).
\end{enumerate}

\end{itemize}

\subsection{Slow Control}

The slow control system of the Spectrometer will be developed to master
all the hardware related: the magnet power supplies, the active detectors
and all the ancillary systems.

According to the experience acquired for the OPERA experiment~\cite{bib:opera-sc}
the system is organized in tasks and data structures developed to acquire
in short time the status of the detector parameters which are important for a
safe and optimal detector running.

The slow control should provide a set of tools which automatize specific
detector operation (for instance ramping up of the detector High Voltage
before the start of a physics run) and lets people on shift control the
different components during data taking.

Finally, the slow control has to generate alarm messages in case of a
component failure and react promptly, without human intervention, to preserve
the detector from possible damages. As an example the RPC High Voltages have
to be ramped down in case of any failure of the gas system.

A possible structure of the slow control can be organized as follows:
\begin{itemize}
\item one databases is the heart of the system; it is used to store both the
slow control data and the detector configurations.
\item the acquisition task is performed by a pool of clients, each serving
a dedicated hardware component. The clients are distributed on various Linux
machines and store the acquired data on the database.
\item the hardware settings are stored in the database and served through a
dynamic web server to all the clients as XML files.
A configuration manager gives the possibility to view and modify the hardware
settings through a Web interface.
\item a supervisor process, the Alarm Manager, which retrieves fresh data
from the database, and is able to generate warnings or error messages in case
of detector malfunctioning.
\item the system is integrated by a Web Server which allows controlling the
global status of data taking, the status of the various components, and
to view the latest alarms.
\end{itemize}

\section{Detectors for the Iron Magnets}\label{sec:rpc}

The Near and Far Spectrometers of NESSiE will be instrumented with large area 
detectors, ND and FD respectively, for precision tracking of muon paths
allowing for high momentum resolution and charge identification capability. According to 
Fig.~\ref{fig:displacement} a resolution of the order of $1\ mm$ is 
needed for the low momentum muons crossing the magnetic field in air. A resolution 
of about $1\ cm$ is instead enough for the large mass detectors 
that we plan to use within the iron slabs of the Spectrometers. In this Section we focus 
on the latter whilst several different options will be described in the 
next Section for the high resolution detectors which have to be used in the ``air" part.

Suitable active detectors for the ND and FD Iron Spectrometers are the RPC
- gas detectors widely used in high energy and astroparticle experiments~\cite{bib:origin} - because: 
\begin{itemize}
\item they can cover large areas;
\item are relatively simple detectors in terms of construction, flexibility in operation and use;
\item their cost is cheaper than other other large area tracking systems;
\item they have excellent time resolution;
\item and large counting rate power (in specific operational modes).
\end{itemize}
Furthermore a large part of the units to be used are already available. In fact by
considering the remainders of the OPERA RPC production, about  $1500\ m^2$ of RPC's can be used.

\subsection{RPC's detectors}\label{subsec:rpc-dect}
We plan to use standard bakelite RPC : two electrodes made of 2 mm plastic laminate kept 2 mm apart by polycarbonate spacers, of 1 cm diameter,
in a 10 cm lattice configuration. The electrodes will have high volume resistivity ($10^{11} - 10^{13} \Omega$ cm). Double coordinate read-out is obtained by 
copper strip panels. 
The strip pitch can be between 2 and 3.5 cm in order to limit the number of read-out channels over a large area. An optimization of the strip size and orientation 
(horizontal, vertical and tilted ones) is required in order to define the best performances of the Spectrometer (track reconstruction resolution and reduction of ghost hits).
RPC's are commonly used in streamer mode operation with a digital read-out as described in the following.

\subsection{Ancillary systems}
For the operation of the RPC's in the Spectrometer, ancillary systems are needed, namely:
\begin{itemize}
\item a Gas distribution system;
\item a High Voltage system with current monitoring
   performed by dedicated nano-amperometers designed by the Electronics
   Workshop of LNF. 
\item monitoring of several environmental/operational parameters (RPC temperatures, gas pressure and relative humidity).
\end{itemize}

\subsection{The Gas system}
Since the overall rate (either correlated or uncorrelated) is estimated to be low (see Sect.~\ref{sec:bck}),
standard gas mixtures for streamer   operation can be used, like for instance the one employed in the OPERA RPC's,
composed of Ar/tetrafluorethane/isobutane and sulfur-hexafluouride in the volume ratios: 75.4/20/4/0.6. 

   Different gas mixtures can be further investigated and an optimization 
is advisable in view of the DAQ system adopted (digital versus analog read-out) and the safety regulations in the experimental halls.
OPERA RPC's are flushed with an open flow system at 1500 l/h corresponding to
   5 refills/day. The installation of a recirculating system could also be considered,
  if the gas flow has to be increased to prevent detector aging.   
   
\subsection{The Tracking Detectors for the Near and Far Spectrometers}

The design and evaluation of the requirements for the two detectors are base on the RPC's  developed for the OPERA experiment at LNGS.
Each of its unit of RPC owns dimensions of $2904\times 1128\ mm^2$. 

The Near Spectrometer will have a magnetic field in air followed by a {\em calorimeter} of interleaved planes of iron and RPC.
In  the ND  RPC's will be arranged in planes of 2 columns and 3 rows for a total exposed surface of about 20 m$^2$. 
A total of 40 planes will instrument the Spectrometer and a total of 720 $m^2$ of detectors are thus required. 
Taking into account the RPC dimensions, this number corresponds to 240 units.

The Far Detector design will be very similar to the one conceived one for the magnetized Iron Spectrometers of the OPERA experiment.
The detector will consist of units arranged in 3 columns and 5 rows to form planes of  about 50 $m^2$. 
As for the Far magnet 40 planes of detectors will be interleaved with iron absorbers for a total of about 1800 $m^2$ (600 units).

\subsection{Production and QC tests}
OPERA RPC's, before the installation, underwent a full chain of quality control tests serialized in the following steps:
\begin{itemize}
\item mechanical tests (gap gas tightness and spacers adhesion);
\item electrical tests;
\item efficiency measurement with cosmic rays and intrinsic noise determination.
\end{itemize}
The setup, still partially available at Gran Sasso INFN Laboratories, was able to validate about 100 m$^2$ of RPC's per week.

\subsection{Costs}
Plastic laminate can be produced in Italy by Pulicelli, a company located near Pavia (the company is currently producing material for the CMS RPC
upgrade system).
The cost of plastic laminate is about 30 \euro/$m^2$.
The RPC chamber assembly can be done in Italy by the renovated General Tecnica company with an estimated cost of about 300 \euro/$m^2$.
The overall cost of a complete new RPC production is foreseen in about 1.5 M\euro.

\noindent As for the read-out strips,  the costs for 5000 m$^2$ is around 500 K\euro.

\section{Detectors for the Air Magnet}\label{sec:hpt}
\begin{figure}[htbp]
\begin{center}
\includegraphics[height=2.32in]{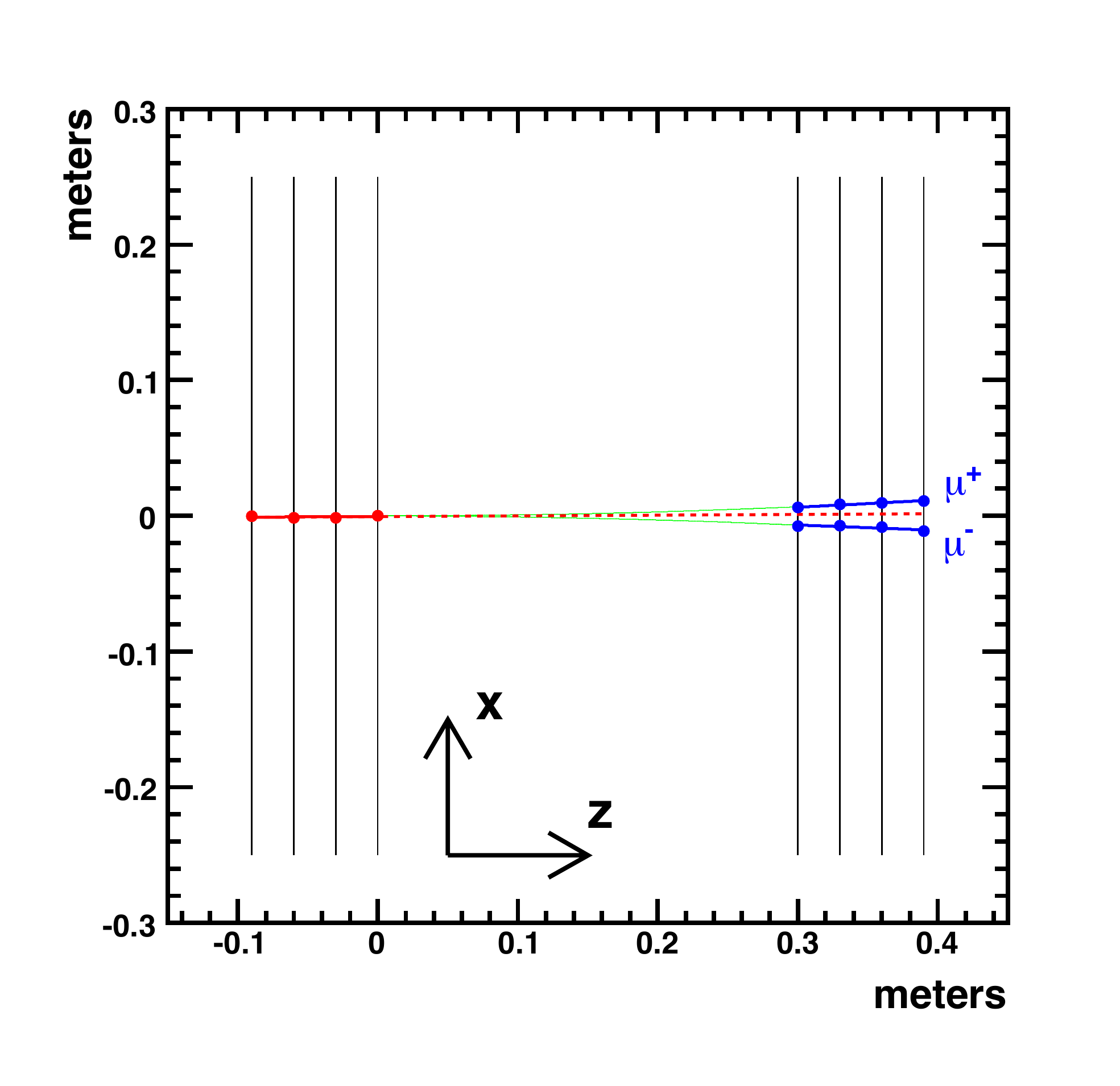}
\includegraphics[height=2.32in]{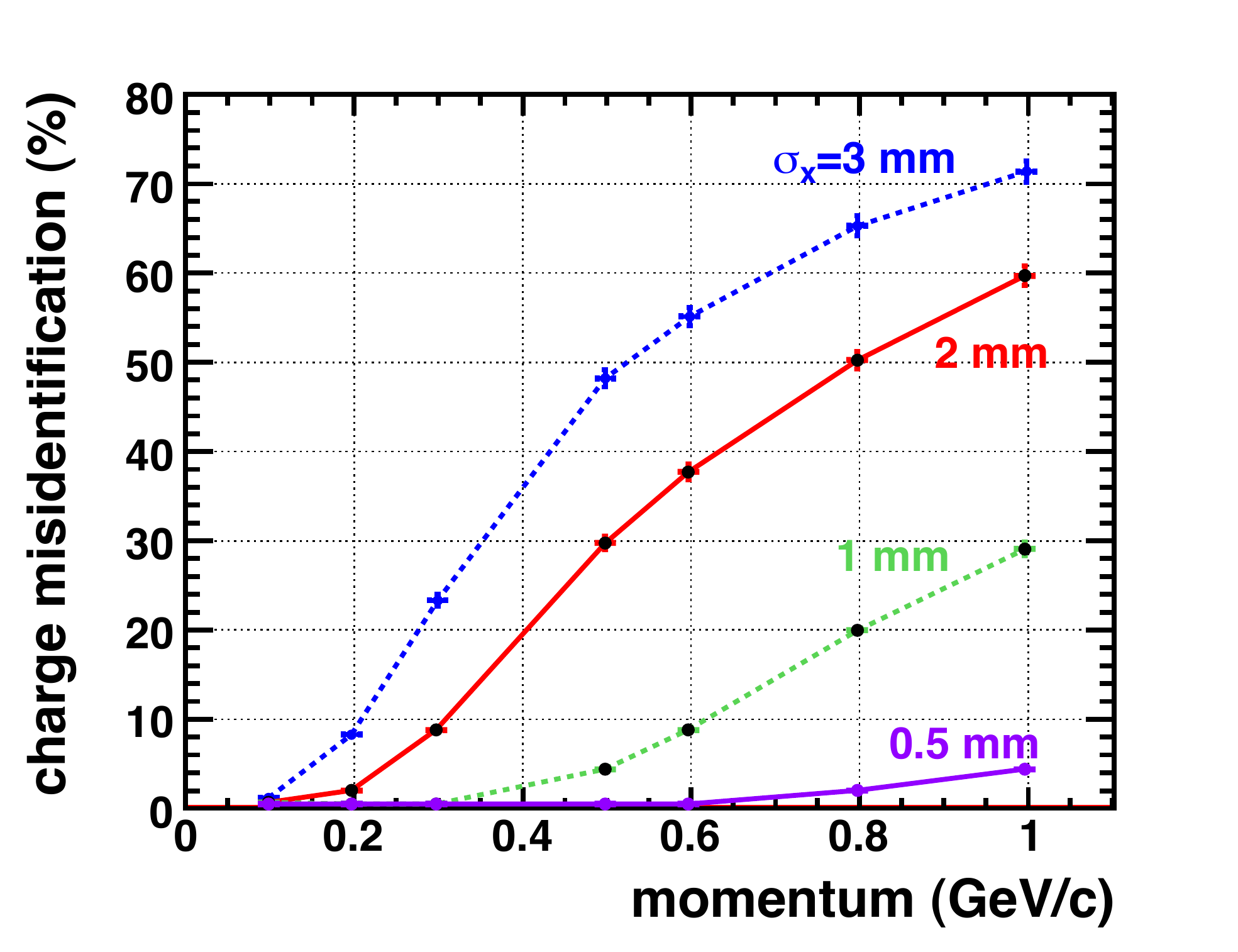}
\caption{Left: Display of a simulated event with momentum $p=500\ MeV/c$ in the air-core magnet.
         Right: the charge mis-identification percentage as a function of
         the muon momentum for different detector resolutions (incoming muons 
         perpendicular to the detector plane). The slope difference was
         used to measure the charge.} \label{fig:display}
\end{center}
\end{figure}

The air-core magnet (Sec.~\ref{sec:air_field}) will be used for the
charge identification of low momentum muons which requires precise measurements
of the muon path. A (4$+$4)-layers tracker was simulated (see left plot in 
Fig.~\ref{fig:display}) assuming a $0.25\ T$ magnetic field of in between. 
Track reconstruction is required only in the bending plane (XZ). 
The identification of the muon charge is optimized 
looking at the change of the track slope before and after the magnetic
field. Different detector resolutions were simulated (see right plot
in Fig.~\ref{fig:display}) and a resolution of the order of $1\ mm$ results as
the proper choice.

Several detector options for such precise measurements are briefly discussed in the following. 
More detailed MC simulations and eventual test-beams 
will be crucial to test the capability of the different detectors to reach
the required performances. The final choice on these precision trackers will
depend also on the cost, on the reliability and on the possibility of re-using
parts from previous experiments.

\subsection{Drift tubes (OPERA-like)}
\begin{figure}[htbp]
\begin{center}
\includegraphics[height=2.0in]{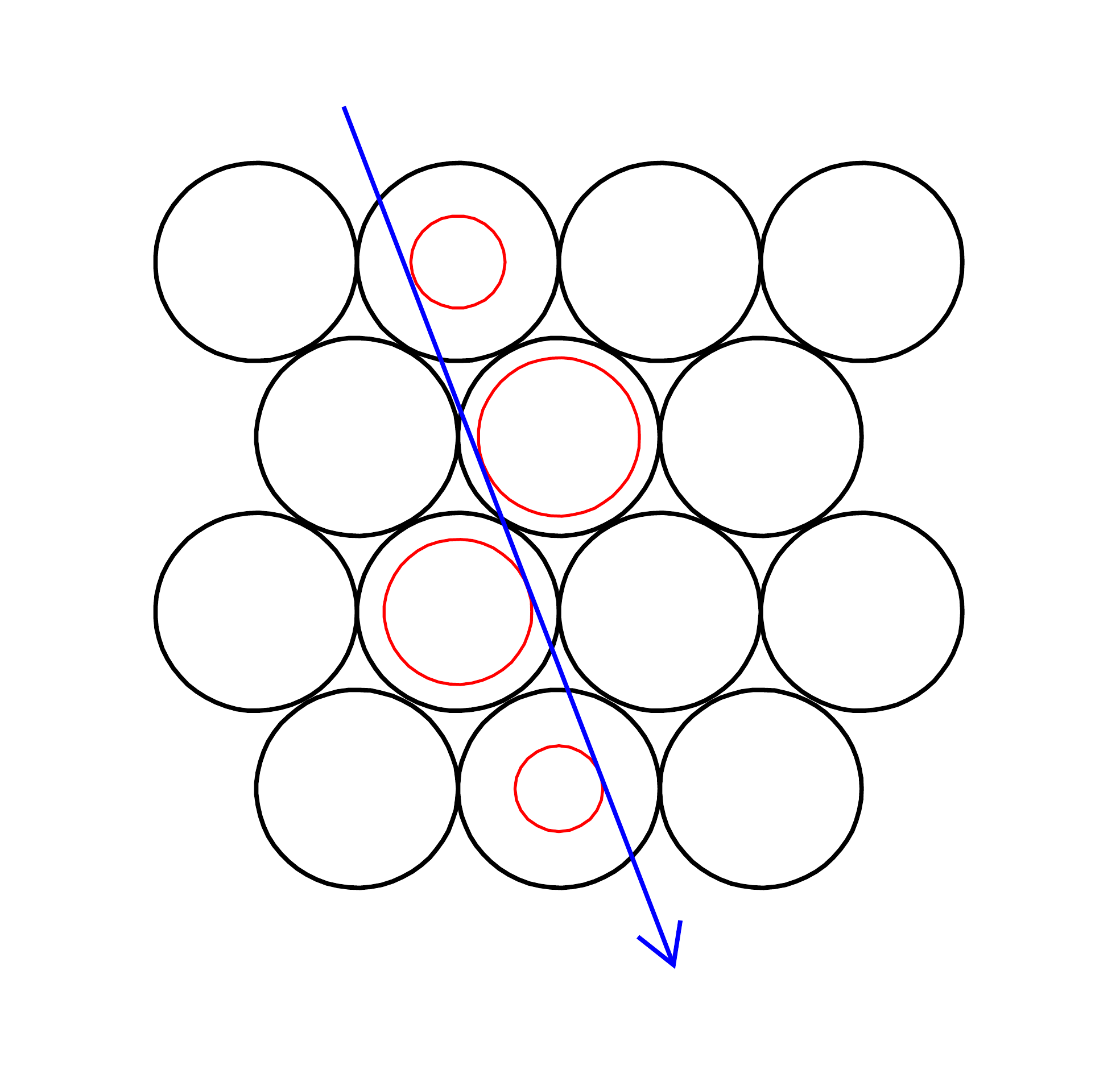}
\caption{Four layers of staggered drift tubes allow reconstructing the track
         without ambiguities.} \label{fig:drift}
\end{center}
\end{figure}

In the OPERA experiment the High Precision Trackers (HPT's) are mainly devoted to 
muon identification and charge measurement. HPT's are drift chambers (aluminum 
tubes each with a central sense wire) arranged in fourfold layers in order
to avoid ambiguities in the track reconstruction and to enhance the acceptance
(see example in Fig.~\ref{fig:drift}). In order to simplify the calibration 
procedures the wires have been located by using cover plates and the wire 
position is decoupled from the position of the tube. Thus a wire position 
accuracy has been achieved with a tolerance of $0.1\ mm$. 
In this configuration and assuming negligible inefficiency, the spatial single
tube resolution (rms) has been measured to be better than $0.3\ mm$.

The main features of these gas devices are mechanical robustness, absence of
glue to retain the gas quality, signal quality guaranteed by a Faraday cage. 
In the OPERA detector the HPT's are grouped in modules 
and the layers are staggered to optimize the acceptance and to minimize the 
left/right ambiguities. The tubes are filled with a gas mixture ($80\%\ Ar$,
$20\%\ CO_2$) and run at a pressure of $1005\ mbar$. TDC units with a Least
Significant Bit of $1.5\ ns$ are used to measure the drift time (the time 
spectrum ranges up to $1.6\ \mu s$ and TDC's cover a double range).

The large size ($8 \times 8\ m^2$) has been the main challenge in designing
the OPERA detector, with vertical tubes of $\sim 8\ m$. The HPT size is not
a problem in designing the NESSiE vertical precise trackers because $3\ m$ and
$5\ m$ (vertical dimensions) are enough for Near and Far detectors, respectively,
and the reduced size could allow the complete re-use of the OPERA HPT's.

\subsection{RPC's with analog read-out}

Resistive Plate Chambers (RPC's) are gas detectors~\cite{bib:origin} widely used
in high energy and astroparticle experiments. A single gas-filled gap delimited by bakelite
resistive electrodes is the simplest set-up commonly used in streamer mode and digital
read-out. We already introduced them in Sect.~\ref{subsec:rpc-dect}.
The most relevant features are the excellent time resolution and the high rate capability. 
Also the position resolution is very good, in particular conditions the centroid of 
the induced charge profile was determined~\cite{bib:120mum} with a FWHM resolution
of $\sim 0.12\ mm$. In the case of the NESSiE experiment such resolution is not necessary
and simpler and cheaper set-up can be used.

\begin{figure}[htbp]
\begin{center}
\includegraphics[height=3.0in]{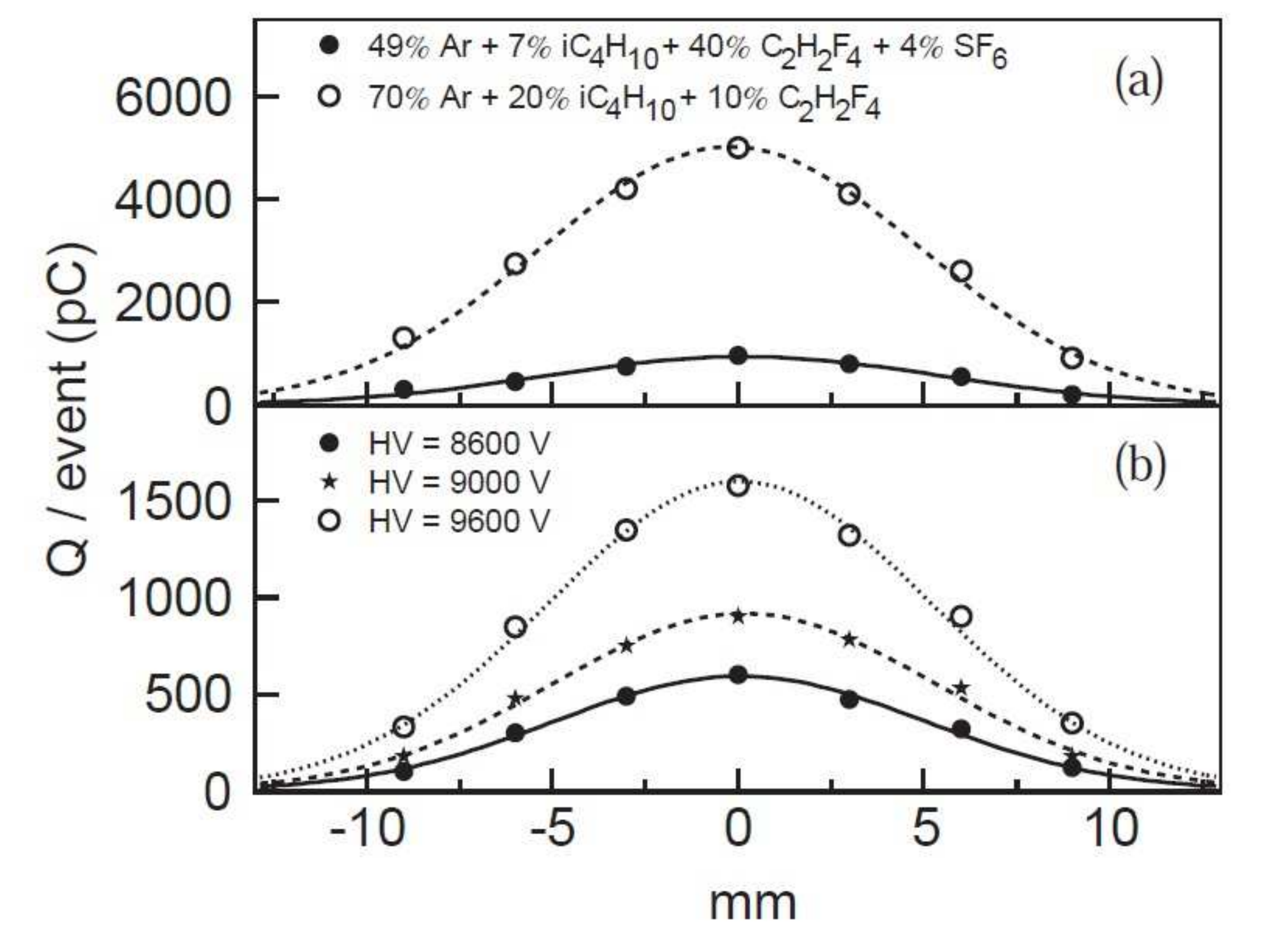}
\caption{Charge profile (normalized per event) observed with RPC working in streamer
mode: (a) for various gas mixtures at nominal running HV, (b) for different high voltages
with the gas mixture ($49\% Ar + 7\% iC_4 H_{10} + 40\% C_2 H_2 F_4 + 4\% SF_6$) foreseen 
for the ALICE detector. The curves are Gaussian fits of the data. The standard deviation 
of these distributions is about $5\ mm$ in all cases~\cite{bib:ALICERPC}.} \label{fig:charge}
\end{center}
\end{figure}

The analog read-out of RPC has been implemented in the last years~\cite{bib:argo} or
  variously proposed\footnote{The analog read-out of RPC's strips was for example 
  proposed as an alternative option for the Target Trackers of the
  OPERA experiment~\cite{bib:Autiero}.}.
With this technique by reading the total amount of charge induced on the strips 
detailed information can be obtained on the streamer charge distribution across the strips and 
better estimate of the track across the detector is thus achieved than in the digital case. 
The charge profile can be approximated by a Gaussian shape whose width ($\sim 5\ mm$) does not depend 
on the gas mixture and operating high voltage, unlike the total charge which is strongly dependent on them 
(see Fig.~\ref{fig:charge}).

A few $mm$ resolution in the charge position determination is obtained by choosing an adequate strip size.
Also the dynamic range is improved, allowing the detection of particles at a density 
of the order of $1000\ particles/m^2$.

\subsection{RPC's with digital read-out in avalanche regime}

Another possibility is to operate the RPC's in avalanche mode: the incoming particles
release primary charges followed by Townsend avalanches in the gas gap.

The bidimensional measurement of the avalanche position with $mm$-accuracy has been
verified in single-gap chambers~\cite{bib:ToFsystem}. As in avalanche regime the charge
cluster size results to be smaller by some order of magnitude with respect to the streamer mode operation
(it depends on HV and gas mixture), it is possible to achieve better space resolution by using smaller digital strips.
A higher rate capability is also attainable due to the lower amount of charge
delivered in the avalanche. The response of standard $2\ mm$ gap RPC's in streamer mode can not be higher than
$3\ MHz/m^2$, while in avalanche mode it can attain $30\ MHz/m^2$. 

It is also possible to use the RPC in {\em saturated avalanche} mode. The advantage is to
have a lower charge signal than in streamer mode but still high enough to remove the need for a pre-amplifier.
Analog read-out with an optimized strip size guarantees an adequate
position resolution.

Anyway the avalanche mode has the inconvenient of being more sensitive to temperature and
pressure variability due to the lower amount of charge produced.

\begin{figure}[htbp]
\begin{center}
\includegraphics[height=2.0in]{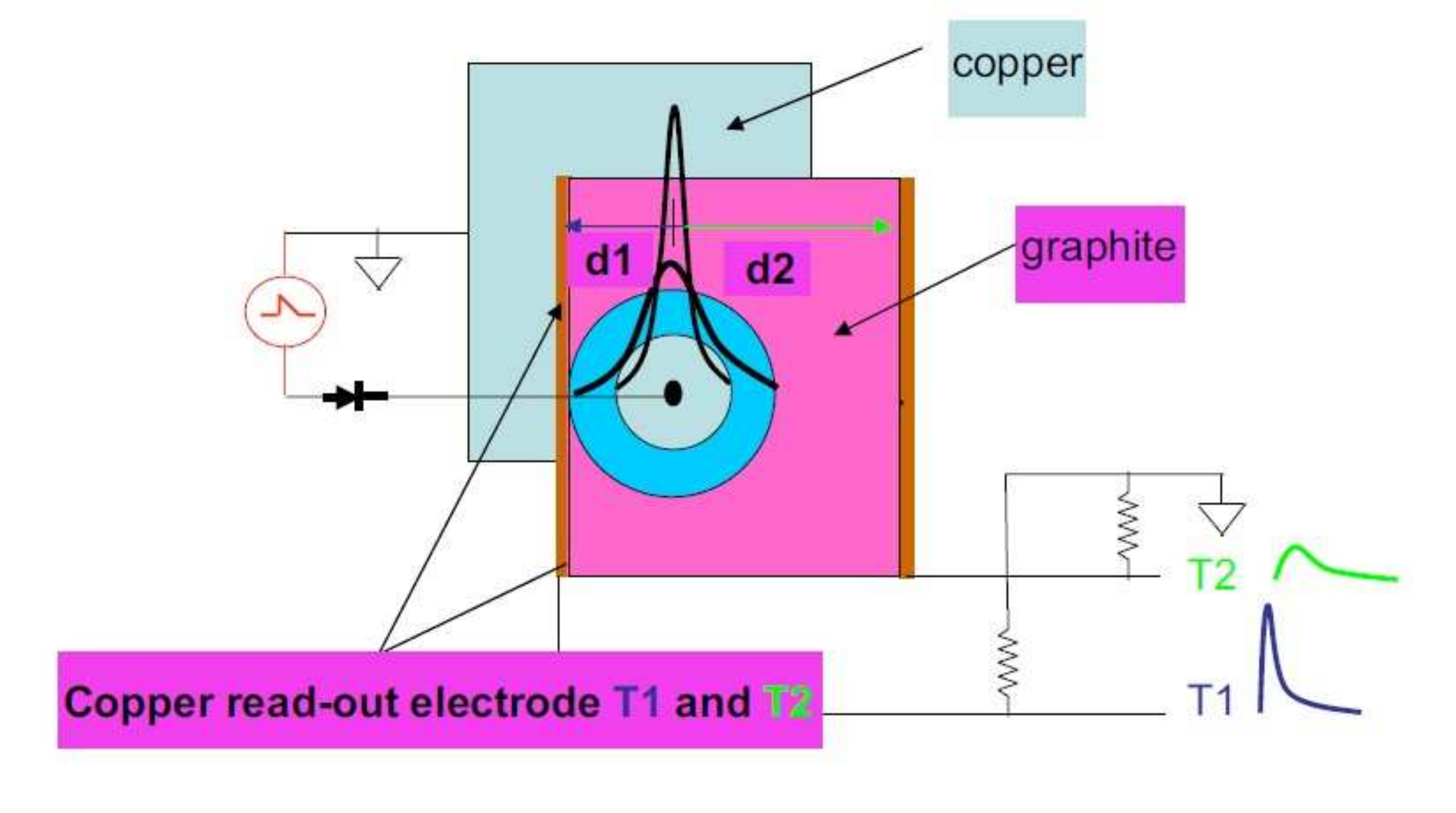}
\caption{Pick-up scheme in the graphite electrode. From~\cite{bib:carda}.} \label{fig:time}
\end{center}
\end{figure}

Finally, regarding the possibility to get high precision position measurement with RPC's
at a sustainable cost, a new procedure has been developed to determine the charge position
by timing measurements. Precision in the sub-millimeter range has been reached~\cite{bib:carda}. 
The method is based on the read out of the signal propagating in the graphite. The graphite
electrode, coupled to the ground reference of the detector read-out panels, is a distributed
capacitance-resistance system (see Fig.~\ref{fig:time}). The occurrence of a discharge in 
the gas produces a point-like perturbation of the steady potential distribution of the system.
The time behaviour of the perturbation can be described, in the approximation of infinite
electrode size, by a two-dimensional Gaussian distribution with time dependent variance 
and amplitude. The distance is related to the time (of the maximum) by a quadratic 
relation which can be used to measure the position from the time measurement. The sharpness of the 
maximum is strongly dependent on the distance. A distribution with a FWHM of $0.8\ mm$ 
has been reached.

\section{Backgrounds}\label{sec:bck}

Assuming neutrino fluxes as described in Sect.~\ref{sec:beam}
around 10-15 events per spill are expected in the Near Detector, a spill being 2.1 $\mu s$ long within a cycle of 1.2 $sec$.
For the Far Detector the rate of events is reduced by about a factor of 20.

The possible background rates are analyzed assuming a tracking system with single-gap RPC's.
We distinguish the uncorrelated background due to detector noise and local radioactivity
(dark counting rate) and the correlated background due to cosmic rays.

\subsection{Uncorrelated background}

The dark counting rate depends on detector features and ambient radioactivity. A typical
value for RPC as measured at sea level is $300\ Hz/m^2$. Therefore the expected rate per plane is
$\lambda_N = 300\ Hz/m^2 \times (3 \times 6\ m^2) \simeq  5.4\ kHz$ on the Near Spectrometer and 
$\lambda_F = 300\ Hz/m^2 \times (5 \times 9\ m^2) \simeq 13.5\ kHz$ on the Far Spectrometer. 
Assuming a read-out time window of $10\ \mu s$ we expect 0.05 (0.14) hits per RPC plane
per event on the Near (Far) detector. The number of fired strips will depend on the strip width
(in OPERA with $2.6$ and $3.5\ cm$ strip-wide the typical cluster size is $\sim 1.5$ strips).

The requirement of three-contiguous-planes coincidence in the beam-spill time (tipically 
$2\ \mu s$) makes the dark noise contribution to the trigger rate negligible
(see last column in Tab.~\ref{tab:volmas}).

\subsection{Cosmic Ray background}
The contribution of Cosmic Rays (CR) to the plane-by-plane background is similar to that 
discussed in the previous Section, but  CR events yield long tracks that constitute a potentially more
dangerous correlated background.

Assuming a trigger majority of at least 3 fired planes a cosmic 
particle can trigger the data-taking when it has enough energy to cross 2
iron slabs ($2 \times 5\ cm$). The CR flux at sea level is essentially due 
to muons (hard component of Extensive Air Showers) and these particles can
trigger the data-taking when their momentum exceeds $250\ MeV/c$.
Then the integrated vertical flux of muons is $J = 97\ Hz/m^2/sr$ at sea 
level~\cite{bib:cr}.

The integrated vertical flux of the soft component (electrons and positrons)
can be represented by $J(> E) = 0.22\ E^{-1.45}\ [Hz/m^2/sr]$, where $E$ is
in $GeV$. Taking into account this contribution and minor ones due to hadrons,
$J = 100\ Hz/m^2/sr$ will be used in the following calculations as integrated
vertical CR flux at sea level.

In the conservative hypothesis that the CR ray flux is isotropic above the
horizon and equal to the vertical flux, the total rate $\lambda_{RC}$ on a detector
shaped as a fully efficient box is

$$ \lambda_{RC} = \frac{\pi}{2} S_{tot} J $$

\noindent where $S_{tot}$ is the surface of the detector. The expected number of CR 
events in a time window of $2\ \mu s$ (the beam-spill time) is
reported in Tab.~\ref{tab:volmas}. They scale with 
the time window $T$ (by a factor $T/2\ \mu s$).
The data in Tab.~\ref{tab:volmas} conservatively ignore that more detailed
trigger conditions allow a significant reduction of the background.

\begin{table}[ht] \begin{center}
\begin{tabular}{|l|ccc|cc|} \hline
     & Dimensions            & Surface &  Total Iron Mass & CR events          & Dark noise ev.     \\
     & ($m\times m\times m$) & ($m^2$) &  ($ton$)         & in $2\ \mu s$      & in $2\ \mu s$      \\ \hline
Near & $6 \times 3 \times 5$ & 126     &  364             & $4 \times 10^{-2}$ & $2 \times 10^{-4}$ \\ 
Far  & $9 \times 5 \times 5$ & 230     &  846             & $7 \times 10^{-2}$ & $3 \times 10^{-3}$ \\ \hline
\end{tabular}
\caption{Dimensions and surfaces of the detectors, iron masses and estimated background events
         (CR and dark noise) for the Near and Far Spectrometers.} \label{tab:volmas}
\end{center} \end{table}

\section{Read-out, Trigger and DAQ}\label{sec:daq}

\subsection{DAQ overview}
The aim of the DAQ is to read the signals produced by the electronic detectors and to create a database of detected events. 
We recall in Table \ref{tab:prot-beam} the characteristics of the CERN PS primary proton beam. These are important inputs
to define properly the data acquisition and flow.

\begin{table}[htbp]
\caption{Characteristics of the PS primary proton beam.}
\label{tab:prot-beam}
\begin{center}
\begin{tabular}{|l|c|c|}
\hline
& PS Parasitic & PS Dedicated \\
\hline
Proton beam momentum & 20 GeV & 20 GeV \\
\hline
Protons per pulse & $2.6\times 10^{13}$ & $3\times 10^{13}$\\	
\hline
Number of bunches & 7 & 8 \\
\hline
Bunch length (4 sigmas) & 65 ns & 65 ns \\
\hline
Bunch spacing & 262 ns & 262 ns \\
\hline
Burst length & 1.8 $\mu$s & 2.1 $\mu$s\\
\hline
Maximum repetition rate & 1.2 s & 1.2 s\\
\hline
Beam energy & 84 kJ & 96 kJ\\
\hline
Average beam power & 70 kW & 80 kW\\ 
\hline
\end{tabular}
\end{center}
\label{default}
\end{table}

\vskip 10pt
The foreseen DAQ 
architecture is composed of three stages: 
\begin{itemize}
\item the front end electronics close to the detector (FEB)
\item the read-out interface which together with the trigger board control the read-out of the FEB
\item the Event Building which reconstructs events using standard workstations.
\end{itemize}
The whole event reconstruction is based on the time correlation of the channels, which depends on the accuracy of the electronic channels time stamping. 
A time resolution in the range $5\div 10\ ns$ is sufficient to correctly associate the different hits to the corresponding events. A common time 
reference with respect to the PS extraction time is used to time-stamp the data and to correlate the events recorded by the Spectrometer to the data 
recorded in LAr.  

\subsection{Data Flow}
The Far detector is designed with 15 RPC/plane $\times$ 20 planes/arm times 2 arms 
for a total of 600 RPC detectors, 3 $m^2$ each.
The Near detector is designed with 6 RPC/plane $\times$ 20 planes/arm times 2 arms for a 
total of 240 RPC's.

Assuming a strip size of 2.6 cm along Y and 3.5 cm along X for the Far
detector each plane will be equipped with $32\times 5=160$ horizontal strips plus 
$112\times 3 = 336$ vertical strips for a total number of about 500 electronic 
channels per plane. The total number of channels is therefore about 20,000.
For the Near detector the number of electronic channels will be about
300 per plane. The total number of channels is about 12,000.


Given the expected background rate due to the RPC single rate and the counts due to cosmic ray within the beam spill 
reported in Sect.~\ref{sec:bck}, the expected data rate is dominated by the beam related events. With a maximum PS beam intensity of 
$3\times10^{13}~\mbox{p.o.t.}$ about 10 to 20 events are expected in the Near detector. Assuming four hit strips per plane and at 
most $16~byte$ of data per hit (channel address, signal and time) the size of the event after zero suppression is expected to be $1~kbyte$.

\subsection{Front-End Electronics}
The role of the electronic read-out is to discriminate the signals coming from the RPC's strips and to 
record the signals above threshold. The start-of-burst signal will be used as a trigger and discriminated signals will thus be recorded 
during the whole burst duration. A time-stamp with a resolution of $5\div 10~ns$ will allow associating hits belonging to the same events.

 The electronic read-out of the RPC's could be developed according to the same 
design scheme adopted for the OPERA experiment, where they are operated in streamer mode and read-out by means of twisted pair cables;
these are in turn connected to
 special interconnection boards allowing the front-end electronics to be placed far from the detector.   The Front-end Boards (FEB) collect signals from
 RPC's strips and deliver them to a Controller Board (CB) acting as an interface  to the DAQ system.
 
   About 32,000 channels will be used to read-out 2,500 $m^2$ 
of RPC detectors with pickup strips of $2.5\ cm$ pitch in the vertical direction (orthogonal to the bending plane) and $3.5\ cm$ pitch in the horizontal 
direction (tracking without bending)\footnote{To get a spatial resolution of about $2\ mm$ strips with a pitch of $1\ cm$ have to be used for the 
measurement in the bending plane. In such case  the total number of channels increases of about a factor 2.}.  
Each FEB board will collect signals from 64 strips providing 
discrimination, time stamping and local buffering sized to store up to 40 events per burst.  
The local buffer will be read-out from the CB, during the inter-spill time, lasting 1.2 $s$.
Serial FEB data transfer to the CB will occur at a clock of 10 MHz, taking a few hundred microseconds. 
A new ASIC front-end chip will be designed to assign the time-stamp to each event and make it possible to record  multiple events in the same burst. 
  The possibility to have a double electronic chain, digital and analog, for each channel will be considered.  In the ASIC project
a 10-bit ADC with a multiplexing and sample-and-hold system allowing for the analog read-out  could be added.  
The design of this new ASIC chip can be provided by the INFN electronics CAD service.

 The cost estimation,  based on a production of 2,500 16-channel chips, amounts to 30 \euro/chip. The production cost of 700 Front-End Boards,
  including additional components and assembly, has been estimated to be about 50 \euro/board. The production of  100 VME Controller-Board has 
been estimated to be  300 \euro/board. Additional costs: cables and connectors 20 K\euro,  low voltage power supply  20 K\euro, crates 30 K\euro. 
 The total cost of the RPC's read-out electronics has been estimated to be 210 K\euro, taking into account 32000 digital channels  plus some contingency.
  
  \vskip 10pt
	
\subsection{DAQ}
The acquisition for each detector (FD/ND) is composed by Front-End
Boards (about 5 FEB per plane), Trigger Boards, Controller Boards
(1 CB per plane) and the Event Builder.
In order to acquire with a rate of 10 to 20 events/spill 
(each spill is 2.1 $\mu s$ long),
Front-End Boards should provide a pipeline system to save all the data
of a spill, provide a FAST\_OR signal for trigger purposes of granularity
in the range between 8 and 32 strips and perform the zero suppression. The
possibility of a digital or analog read-out must be envisaged.
The start of the pipeline should be given by the start of the spill and
the time resolution should be in the range 5 - 10 $ns$. 

The Trigger Boards should process the FAST\_OR signals from the Front-End
Boards and provide an external trigger condition to the Front-End Boards.
This signal could also be used as an external trigger for the High
Precision Tracker read-out ($t_0$ signal for the TDC).
A Trigger Bus should be implemented for the distribution of the trigger. 
Given the expected overall rate (beam events, cosmic, single rate) and
due to the Front-End Boards buffer a trigger signal based only
on the beam spill could be envisaged. 

The Controller Boards should read-out the signals from the Front-End Boards
and propagate them to the Event Building which is receiving data from the
Spectrometer and the High Precision Tracker systems. The Event Building is
using standard network protocols. 

Data read-out and Event Building should be performed in the time between 
two spills ($\sim 1.2 s$). The Event Builder should be based on 
standard commercial workstation. Data spying and monitoring process will
also be implemented at this level. \\
The correlation between Spectrometer and LAr data is achieved using
a common clock signal to time-stamp the events. Data merging is
performed offline at the reconstruction level.

\section{CERN Logistics}
The two experimental halls identified in the LAr proposal (see Sect. 3 of~\cite{Icarus-PS}) to host the Near and Far detectors,
B181 and B191, respectively, are well suited for the two Spectrometers too. The two halls have been chosen to be in the line of
the PS neutrino beam and to fit the size of the detectors. Some discussion is undergoing with CERN for the use of the chosen pit 
in hall B181, already exploited by some servicing. As the pit would not actually fit both the LAr and the Spectrometer detectors
and some excavation works will be anyhow needed, a different solution is in addition under study. A new pit in a new hall in front or
behind B181 may be constructed.

\begin{figure}[htbp]
\centering
\includegraphics[scale=0.6]{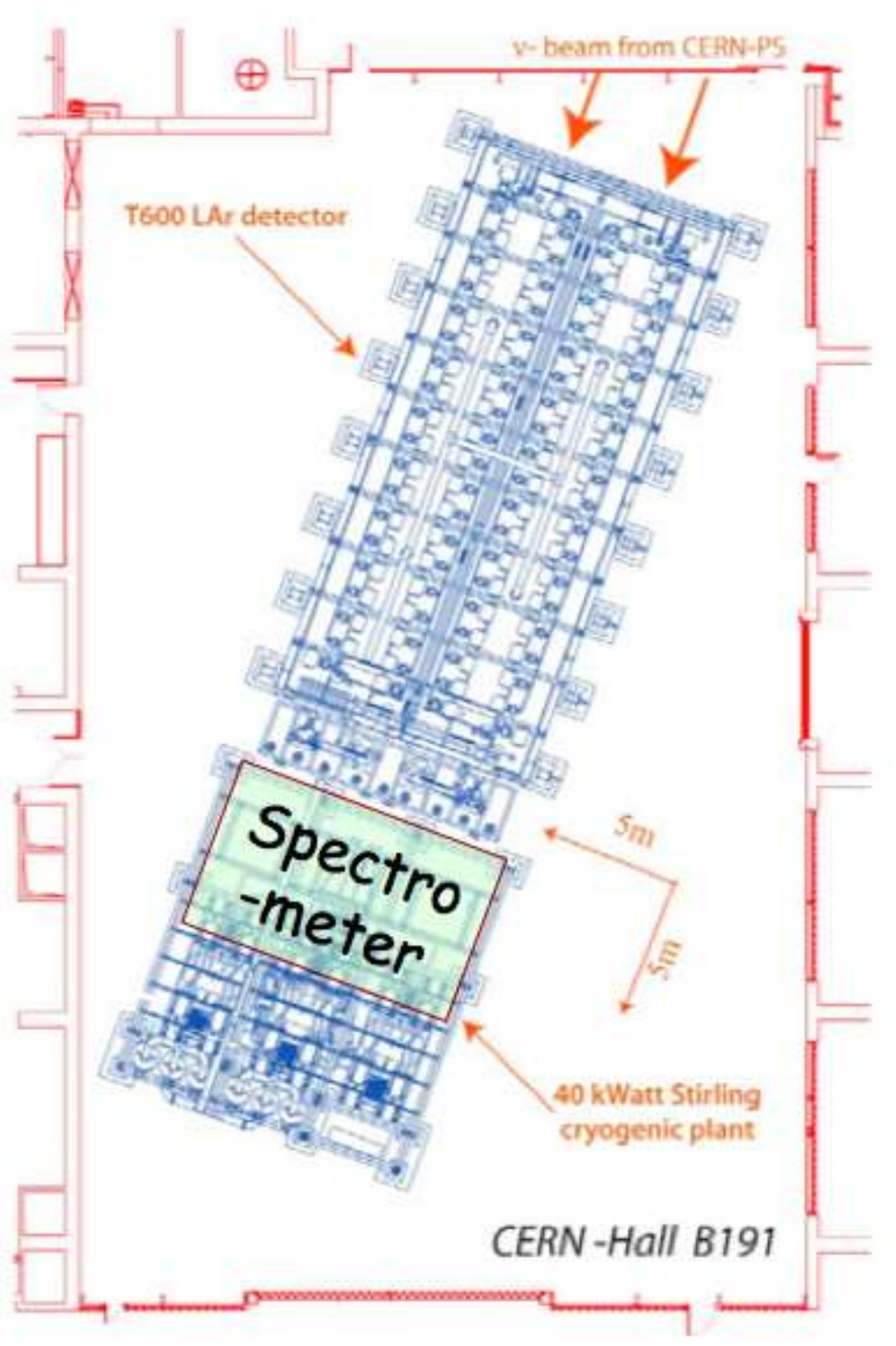} 
\caption{The general layout of the Hall B-191 at CERN that is supposed to host the Far detectors, LAr and Spectrometer.
The Spectrometer is placed behind the LAr tanks and overlays the cryogenic plant, which has to be moved e.g. on the right side.
The direction of the neutrino beam is also indicated on the top of the picture.}
\label{fig:hallB191}
\end{figure}

For the Far site (see Fig.~\ref{fig:hallB191}) the Spectrometer detector can easily be allocated behind the two Icarus tanks in case the cryogenic
equipment be allocated laterally. The total weight for the magnet will exceed 1 K$ton$, on an extension of less than 40 $m^2$.
Therefore the floor has to be eventually adapted. 
As described in Sect.~\ref{sec:mech-struc} the installation of the magnets has to follow a precise time sequence which 
has to be compatible with the LAr installation and contemporary allow for the closest positioning of Spectrometer and LAr.
Detailed engineering studies will take care of this subject.

\section{Schedule and Costs}

The choice of performing a  design study that is {\em reliable} under several aspects led to make conservative, well controlled 
and realistic options. In particular the choice to use detectors like the RPC ones and the development of dipole magnets
which had, at least partly, already been constructed and are in use, allows us to keep under control both the time schedule and the 
costs.

With respect to the schedule, which is reported in Tab.~\ref{tab:time}, it should be underlined that is based on the  
deep experience acquired with the OPERA
Spectrometers, built up from 2005 to 2006 under critical conditions\footnote{During the period 2005-2007 LNGS laboratory
underwent restoration and safety works enforced by Italian Government, following the temporary seal of May 2005.}.  

\begin{table}[h!]
\begin{center}
\begin{tabular}{ll}
Year(portion)&Action\\
\hline
2012&Design optimization\\
\hline
1$^{rst} $ half 2013&Define tenders/contracts \\
&                           Setting up Detectors Test-stands   \\
\hline
2$^{nd}$ half 2013&Mechanical Structure construction\\
&                          Detector production \\
&                          Start Detector test \\
&                           Magnet (Air) preparation \\
\hline
1$^{rst}$ half 2014&Start Magnet installation \\
&                          Start detectors installation \\
\hline
2$^{nd}$ half 2014&End installation \\
\hline
\end{tabular}
\end{center}
\caption{ \label{tab:time}
Tentative time schedule for the 2012-2014 years which will include the final optimization of the detectors,
their construction and installation at CERN.
}
\end{table}           

With respect to the cost estimate, the expenses needed for the major items are reported in Tab.~\ref{tab:costs}.

\begin{table}[h!]
\begin{center}
\begin{tabular}{ll}
Item&Cost (in M~\euro ) \\
\hline
Far & \\ \hline
Magnet & 2.5\\
Detectors & 1.0 \\
Strips & 0.5 \\
Data Acquisition & 1.0 \\ 
\hline\hline
Near & \\ \hline
Magnet & 2.0\\
Detectors & 0.5 \\
Strips & 0.3 \\
Data Acquisition & 0.5 \\ 
\hline\hline
Total& 8.3 \\
\hline
\end{tabular}
\end{center}
\caption{ \label{tab:costs}
Estimate of the costs of the major items.
}
\end{table} 

The partial re-use of systems developed for the OPERA experiment may be envisaged in case that experiment 
would not need them anymore at the time of the construction. We note that the OPERA Spectrometers have been
fully funded by INFN, except the Precision Trackers, which is therefore committed to their dismantling and entitled to 
possibly re-use as well. To this respect, it is clear that e.g. the iron raw material will be available sooner or later to INFN
which may then take it into account for company arrangements.

\section{Conclusions}
Existing anomalies in the neutrino sector  may hint to the existence of  one or more additional {\em sterile} neutrino families. 
We performed a detailed study of the physics case in order to set a Short-Baseline experiment at a refurbished 
CERN-PS neutrino beam, able to either prove or reject the existence of sterile neutrinos.

The already submitted proposal based on the technology of imaging in ultra-pure cryogenic Liquid Argon (LAr)
may suffer from some experimental limitations which we deem critical: the measurement of the muon
charge on event-by-event basis extended to the lowest achievable energy range would be mandatory.
Indeed the muon leptons from Charged Current (CC) (anti)-neutrino interactions play an important role in disentangling different phenomenological 
scenarios, provided their charge state is determined. Since CC muon events induced by the primary beam would be very abundant, 
a search based on muon appearance-disappearance will benefit of large statistics.

The best option in terms of physics reach and funding constraints is provided by two Spectrometers based on dipoles magnets
mostly in iron, at the Near and Far sites (located at 127 and 850 $m$ from the CERN PS neutrino beam, respectively), 
to be placed behind the LAr detectors. 

In order to measure the momentum and identify charge with high sensitivity in an extended energy range (from few hundred $MeV$ to above 5 $GeV$)
we complement the iron-core dipole by a magnetic field in air in a limited region just in front.

The selected detectors, mainly Resistive Plate Chambers, would exploit well known technologies and make it possible to re-use part of existing ones 
(should they become available; if not, it would imply an increase of the costs with no additional delay). 

\vskip 10pt

The Near and Far Spectrometers by complementing the capabilities of the LAr detectors will fully exploit the very rich 
and exciting physics information in case sterile neutrinos indeed exist. In fact the Spectrometers
by measuring the momentum and identifying the charge of the muons will provide valuable 
information in:
\begin{itemize}
	\item
measuring \numu disappearance in the full momentum range which is a key ingredient in 
rejecting the anomalies or measuring the whole parameter space of oscillations involving sterile neutrino;
	\item
measuring the neutrino flux in the Near detector, in the full muon momentum range, which 
is quite relevant to keep the systematic errors low.
\end{itemize}
The measurement of the muon charge will furthermore provide valuable information in: 
\begin{itemize}
\item
separating \numu from \nubarmu in the antineutrino beam where the 
\numu contamination is quite important. 
This measurement is critical to fully exploit the experimental capability of measuring a 
difference between \numunue and \nubarmunubare difference
in view of a possible signature of CP violation.
\end{itemize}

Results of our study are reported in detail in this proposal where our
experiment is identified with the acronym NESSiE (Neutrino Experiment with SpectrometerS in Europe).

\newpage

\bibliographystyle{my-h-physrev}
\bibliography{MyBib}

\begin{thebibliography}{100}
\bibitem{Icarus-PS}
LAr Proposal SPSC-M-773, submitted on March 9$^{th}$ 2011.
\bibitem{Aguilar:2001ty}
LSND Collaboration, A.~Aguilar {\em et~al.},
\newblock {\em ``Evidence for Neutrino Oscillations from the Observation of
  Anti-Neutrino(electron) Appearance in a Anti-Neutrino(muon) Beam"},
\newblock Phys. Rev.  {\bf D64}, 112007 (2001), hep-ex/0104049.
\bibitem{Giunti:2010zu}
C.~Giunti, M.~Laveder,
\newblock{ \em  ``Statistical Significance of the Gallium Anomaly''}
\newblock  Phys. Rev. {\bf C83}, 065504 (2011) arXiv:1006.3244.
\bibitem{Mueller:2011nm}
T.~Mueller {\em et~al.},
\newblock {\em ``Improved Predictions of Reactor Anti-Neutrino Spectra"},
\newblock Phys. Rev. {\bf C} , 1101.2663 (2011).
\bibitem{CDHS}
CDHS Collaboration, F. ~Dydak {\em et al.}, 
{\em ``A Search for $\nu_{\mu}$ Oscillations in the $\Delta m^2$ range 0.3$\div$90 $eV^2$"},
Phys. Lett. {\bf B134}, 281 (1984).
\bibitem{Giunti:2010wz}
C.~Giunti and M.~Laveder,
\newblock {\em ``Short-Baseline Electron Neutrino Disappearance, Tritium Beta
  Decay and Neutrinoless Double-Beta Decay"},
\newblock Phys. Rev.  {\bf D82}, 053005, (2010) arXiv:1005.4599.
\bibitem{Mention:2011rk}
G.~Mention {\em et~al.},
\newblock {\em ``The Reactor Anti-Neutrino Anomaly"},
\newblock (2011) hep-ex/arXiv:1101.2755.
\bibitem{Mangano:2006ur}
G.~Mangano, A.~Melchiorri, O.~Mena, G.~Miele, and A.~Slosar,
\newblock {\em ``Present bounds on the Relativistic Energy Density in the
  Universe from Cosmological Observables"},
\newblock JCAP  {\bf 0703}, 006 (2007), astro-ph/0612150.
\bibitem{Melchiorri-beyond}
A.~Melchiorrri,
{\em ``Sterile neutrino constraint from cosmology"},
\newblock {Talk at Beyond3nu, LNGS, 3-4/05/2011, see slide \# 23}.
\bibitem{Giunti:2009zz}
C.~Giunti and M.~Laveder,
\newblock {\em ``VSBL Electron Neutrino Disappearance"},
\newblock Phys. Rev. {\bf D80}, 013005, 0902.1992 (2009).
\bibitem{AguilarArevalo:2007it}
MiniBooNE Collaboration, A.~Aguilar-Arevalo {\em et~al.},
\newblock {\em ``A Search for Electron Neutrino Appearance at the $\Delta m^{2}
  \sim 1$eV$^{2}$ scale"},
\newblock Phys. Rev. Lett.  {\bf 98}, 231801 (2007), arXiv:0704.1500.
\bibitem{AguilarArevalo:2009xn}
MiniBooNE Collaboration, A.~Aguilar-Arevalo {\em et~al.},
\newblock {\em ``A Search for Electron Anti-Neutrino Appearance at the Delta m**2
  ~ 1-eV**2 Scale"},
\newblock Phys. Rev. Lett.  {\bf 103}, 111801 (2009), arXiv:0904.1958.
\bibitem{Kopp:2011qd}
J.~Kopp, M.~Maltoni, and T.~Schwetz,
\newblock {\em ``Are there Sterile Neutrinos at the eV scale?"},
\newblock arXiv:1103.4570, (2011).
\bibitem{Giunti:2010zs}
C.~Giunti and M.~Laveder,
\newblock {\em ``Hint of CPT Violation in Short-Baseline Electron Neutrino
  Disappearance"},
\newblock Phys. Rev. {\bf D82}, 113009 (2010), arXiv:1008.4750.
\bibitem{Armbruster:2002mp}
KARMEN Collaboration,  B.~Armbruster {\it et al.},
 {\em ``Upper limits for Neutrino Oscillations \pnubarmunubare from Muon Decay at rest"},
  Phys. Rev. {\bf D65 }  112001, (2002),  hep-ex/0203021.  
\bibitem{MiniBooNE-numu}
MiniBooNE Collaboration, A.~Aguilar-Arevalo {\em et~al.},
\newblock {\em ``A Search for Muon Neutrino and Anti-Neutrino Disappearance in MiniBooNE"},
\newblock Phys. Rev. Lett. {\bf 103}, 061802 (2009).
\bibitem{atmo}
M.~Maltoni, T.~Schwetz and J.W.F.~Valle,
{\em ``Cornering (3+1) Sterile Neutrino Schemes''},
Phys.Lett. {\bf B518} 252 (2001), hep-ph/0107150.
\bibitem{Alexander:1991vi}
LEP Collaboration (ALEPH, DELPHI, L3, OPAL),
Alexander, G. and others,
\newblock{\em ``Electroweak Parameters of the $Z^0$ Resonance and the
                        Standard Model"},
\newblock Phys. Lett., {\bf B276}, 247 (1992).
\bibitem{Maltoni:2007zf}
M.~Maltoni and T.~Schwetz,
\newblock {\em ``Sterile Neutrino Oscillations after first MiniBooNE results"},
\newblock Phys. Rev. {\bf D76}, 093005 (2007), 0705.0107.
\bibitem{Schreckenbach:1985ep}
K.~Schreckenbach, G.~Colvin, W.~Gelletly, and F.~Von~Feilitzsch,
\newblock {\em ``Determination of the Anti-Neutrino Spectrum from U-235 Thermal
  Neutron Fission Products up to 9.5-MeV"},
\newblock Phys.Lett. {\bf B160}, 325 (1985).
\bibitem{Karagiorgi}
G.~Karagiorgi,
\newblock {Talk at Laguna Meeting of 03/04/2011}.
\bibitem{bugey} BUGEY Collaboration, Y. Declais, H. De Kerret, B. Lefievre, M. Obolensky and A. Etenko {\em et al.},
{\em ``Study of Reactor Anti-Neutrino Interaction with proton at Bugey nuclear power plant''},
Phys. Lett. {\bf B338} 383 (1994).\\
Y. Declais, J. Favier, A. Metref, H. Pessard and B. Achkar {\em et al.},
{\em ``Search for Neutrino Oscillations at 15-meters, 40-meters, and 95-meters from a nuclear power reactor at Bugey''},
Nucl. Phys. {\bf B434} 503 (1995).
\bibitem{Chooz-final} CHOOZ Collaboration, M.  Apollonio {\em et al},
{\em ``Search for Neutrino Oscillations on a Long Base-Line at the CHOOZ nuclear power station''},
 Eur. Phys. J. {\bf C27}, 331 (2003), hep-ex/0301017.
\bibitem{Giunti:2011gz}
Giunti, C. and Laveder, M.,
\newblock {\em ``3+1 and 3+2 Sterile Neutrino Fits"},
\newblock arXiv:1107.1452 (hep-ph), 2011.
\bibitem{ref_GEN} C.Andreopoulos {\em et al.}, 
{\em ``The GENIE Neutrino Monte Carlo Generator}, Nucl. Instrum. Meth. {\bf A614}, 87 (2010). 
\bibitem{nomad} NOMAD Collaboration, P. Astier {\em et al.},
{\em ``Search for \numu $\rightarrow$ \nue Oscillations in the NOMAD Experiment''}, 
Phys. Lett. {\bf B570} 19 (2003), hep-ex/0306037.
\bibitem{Akhmedov:2011zz}
E.~K.~.Akhmedov, T.~Schwetz,
\newblock{ \em ``New MiniBooNE Results and non-standard Neutrino Interactions''},
\newblock Nucl. Phys. Proc. Suppl.  {\bf 217} 217 (2011).
\bibitem{Giunti:2010jt}
C.~Giunti, M.~Laveder,
\newblock{``Short-Baseline \nubarmu $\rightarrow$  \nubare Oscillations''},
\newblock  Phys. Rev. {\bf D82} 093016 (2010), arXiv:1010.1395.
\bibitem{Giunti:2010uj}
C.~Giunti, M.~Laveder,
\newblock{\em ``Large Short-Baseline \nubarmu Disappearance''},
\newblock  Phys. Rev.  {\bf D83}  053006 (2001) arXiv:1012.0267.
\bibitem{BEBC} M. Baldo-Ceolin {\em et al.}, CERN/PSCC/80-130, PSCC/P33, 30-10-1980;\\
C. Angelini et al. [PS180 Coll.], Phys. Lett. {\bf B179} 307 (1986).
\bibitem{I216_99} M. Guler {\em et al.} (I216/P311 Proposal) 1999, {\em ``Search for \nue Oscillation at the CERN PS''}, Tech. Rep. CERN-SPSC/99-26, SPSC/P311.
\bibitem{I216_97} I216/P311 LoI, CERN/SPSC/97-21.
\bibitem{fluka}
G.~Battistoni {\em et al.},
{\em ``The FLUKA code: Description and Benchmarking''},
AIP Conf.\ Proc.\  {\bf 896}, 31 (2007). 
\bibitem{mySPLarticle} A.~Longhin. {\em ``A new Design for the SPL-Fr\'ejus Super-Beam''}, arXiv:1106.1096.
\bibitem{roberta} R.~Santacesaria, private communication.
\bibitem{numi}
K.~Anderson {\em et al.}, FERMILAB-DESIGN-1998-01(1998). 
\bibitem{PSrefur} http://info-psnf.web.cern.ch/info-PSNF/Presentations.
\bibitem{Edda}E.~Gschwendtner, {\em ``The CNGS, Operation and Perspectives''}, CERN, presentation at SBNW11, 12-14 May 2011, Fermilab.
\bibitem{scott1949} W.T.~Scott, {\em ``Correlated Probabilities in Multiple Scattering''}, Phys. Rev. {\bf 76} 212 (1949).
\bibitem{gluckstern1963} R.L.~Gluckstern, {\em ``Uncertainties in Track Momentum and Direction due to Multiple Scattering and Measurement Errors''}, 
Nucl. Instr. and Meth. {\bf 24} 381 (1963).
\bibitem{innes1993} W. Innes, {\em ``Some Formulas for Estimating Tracking Error''}, Nucl. Instr. and Meth. in Phys. Res. {\bf A} 238 (1993).
\bibitem{comsol} COMSOL is a commercial product, see www.comsol.com
\bibitem{sciboone} MiniBooNE-SciBooNe Collaborations, K. B.M. Mahn {\em et al.},
{\em ``Dual Baseline Search for Muon Neutrino Disappearance at $0.5 eV^2 < \Delta m^2 < 40 eV^2$''},
arXiv:1106.5685 [hep-ex], submitted to PRL (2011).
\bibitem{ccfr} CCFR Collaboration, I. E. Stockdale and A. Bodek, F. Borcherding, N. Giokaris, K. Lang, K. {\em et al}, 
{\em ``Limits on Muon Neutrino Oscillations in the Mass Range $55\ eV^2 < \Delta m^2 < 800\ eV^2 $''},
Phys.Rev.Lett. {\bf 52} 1384 (1984).
\bibitem{bopera}{The OPERA collaboration, {\em ``The OPERA experiment in the CERN 
to Gran Sasso neutrino beam''}, JINST {\bf 4} 4018 (2009)}.
\bibitem{bib:opera-sc}
A. Bergnoli {\em et al.},  {\em `The OPERA Spectrometer Slow Control System"}, IEEE
Trans. Nucl. Sci. {\bf 55} 349 (2008).
\bibitem{bib:origin} R. Santonico, R. Cardarelli, {\em ``Development of Resistive Plate Counters''}, Nucl. Instr. and Meth. {\bf A187} 377 (1981). \\
                     R. Santonico, R. Cardarelli, {\em ``Development of Resistive Plate Counters''}, Nucl. Instr. and Meth. {\bf A263} 20 (1988).
\bibitem{bib:120mum} E. Ceron Zeballos et al, {\em ``Resistive Plate Chambers with secondary electron emitters and Microstrip Readout''}, 
Nucl. Instr. and Meth. {\bf A392} 150 (1997).
\bibitem{bib:argo}      G. Aielli et al (ARGO-YBJ Collaboration), {\em ``Calibration of the RPC charge readout in the ARGO-YBJ experiment''}, 
                  Nucl. Instr. and Meth. {\bf A} (2010), doi:10.1016/j.nima.2010.09.066.
\bibitem{bib:Autiero} D. Autiero et al, {\em ``Design and Prototype Tests of the RPC system for the OPERA spectrometers''}, RPC 2001 Workshop, 
   Coimbra, November 2001.
\bibitem{bib:ALICERPC}  R. Arnaldi {\em et al.}, {\em ``Performances of a Prototype for the ALICE Muon Trigger at LHC''}, Nucl. Instr. and Meth. {\bf A490} 51 (2002).
\bibitem{bib:ToFsystem}  A. Blanco et al, RPC 2001 Workshop, Coimbra, November 2001.\\
                        A. Blanco et al, {\em ``Development of Large Area and of Position-sensitive timing RPCs''}, Nucl. Instr. and Meth. {\bf A478} 170 (2002).
\bibitem{bib:carda}     R. Cardarelli et al, {\em ``Track Resolution in the RPC chamber''}, Nucl. Instr. and Meth. {\bf A572} 170 (2007).
\bibitem{bib:cr} P.K.F. Grieder "Cosmic Rays at Earth", Elsevier (Amsterdam, 2011).



\end{thebibliography}



\end{document}